# Unmasking the resolution–throughput tradespace of focused-ion-beam machining

Andrew C. Madison[1] John S. Villarrubia[1] Kuo-Tang Liao[1,2] Craig R. Copeland[1] Joshua Schumacher[3]
Kerry Siebein[3] B. Robert Ilic[1,3] J. Alexander Liddle[1] and Samuel M. Stavis[1,*]

[1.] National Institute of Standards and Technology, Gaithersburg, MD, 20899, United States
[2.] Maryland Nanocenter, College Park, MD 20740, United States
[3.] CNST NanoFab, National Institute of Standards and Technology, Gaithersburg, MD, 20899, United States
[*] e-mail: samuel.stavis@nist.gov

## Abstract
Focused-ion-beam machining is a powerful process to fabricate complex nanostructures, often through a sacrificial mask that enables milling beyond the resolution limit of the ion beam. However, current understanding of this super-resolution effect is empirical in the spatial domain and nonexistent in the temporal domain. This article reports the primary study of this fundamental tradespace of resolution and throughput. Chromia functions well as a masking material due to its smooth, uniform, and amorphous structure. An efficient method of in-line metrology enables characterization of ion-beam focus by scanning electron microscopy. Fabrication and characterization of complex test-structures through chromia and into silica probe the response of the bilayer to a focused beam of gallium cations, demonstrating super-resolution factors of up to $6 \pm 2$ and improvements to volume throughput of at least factors of $42 \pm 2$, with uncertainties denoting 95 % coverage intervals. Tractable theory models the essential aspects of the super-resolution effect for various nanostructures. Application of the new tradespace increases the volume throughput of machining Fresnel lenses by a factor of 75, which we introduce as projection standards for optical microscopy. These results enable paradigm shifts of sacrificial masking from empirical to engineering design, and from prototyping to manufacturing.

## Introduction

Nanoscale milling with a focused ion beam enables the direct formation[1] of nanostructures with complex surfaces in three dimensions. This process is broadly applicable to the fabrication of electronic[2], mechanical[3], photonic[4], and fluidic[5] devices, and is becoming increasingly important to prepare samples for materials characterization[6] and biological imaging[7]. Despite the widespread use of the focused ion beam as a machine tool, its limits of resolution and throughput remain unclear, masking its ultimate utility.

Understanding the limits of focused-ion-beam machining begins with focus. A narrow focus of an ion beam with a low current enables patterning of fine features, whereas a high current enables rapid milling of deep features across wide areas. Although the de Broglie wavelength of gallium cations in a conventional focused-ion-beam system is at the femtometer scale, aberrations intrinsic to electrostatic lenses and space-charge effects from Coulomb interactions broaden the focus of such an ion beam into an approximately Gaussian profile at the nanometer scale[8]. For dielectric substrates, exposure to an ion beam results in electrostatic charging and repulsion of ions from the surface, further complicating the process of focusing the ion beam and degrading lateral and vertical resolution. The lateral extent of the ion-beam profile generally follows a power-law dependence on ion-beam current, with effective lateral resolution requiring characterization[9] for specific ion sources and substrate materials. Accordingly, reducing the lateral extent of the focused ion beam, simply by decreasing the current, generally improves lateral resolution and decreases the accumulation of charge. However, a lower current also results in a longer time to mill through the vertical range, prolonging any drift of the system and wear of the aperture that limits current. These issues can degrade lateral resolution[10] to an intolerable extent, while operating costs can accrue to an unsustainable level. Depending on the materials and dimensions, a single workpiece can take tens of hours to machine.

This inherent coupling of lateral resolution and volume throughput yields a tradespace of spatial and temporal constraints. Although an explicit description of this tradespace is absent from the literature, there is a general tendency to consider the focused ion beam as being slow, costly, and more useful for prototyping than for manufacturing, with excursions into the latter domain often relying on pattern replication[5d, 11]. In contrast, previous studies have described a tradespace for lithographic processes, finding a power-law relation between lateral resolution and areal throughput[12] extending into the commercial domain of manufacturing[13]. These studies explicitly excluded the focused ion beam from analysis, for several reasons. First and foremost, direct milling conventionally involves a volume throughput rather than an areal throughput, without a separate process of pattern transfer. Second, focused beams of metal ions implant remnants of the machine tool in the workpiece, requiring additional study of the milling process and material responses. Third, conventional processes at the state of the art of focused-ion-beam machining are commercially viable for only a few products of high value, such as editing circuits[14] and repairing photomasks[15].

Constraints on lateral resolution motivate the use of sacrificial films[4d, 5b, 16] to dissipate charge from insulating substrates[17], protect workpieces from redeposition and damage along the edges of device features[5b, 16b], enable super-resolution of such edges down to the scale of one to ten nanometers[5b], and prepare samples for microscopy[6-7] and tomography[10b]. In each of these applications, a sacrificial film functions as a physical barrier that masks the workpiece from the diffuse periphery, or tail, of the focused ion beam. This masking process effectively reduces the radius of the ion beam near boundaries of the milling pattern and yields edge profiles that are narrower than, and differ in shape from, those resulting from direct milling. In this way, sacrificial masking manifests as a super-resolution effect, in which a multiplicative factor quantifies the reduction of edge width and corresponding improvement from a conventional resolution to an unconventional resolution. Despite its common use, current understanding of spatial masking is incomplete, beginning with material selection. Milling rates of polymeric masks[16b, 16c] exceed those of common substrates such as silica[18], resulting in soft masks with low values of physical selectivity, a fundamental property of the mask–substrate bilayer. Metallic masks[4d, 5b, 16a, 16d], such as aluminum, platinum, and chromium, have lower milling rates but tend to be polycrystalline, with grain sizes ranging from tens to hundreds of nanometers. Additionally, milling rates depend on grain orientation[19], transferring surface and line-edge roughness into the substrate[16d], degrading lateral and vertical resolution, and potentially limiting the accuracy of models of ion transport[20] to understand the bilayer response.

Reductions of electrostatic charging and edge defects are obvious advantages of a conductive sacrificial mask, and the super-resolution effect implies some advantage. However, it is nonobvious that a sacrificial mask presents any fundamental advantage to super-resolve nanostructure edges, in comparison to simply reducing ion-beam current to improve lateral resolution. To answer this open question, which is central to our study, we unmask the resolution–throughput tradespace of focused-ion-beam machining, discovering that a sacrificial mask enables patterning with the effective resolution of a low value of ion-beam current and the volume throughput of a high value of ion-beam current. This dramatic improvement demonstrates that the dominant advantage of the super-resolution effect occurs in the temporal domain, rather than in the spatial domain.

To substantiate this surprising finding, we perform the first comprehensive and systematic study of this topic, integrating five new concepts for the most widely available type of electron–ion beam system (**Figure 1**). First, we introduce a chromia, $Cr_2O_3$, film for electron microscopy and sacrificial masking of silica, characterizing the advantageous properties of this multifunctional masking material (Figure 1a). Second, we demonstrate a new method for in-line metrology of effective lateral resolution, enabling new efficiency and reproducibility (Figure 1b). Third, we pattern complex nanostructures through chromia and into silica, measuring and simulating responses of the bilayer to ion exposure, and quantifying vertical resolution and lateral super-resolution (Figure 1c). Fourth, we develop a model of the super-resolution effect, achieving good agreement with experimental data, simulating the super-resolution of conventional test-structures, and elucidating the temporal efficiency of the process (Figure 1d). Fifth, we test our model predictions in an application of sacrificial masking to the fabrication of arrays of Fresnel lenses, increasing volume throughput by a factor of 75 relative to a single, functionally equivalent lens that we fabricate directly into silica with a low value of ion-beam current (Figure 1e). These results demonstrate paradigm shifts of focused-ion-beam machining through sacrificial masks, from empirical to engineering design of processes for device fabrication and sample preparation, and from prototyping single devices to manufacturing device arrays and otherwise surpassing the volume throughput limit of conventional resolution.

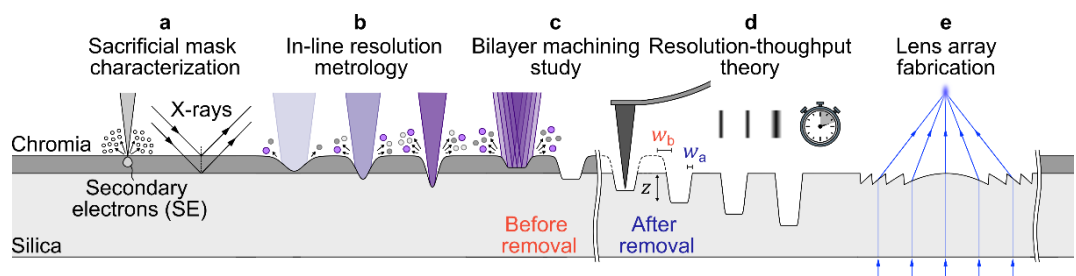

**Figure 1**. Experimental overview. (a) Materials characterization reveals the structure and solid state of a sacrificial mask of chromia on a silica substrate. (b) In-line resolution metrology enables reproducible and optimal focus of a beam of gallium cations. (c) Study of the bilayer structure before and after removal of the chromia elucidates milling responses. (d) Theory predicts spatial and temporal advantages of masking the periphery of the ion beam with a sacrificial film. (e) Application of the resolution–throughput tradespace increases volume throughput of the fabrication of Fresnel lens arrays.

## Results and Discussion
### Materials characterization
The semiconducting behavior[21], nanometer roughness[22], gigapascal hardness[22b, 23], and high selectivity in chemical etching of chromia against silica make this material a good candidate for our application. We form a silica film by thermal oxidation of a silicon substrate, over which we form a chromia mask by sputter deposition[24] (Note S1). To comprehensively characterize our bilayer and provide input quantities for theoretical models, we combine atomic force microscopy, scanning electron microscopy, transmission electron microscopy and X-ray diffraction (Note S2).

*Surface Structure*
The silica surface has a subnanometer root-mean-square roughness and sparse asperities with a height of 1.6 nm ± 0.2 nm (Figure S1). We report all uncertainties as 95 % coverage intervals, or we note otherwise. Atomic force micrographs provide estimates of the lower bound[25] of root-mean-square roughness of 0.3 nm ± 0.2 nm for silica and 0.6 nm ± 0.2 nm for chromia (**Figure 2**, Figure 2a). X-ray diffractometry data are generally consistent with these measurements, yielding estimates of root-mean-square roughness of 0.4 nm ± 0.4 nm and 1.5 nm ± 0.4 nm (Figure 2g, Table S1). Scanning electron micrographs of the chromia surface indicate negligible charging during imaging and show lateral roughness ranging in scale from 10 nm to 40 nm (Figure 2b). We report quantitative ranges of experimental data as 95 % coverage intervals. Discrete Fourier transforms of atomic force micrographs (Figure 2d) and of scanning electron micrographs (Figure 2e) indicate an aperiodic structure of the chromia surface.

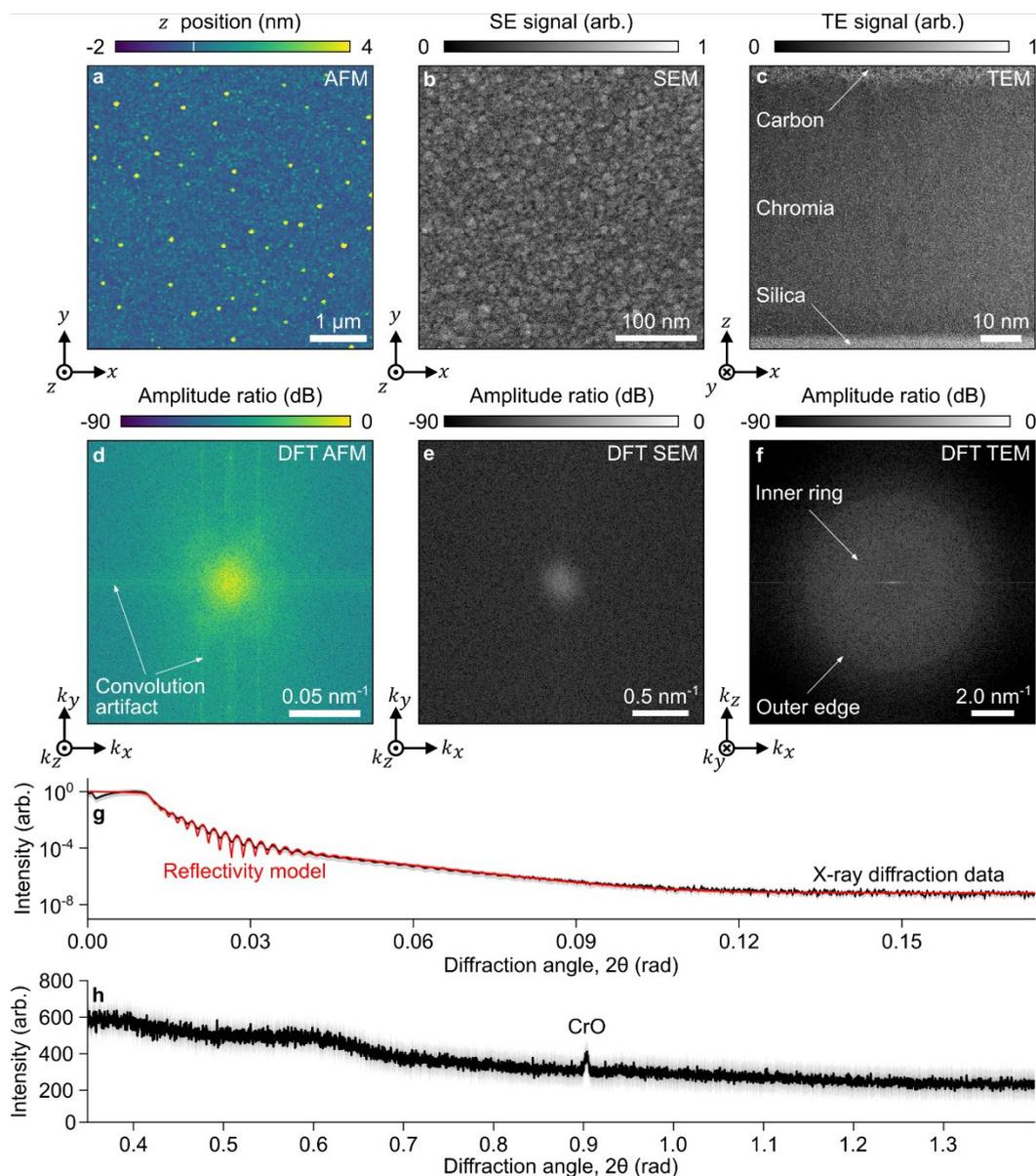

**Figure 2.** Chromia characterization. (**a**) Atomic force and (**b**) scanning electron micrographs showing the surface structure of the chromia mask on silica before milling with a focused ion beam. (**c**) Transmission electron micrograph showing a cross section of the chromia mask. The carbon is an artifact of sample preparation. (**d-f**) Plots showing discrete Fourier transforms (DFT) of the micrographs in (a-c). The origin, corresponding to spatial frequencies of 0 nm$^{-1}$, occurs at the center of each image. Horizontal and vertical lines in (d) indicate the presence of a convolution artifact of the probe tip in atomic force micrographs. (**g**) Plot showing (black) a grazing-incidence 2θ scan of X-ray diffraction and (red) a fit of an X-ray reflectivity model[26] with a reduced chi-square statistic, $\chi_\nu^2$, of 0.4. (**h**) Plot showing a conventional 2θ scan of X-ray diffraction. The gray regions around the black lines in (g) and (h) indicate the 95 % coverage interval of diffraction data.

*Volume Structure*
Ellipsometry data show a silica thickness of 488 nm ± 2 nm. Consistent with our measurements of surface roughness by X-ray diffraction, transmission electron micrographs show a chromia thickness of 63 nm ± 2 nm (Figure 2c, Table S1). In discrete Fourier transforms of these transmission electron micrographs (Figure 2f), a diffuse inner region with a sharp outer edge around 4.0 nm$^{-1}$ indicates a generally amorphous composition, which is consistent with X-ray diffraction data. A fit of a reflectivity model[26] to the X-ray diffraction data yields a chromia thickness of 65 nm ± 3 nm and density of 5.3 g cm$^{-3}$ ± 0.1 g cm$^{-3}$ (Figure 2g). The X-ray diffraction pattern of the chromia (Figure 2h) shows a broad peak of low intensity below angles of 0.7 rad (40°), indicating that the chromia is generally amorphous, and a narrow peak of low intensity at 0.901 rad ± 0.003 rad (51.6° ± 0.2°), suggesting the presence of crystalline domains of CrO[27] with a size of approximately 40 nm.

In-line Resolution Metrology

Having comprehensively characterized our bilayer, we load a sample into an electron–ion beam system and focus both beams. Scanning electron microscopy offers the potential to measure and optimize the effect of the focused ion beam prior to nanofabrication. In such a process, the system operator mills an open cavity or pit, the simplest test structure, by dwelling the focused ion beam in a single position for a certain amount of time. Measurement of the radius of the pit by scanning electron microscopy quantifies the effective lateral resolution of the focused ion beam. Manual focusing of the ion beam proceeds by iteratively adjusting the voltage of the final lens in the ion-beam column, milling a new pit, and measuring its radius. This process can occur in real time if a fast method for dimensional metrology is available, and concludes with the estimation of a minimum radius, indicating that the focus of the ion beam is near optimal.

There are two main challenges of such in-line measurements. The first is physical, as scanning electron microscopy is sensitive to dielectric charging, which our chromia mask mitigates. The second is analytical, as scanning electron microscopy yields an indirect measure of surface topography, with image contrast resulting primarily from local tilt, shadow, and material contrast[28]. Previous studies have taken different approaches to analyze the resulting images, yielding different combinations of efficiency and accuracy. Simple approximations without uncertainty evaluation[9] are unsuitable for dimensional metrology, whereas accurate physical models[29] of scanning electron microscopy require energy-loss functions of the constituent materials. Historically, these functions have required synchrotron measurements to obtain. Intermediate approaches can achieve both efficiency and accuracy, such as by using reference data from a physical model to calibrate an empirical model[30].

*Image Formation*

We introduce an efficient method of measuring pit radius by scanning electron microscopy, in comparison to independent measurements of surface topography by atomic force microscopy (**Figure 3**, Figure 3a-b, Table S2). We make the novel observation of a near equality of pit radii corresponding to the maximum signals of secondary electrons in scanning electron micrographs (Figure 3d), and the maximum convexity of surface profiles in atomic force micrographs, where the negative value of the second derivative of the surface profile with respect to position is maximal (Figure 3e). This empirical correlation implies that the curvature of our test pits affects image formation, which we interpret by topographic calculations of tilt and shadow contrast (Note S3, Figure S2). This new analysis suggests that the implantation of gallium, sputtering and redeposition of chromia and silica, and resulting material contrast near the region of maximum convexity contributes to signal peaks in scanning electron micrographs. Further study is necessary to fully understand the cause of this empirical correlation, which we nonetheless put to good use.

*Correlative Measurements*

For a representative test pit, we compare a section from a scanning electron micrograph (Figure 3c) to a section of the negative of the second derivative from an atomic force micrograph (Figure 3d). The micrographs show pit asymmetry due to several non-ideal conditions of milling and microscopy, including astigmatism and drift of the focused ion beam during patterning and an artifact from probe-sample convolution in atomic force micrographs (Note S4, Figure S3, Figure S4). As a result, scanning electron microscopy yields a relatively narrow and approximately normal distribution of pit radius, whereas atomic force microscopy yields a relatively broad and asymmetric distribution of pit radius (Figure 3e). Regardless, the two measurements of mean pit radius agree to within a root-mean-square error ranging from 4 nm to 7 nm (Figure 3f-j, Table S4). These values are comparable to our estimate of statistical uncertainty for atomic force microscopy and range from 3 % to 5 % of the pit radii. Even better, our estimate of statistical uncertainty for scanning electron microscopy has a 95 % coverage interval of 1 nm, whereas accurate measurements with physical models can achieve a total uncertainty of approximately 1 nm, including systematic errors[29]. These results show that our new approach to in-line metrology of effective lateral resolution is usefully precise, although further study is necessary to quantify systematic effects that limit accuracy. These results build confidence in our method of minimizing the radii of test pits, which we use to focus our ion-beam prior to nanofabrication of complex test-structures.

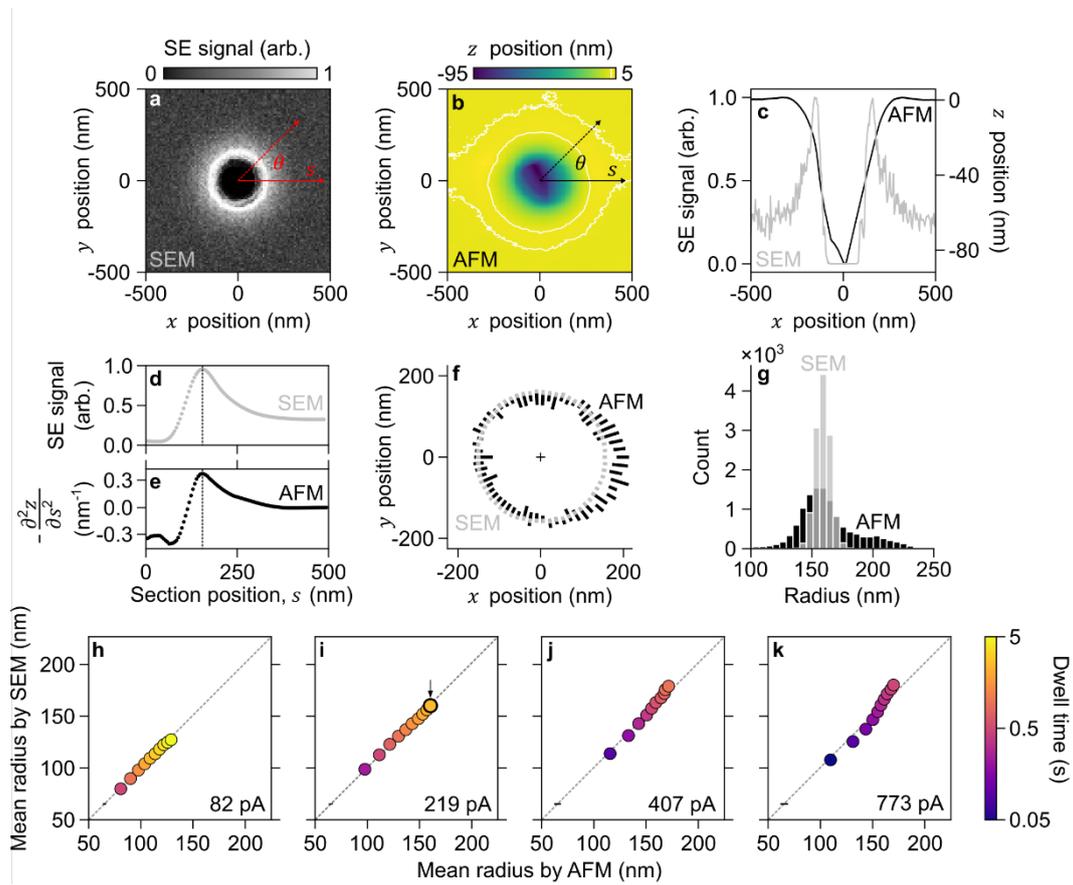

**Figure 3.** In-line resolution metrology. (**a**) Scanning electron micrograph and (**b**) atomic force micrograph showing a representative pit in the chromia–silica bilayer. (**c**) Plots showing (gray) secondary electron signals from (a) and (black) z position from (b) for horizontal line scans through the center of either image. (**d**) Plot showing the secondary electron scattering intensity along a radial section from (a) in the direction of the largest radius of the pit. (**e**) Plot showing the negative of the second derivative of the z position of the pit along a radial section from (b) in the direction of the largest radius of the pit. (**f**) Plots showing angular dependence of pit radius for all values of section angle from (gray) scanning electron microscopy (SEM) and (black) atomic force microscopy (AFM) after smoothing. Uncertainties are 95 % coverage intervals. (**g**) Histograms showing pit radius for all values of section angle of ten replicates from (gray) scanning electron microscopy and (black) atomic force microscopy. (**h-k**) Plots showing correlative measurements of mean radius of pits by atomic force microscopy and scanning electron microscopy for one decade of ion-beam current and two decades of dwell time. Panels (a-g) correspond to the black circle in (i). The lone black bars near the lower left corner of the plots in (h-k) represent 95 % coverage intervals of mean radius. More details are in Table S2, Table S3, Note S3, Note S4, Figure S3, Figure S4, and Table S4.

## Complex test-structures

*Ion Exposure*

In contrast to lithographic patterning, which requires pattern transfer and is generally serial for each vertical feature dimension, focused-ion-beam milling directly forms complex nanostructures. To quantify the relationships between milling depth, mask thickness, effective lateral resolution, and volume throughput, we mill checkerboard patterns through the chromia mask and into the underlying silica (**Figure 4**, Figure 4a,b, Figure S5). The depths vary from 0 nm to 200 nm before removal of the chromia and from 0 nm to 130 nm after removal of the chromia by selective chemical etching. This complex pattern tests lateral resolution at the edges of square structures, which are more robust than pit or line structures to artifacts from probe-sample convolution in atomic force micrographs (Note S5). Subsequent simulations of conventional line–space arrays supplement our experimental test-structures.

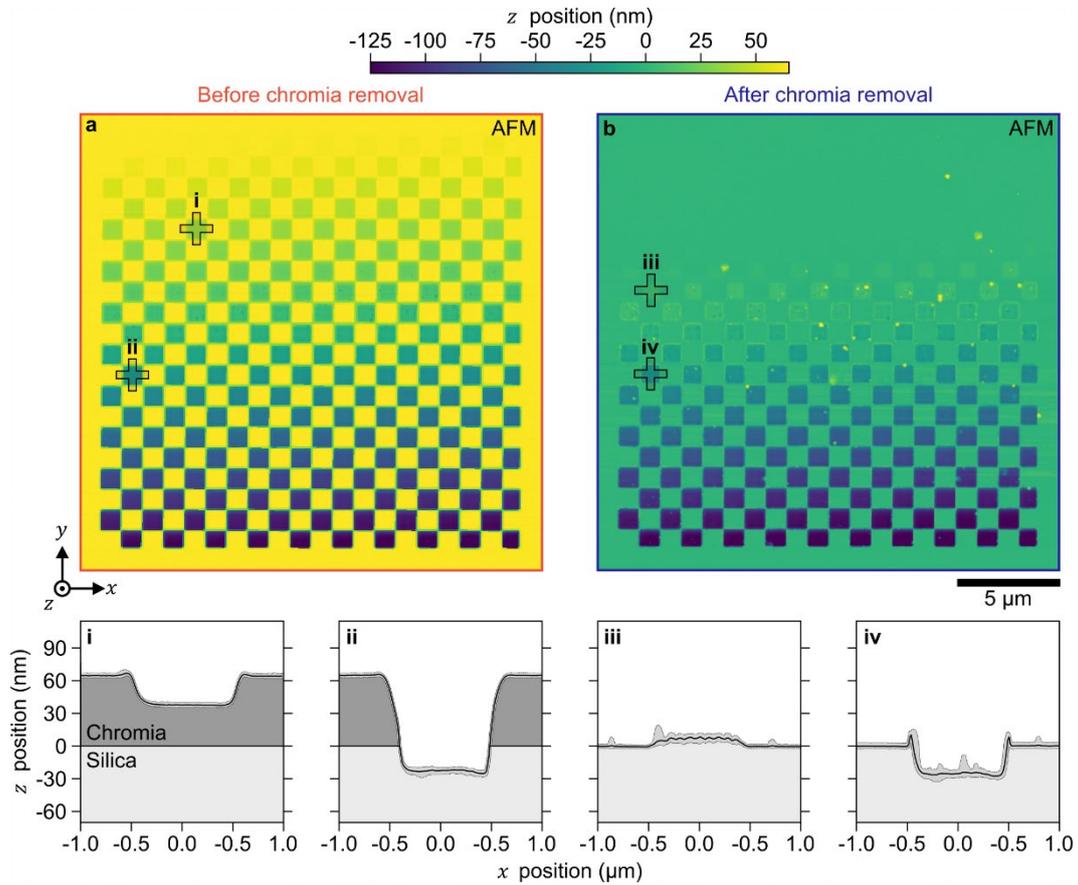

**Figure 4.** Complex test-structures. (**a**, **b**) Atomic force micrographs showing representative checkerboard patterns (**a**) before and (**b**) after removal of the chromia mask. (**i, ii, iii, iv**) (black line) representative features before and after removal of the chromia mask. The shaded regions around the black line indicate the 95 % coverage interval of the data from sections of atomic force micrographs. Panels (**ii**) and (**iv**) show the same feature before and after removal of the chromia mask by selective chemical etching, respectively. We define the zero plane to be top of the silica surface. The periodic surface topography within some squares may be due to an aliasing artifact[5a].

*Vertical Responses*

Vertical responses of the bilayer include variable milling rates and apparent modification of the material interface. Milling rates increase monotonically through chromia into silica (**Figure 5**, Figure 5a, Figure S6, Table S5). A piecewise-linear approximation of the milling response fits four apparent milling rates. The initial milling rate of chromia is low, which is relevant to patterning near the top surface. At higher doses, the milling rate of chromia triples to a value representative of milling bulk chromia. Milling through bulk chromia, gallium ions penetrate the chromia–silica interface, and the milling rate increases by approximately one third. Milling through the interfacial layer, bulk silica shows a milling rate that again increases by one third, indicating that any initial response of silica, such as a lower milling rate[5a], occurs through and under the chromia. The ratio of the bulk milling rate of silica to that of chromia exceeds unity, resulting in a physical selectivity of 1.63 $^{+0.61}/_{-0.44}$ (Table S6). This fractional notation denotes 95 % coverage intervals that are asymmetric. In addition to the critical parameter of physical selectivity, which exceeds unity for this bilayer, crystalline grains and surface roughness of hard masks should be minimal to reduce lateral and vertical anisotropy of the milling process. Our chromia mask performs well in this regard, lacking granular structure and having surface roughness of less than 2 nm in comparison to the roughness of the underlying silica surface, which is less than 1 nm (Figure 2a-c).

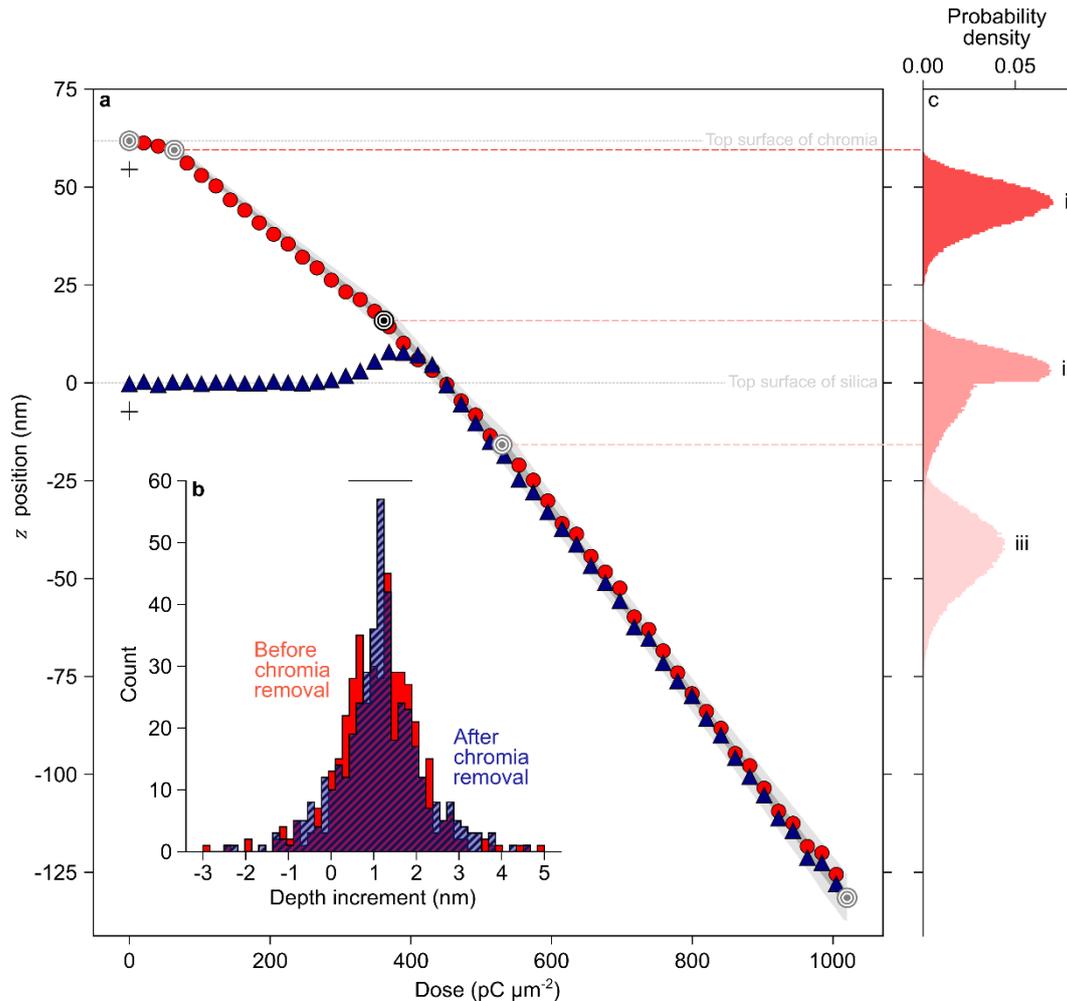

**Figure 5.** Vertical responses. (**a**) Plot showing milling responses of chromia on silica for an ion-beam current of 227 pA ± 1 pA (red circles) before and (blue triangles) after chromia removal. Uncertainties of ion-beam currents are conservative estimates of 100 % coverage intervals. The light and dark gray regions respectively indicate the 95 % coverage interval and the interval between the lower and upper quartiles of the piecewise-linear model of milling responses. The lone black crosses indicate representative uncertainties of dose and z position, which we plot as 95 % coverage intervals. (**b**) Histograms showing depth increments (red) before and (blue hatch) after removal of the chromia mask for all ion-beam currents. The lone black bar indicates a representative uncertainty of z position. (**c**) Histograms showing simulations of stopping range as a function of z position for depths that correspond to (roundels) inflection points in the milling rates of chromia on silica, (**i**) (69 pC µm$^{-2}$, 59 nm), (**ii**) (366 pC µm$^{-2}$, 16 nm), and (**iii**) (531 pC µm$^{-2}$, -16 nm). The lone black roundel indicates the onset of a change in milling rate due to the underlying chromia–silica interface. More details are in Figure S6, Table S5, Table S6, and Table S7.

At intermediate doses, nanostructures rise above the zero plane of the chromia–silica interface after exposure to chromium etchant (Figure 4, iii). In a previous study of a similar substrate, thermal oxide did not swell significantly at low doses [5a], suggesting that gallium cation penetration modifies the chromia–silica interface, potentially forming a silicate with lower susceptibility to chromium etchant than chromia. Higher doses mill these nanostructures back down through the zero plane (Figure 5a). We achieve our target depth increment of 1 nm in silica, with standard deviations that are within 1 nm of the surface roughness of the silica, down to a depth of 130 nm (Figure 5b, Figure S6). Depth increment distributions broaden slightly after chromia removal (Table S5).

To better understand these material responses, we perform three simulations of gallium cation penetration through chromia and silica, elucidating a primary interaction of the machine tool and workpiece. The suitability of amorphous materials for calculation of stopping ranges of ions in matter, and the input parameters from our materials characterization, build confidence in the simulation results. Three histograms of stopping range as a function of depth correspond to bulk chromia, the chromia–silica interface, and bulk silica (Figure 5c, i-iii). The mean stopping range of gallium cations in bulk chromia agrees to within 2 nm with the z position onset of a higher milling rate of the chromia–silica interface (Figure 5c, i-iii, Table S7). This consistency supports our experimental measurements, piecewise-linear approximations of vertical responses, and simulations of ion penetration.

*Edge Profiles*

Completing our characterization of the test structures, we quantify widths of nanostructure edges before, $w_b$, and after, $w_a$, removal of the sacrificial mask. Edge and line widths are both straightforward to measure and are both proportional to the radius of the ion beam, which ultimately limits lateral resolution. Accordingly, we use edge width as a metric of effective lateral resolution and super-resolution. Focused-ion-beam milling of the bilayer yields edge profiles that are approximately sigmoidal. Removal of the sacrificial mask changes the shape of the profile, abruptly truncating the sigmoid, which can be advantageous for nanofluidic channels[5a, 5b], waveguides[16a], or for the process of replica molding[11]. Accordingly, the reduction of data from profiles in atomic force micrographs to quantities of edge width requires two different models. To extract widths and depths, we fit a common error function to edge profiles before removal of the sacrificial mask, and we truncate an error function to fit to edge profiles after removal (Figure S7)[31]. These approximations originate from integration of a Gaussian function, which is appealing for nanostructure edges resulting from exposure to a focused beam with a profile that is approximately Gaussian. However, edge profiles do include systematic effects from the dependence of milling rate on incidence angle, discontinuities in milling rate within the bilayer, and an overshoot artifact from atomic force microscopy. Even so, a novel Monte-Carlo statistical analysis with jack-knife resampling shows that measurement uncertainty dominates our estimates of uncertainty for $w_b$, $w_a$, and resulting super-resolution factors, $\mathcal{F}_{SR} = w_b w_a^{-1}$ (Figure S7a, Figure S8, Figure S9, Table S8, Table S9, Table S10). After chromia removal, edge widths decrease by factors ranging from 2 for deep features to 6 for shallow features. Shallow features in chromia exhibit wider edges than features of similar depth in silica after removal of the chromia (Figure S8, Table S8), indicating that the sacrificial mask improves lateral resolution for features of similar depth. These results affirm the use of error functions to accurately extract edge widths, building confidence in the reliability of these measurement results and the validity of our subsequent use of error functions in models of spatial masking.

## Resolution–throughput tradespace

*Theoretical basis*

Characterization of the bilayer response enables a new study of the tradespace of lateral resolution and volume throughput. The effective lateral resolution, $\mathcal{R}$, can be either a conventional resolution, $\mathcal{R}_R$, or an unconventional super-resolution, $\mathcal{R}_{SR}$ (Note S9). Effective lateral resolution improves as ion-beam current decreases[32], but reductions in ion-beam current, $I$, incur a non-linear cost of milling time, degrading volume throughput and potentially lateral resolution by drift of the system. To elucidate how sacrificial masks affect this tradespace, we express effective lateral resolution and super-resolution in terms of ion-beam current and equate them. Conventional lateral resolution results from milling a feature of a certain depth directly into a substrate with a low value of ion-beam current, $I_{low}$, whereas super-resolution results from milling a similar feature through a sacrificial mask and into an underlying substrate with a high value of ion-beam current, $I_{high}$. The super-resolution factor, $\mathcal{F}_{SR}$, relates the two quantities,

$$\mathcal{R} = \alpha I_{low}^{\beta} = \mathcal{F}_{SR}^{-1} \alpha I_{high}^{\beta}, \qquad (1)$$

where we apply the approximation of a power-law relation of the radius of a focused ion beam with coefficient, $\alpha$, and exponent, $\beta$, the latter of which typically ranges from 0.25 to 1.00 for liquid-metal ion sources with ion-beam currents of less than 10 nA[32].

To test Equation (1) and further characterize our effective lateral resolution, we fit a power-law model to pit radius (Figure 3h-k) as a function of total charge, and to edge width before and after removal of the chromia mask as a function of ion-beam current (Figure S10). The trends for the different test structures are generally consistent and yield reasonable values of model parameters (Table S11), corroborating our results from in-line resolution metrology and building confidence in Equation (1) as the basis of the resolution–throughput tradespace.

An important issue to consider is that Equation (1) presents a theoretical perspective in which ion-beam current is a continuous variable. However, in typical focused-ion-beam systems, apertures limit the selection of ion-beam current to discrete values. Nonetheless, this theoretical perspective yields important insights into the spatiotemporal domain, while also accommodating experimental validation in the spatial domain. To this end, Equation (1) permits a substitution of terms, in which the widths of nanostructures resulting from milling with a single value of ion-beam current, before and after removal of the chromia mask, take the place of the power-law expressions for the radii of ion beams with low and high values of current. The resulting proportionality in Equation (2) avoids the experimental limitation of discrete values of ion-beam current, enabling validation of the kernel of our model of super-resolution with measurements of edge widths before and after removal of the chromia mask,

$$\mathcal{R} \propto w_a = \mathcal{F}_{SR}^{-1} w_b. \qquad (2)$$

On the basis of Equations (1-2), we introduce new theory, in three aspects, to elucidate the super-resolution effect. First, a model of spatial masking maps patterns of ion dose to resulting nanostructures in chromia and silica to calculate factors of lateral super-resolution, predicting our experimental results without adjustable parameters and fitting other experimental results with adjustable parameters. Second, a model of temporal efficiency uses Equation (1) to account for the different durations of milling a structure of a certain depth by a low or high current, elucidating the temporal advantage of the super-resolution effect. Third, a comparison of the spatial and temporal effects of the sacrificial mask shows dramatic improvements depending on the mask selectivity and thickness, and a figure of merit quantifies performance within the tradespace to guide future optimization.

*Super-resolution*

The common approximation of the current density distribution of a focused ion beam by a Gaussian or bi-Gaussian function enables derivation of a closed-form solution to $\mathcal{F}_{SR}$ in terms of the complementary error function and geometric and physical parameters of the bilayer. We use the Gaussian approximation for simplicity, and we establish the validity of either approximation in our model of lateral super-resolution (Note S6, Figure S11). Our model relates the ratio of edge widths, $w_b$, and $w_a$, to the milling depth after normalization by the mask thickness, $\zeta = z_s z_m^{-1}$, and the physical selectivity of the bilayer, $\mathcal{S} = \bar{m}_s \bar{m}_m^{-1}$, where $z_s$ is the milling depth into the substrate, $\bar{m}_s$ is the bulk milling rate of the substrate, $z_m$ is the thickness of the mask, and $\bar{m}_m$ is the bulk milling rate of the mask (Table S11, Figure S12). The simplifying assumption of bulk milling rates ignores the initial milling response, the transition in milling chromia and then silica, the dependence of milling rate on angle of incidence[33], and redeposition of the working material[34], yielding an analytic function,

$$\mathcal{F}_{SR}(\zeta, \mathcal{S}) = \frac{w_b}{w_a} = \frac{2}{\frac{1}{\sqrt{2}} \operatorname{erfc}^{-1}\left(\frac{2\mathcal{S}}{\mathcal{S}+\zeta}\right) + 1}. \tag{3}$$

Despite the Gaussian approximation of the ion-beam profile and the simplifying assumption of bulk milling rates, the predictions of our model are generally consistent with our experimental results and nearly within our uncertainty estimates. Experimental results from milling with an ion-beam current of 227 pA ± 1 pA show the best agreement with the model, which predicts higher super-resolution factors for shallower features and a monotonic decrease in super-resolution factor for deeper features (**Figure 6**). Similar trends result for ion-beam currents ranging from approximately 80 pA to 800 pA (Figure S13). In comparison to mean values of the spatial masking model, mean values of experimental super-resolution factors from all experiments show a root-mean-square error of 0.5 ± 0.1, corresponding to a mean relative error of approximately 20 %. At the cost of analytic utility, a computational model without our simplifying assumptions might improve agreement of theory and experiment. However, our simple model captures the general effect of spatial masking, enabling predictions of processes using only two parameters.

Beyond this primary comparison of theory and experiment, two variants of our spatial masking model allow secondary tests of its reliability. The first variant models the milling of a single line, matching that of an experimental study which achieved super-resolution at the scale of one to ten nanometers[5b]. This line variant fits the experimental data with $\chi_\nu^2$=5.3 and extracts reasonable parameter values from decreasing widths of nanochannels in quartz, resulting from line scans of a focused ion beam through sacrificial chromium masks of increasing thickness (Note S7, Figure S14). The second variant models the pattern of lines and spaces of variable pitch and duty cycle, simulating a conventional structure to test resolution, albeit with edge profiles that are characteristic of focused-ion-beam milling. This line–space variant shows trends that are comparable to our experimental results (Figure S15, Figure S16), indicating that reductions in edge width translate to improvements of the conventional resolution metric of line–space density or half-pitch of periodic features.

The general agreement of our spatial masking model, various experimental results, and test-structure simulations substantiates our theory as a useful construct, which we extend into the temporal domain to explore the limits of the process of spatial masking.

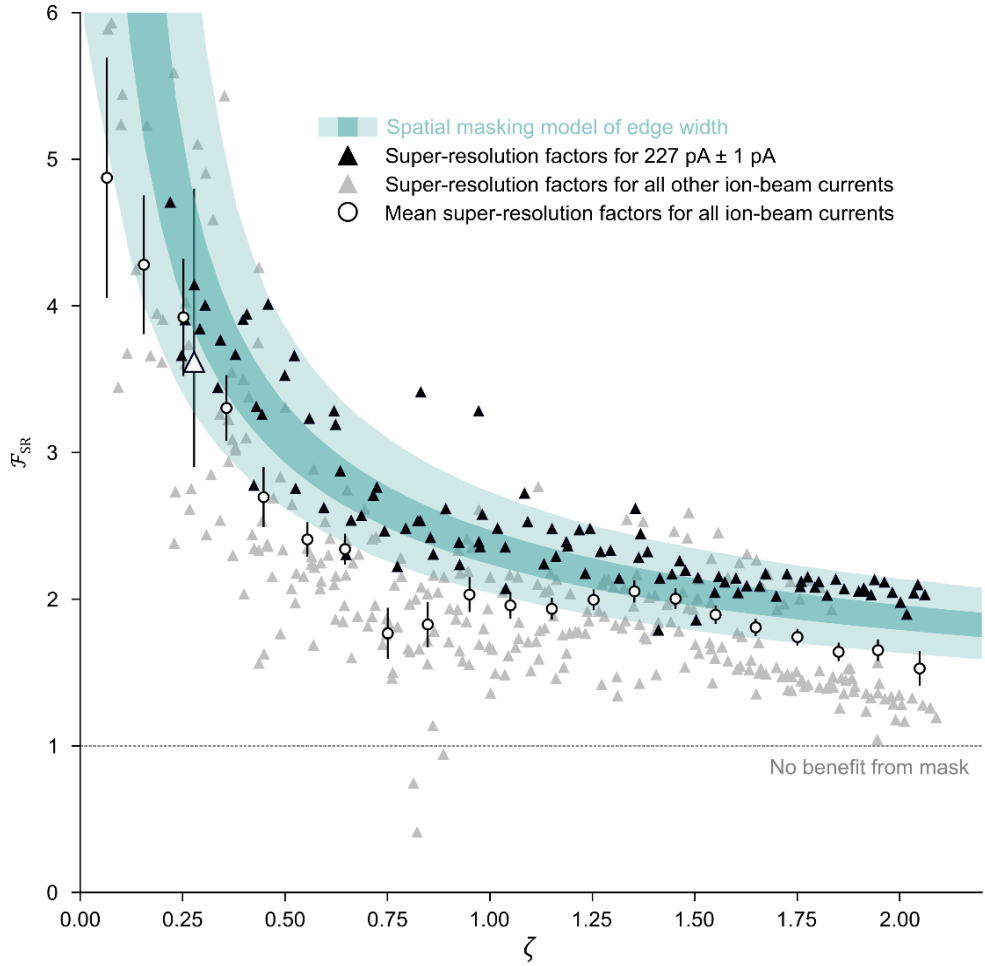

**Figure 6.** Super-resolution. Plot showing lateral super-resolution factors as a function of the ratio of milling depth to mask thickness. The light and dark regions of cyan respectively indicate the 95 % coverage interval and the interval between the lower and upper quartiles of the spatial masking model. The black triangles represent all features below the zero plane that we mill with an ion-beam current of 227 pA ± 1 pA. The white triangle corresponds to a representative feature with a depth of approximately 18 nm and edge widths that decrease from 195 nm ± 18 nm before removal of the mask to 54 nm ± 12 nm after, yielding a super-resolution factor of 3.6 $^{+1.2}_{-0.7}$, which the black bars indicate. The distribution width in the $\zeta$ direction of ± 0.01 is smaller than the data marker. The gray triangles represent features that we mill with all other ion-beam currents, which range from approximately 80 pA to 800 pA. The white circles represent a subset of mean super-resolution factors from the entire data set, with downsampling for clarity. The black bars are 95 % coverage intervals of the mean values and are smaller than the data markers in the $\zeta$ direction. More details are in Figure S8, Table S8, and Figure S13.

*Temporal Efficiency*

Having demonstrated the capability of our analytic model to describe generally the spatial advantage of a sacrificial mask, we extend the model into the temporal domain to investigate the temporal efficiency of the fabrication process. We argue that spatial masking is rational under two conditions – first, if patterning with a high value of ion-beam current through a sacrificial mask requires less time than patterning with a low value of ion-beam current directly into the substrate, and second, if the patterning resolution at a minimal value of ion-beam current, typically around 1 pA for commercial systems, exceeds critical dimensions of the nanostructure design. The temporal efficiency of spatial masking, $\eta_\tau$, describes the extent to which patterning with a high value of ion-beam current through a sacrificial mask to achieve a certain effective lateral resolution saves time,

$$\eta_\tau = \frac{t_s}{t_m + t_s} = \frac{V_s \bar{m}_s^{-1} I_{low}^{-1}}{V_m \bar{m}_m^{-1} I_{high}^{-1} + V_s \bar{m}_s^{-1} I_{high}^{-1}} = \left(\frac{I_{high}}{I_{low}}\right) \frac{\zeta}{S + \zeta} = \mathcal{F}_{SR}(\zeta, S)^{\frac{1}{\beta}} \frac{\zeta}{S + \zeta}, \qquad (4)$$

where $V_s$ and $V_m$ are, respectively, the volumes of the substrate and mask that the ion-beam mills (Note S8). For simplicity, we neglect the time to deposit and remove the sacrificial mask, which is a fraction ranging from $10^{-3}$ to $10^{-1}$ of the time to mill with a low value of ion-beam current. Equation (4) predicts that the temporal efficiency of spatial masking increases as physical selectivity increases and decreases as milling depth increases with respect to mask thickness (**Figure 7**, Figure 7a). Moreover, Equation (4) implies that the limits of milling depth for spatial masking occur where the temporal efficiency falls below unity. Solving for this condition numerically reveals that the efficiency limit is proportional to the physical selectivity of the mask and substrate (Figure 7b). Applying this model, while accounting for all measurement uncertainties, yields a remarkable prediction for our bilayer system – patterning a nanostructure with a certain edge width, using a high value of ion-beam current through a sacrificial mask, would require less time than directly milling a similar nanostructure with a low value of ion-beam current to depths ranging from 35 to 135 times the thickness of the chromia mask, or approximately 2 μm to 9 μm. In this way, even modest improvements in physical selectivity yield relatively large improvements in the milling range, $\zeta_{max}$. To better understand the intriguing implication of these results, we combine our experimental measurements of super-resolution factors with experimental values of temporal efficiency and revisit the resolution–throughput tradespace.

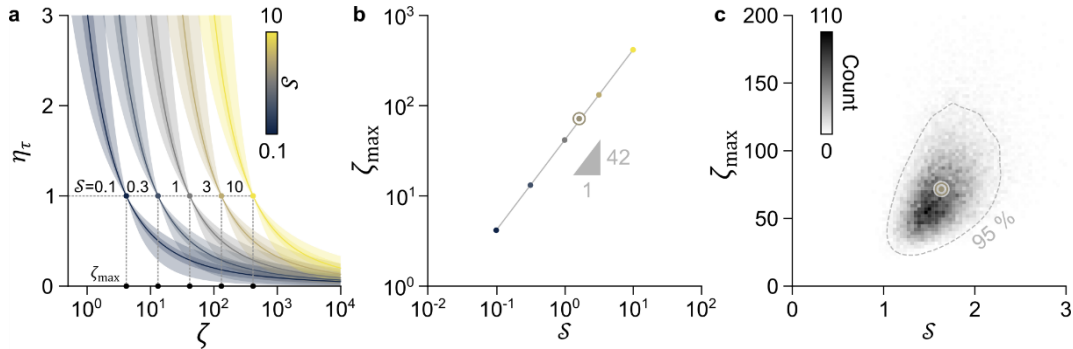

**Figure 7.** Temporal efficiency. (**a**) Plots showing theoretical mean values of temporal efficiency, $\eta_\tau$, as a function of milling depth for a range of values of physical selectivity and a scaling exponent of 0.15 $^{+0.06}/_{-0.07}$. For clarity, we consider uncertainty only from the scaling exponent. The light and dark regions respectively indicate the 95 % coverage interval and the interval between the lower and upper quartiles of temporal efficiency. The circles are positions of maximum milling depth, where the temporal milling efficiency equals unity. (**b**) Plot showing maximum milling depth as a function of physical selectivity. The circles correspond to theoretical values of maximum milling depth in (a). Uncertainties of the data, which are 95 % coverage intervals of the mean values, are smaller than the data markers. (**c**) Joint histogram showing physical selectivity and maximum milling depth with (dash line) contour indicating the 95 % coverage region. The roundels in (b) and (c) correspond to our experimental values that predict a maximum milling depth at which super-resolution patterning remains temporally efficient, $\zeta_{max}$, of 71 $^{+64}/_{-36}$.

*Paradigm Shift*
Lithographic patterning has a characteristic tradespace[12a], with lateral resolution following a power-law dependence on areal throughput for different processes. The tradespace of focused-ion-beam milling is analogous but requires consideration of volume throughput. A key finding of our study is that a sacrificial mask transforms the coupling of lateral resolution and volume throughput. This is evident in the improvement of conventional resolution to super-resolution by factors of 1.8 to 2.7 after removal of the mask. These experimental results are independent of our theoretical model of super-resolution (**Figure 8**, Figure 8a). Power-law models of each resolution–throughput trend enable evaluation of the tradespaces to determine equivalent values of conventional resolution and super-resolution but dramatically different values of volume throughput that result from different ion-beam currents (Table S13, Note S6). A minimal extrapolation of experimental values of volume throughput for equivalent values of conventional resolution and super-resolution indicates an improvement by a factor of 42 ± 2 for our highest value of ion-beam current. Extrapolation to lower values of ion-beam current implies even more impressive improvements, ranging from two to three orders of magnitude (Table S13), while extrapolation to higher values of ion-beam current suggests potential limits of such improvements. To better understand the spatiotemporal advantages of spatial masking, we combine metrics of resolution and volume throughput to empirically investigate this limiting behavior.

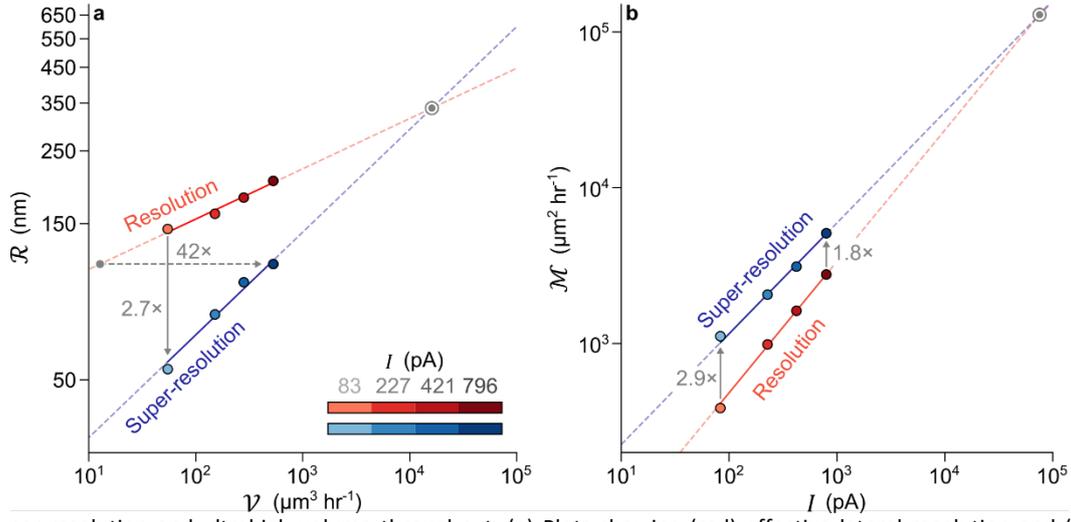

**Figure 8.** Super-resolution and ultrahigh volume throughput. (**a**) Plots showing (red) effective lateral resolution and (blue) super-resolution as a function of volume throughput. The horizontal dash arrow indicates a factor of improvement of 42 in volume throughput for an equal value of lateral resolution with minimal extrapolation outside of our experimental range. The vertical solid arrow indicates a mean factor of improvement of 2.7 in resolution for the volume throughput that corresponds to our lowest value of ion-beam current. (**b**) Plots showing figures of merit (red) before and (blue) after removal of the chromia mask. The vertical arrows indicate mean factors of improvement in figure of merit of 2.9 for our lowest value of ion-beam current and 1.8 for our highest value of ion-beam current. For both plots, solid lines are fits of power-law models to experimental data, dash lines extrapolate outside of the experimental range, roundels mark the intersection of conventional resolution and super-resolution trends, and uncertainties are smaller than data markers.

We introduce a figure of merit, $\mathcal{M}$, as a function of ion-beam current to elucidate the effect of spatial masking on the tradespace of lateral resolution, $\mathcal{R}$, and volume throughput, $\mathcal{V}$,

$$\mathcal{M} = \frac{\mathcal{V}}{\mathcal{R}} = \mathcal{F}_{\text{SR}}(\zeta, \mathcal{S}) \frac{I^{1-\beta}}{\alpha} \left( \frac{V_{\text{m}} + V_{\text{s}}}{V_{\text{m}}\overline{m}_{\text{m}}^{-1} + V_{\text{s}}\overline{m}_{\text{s}}^{-1}} \right) \cong \mathcal{F}_{\text{SR}}(\zeta, \mathcal{S}) \frac{I^{1-\beta}}{\alpha} \left( \frac{z_{\text{m}} + z_{\text{s}}}{z_{\text{m}}\overline{m}_{\text{m}}^{-1} + z_{\text{s}}\overline{m}_{\text{s}}^{-1}} \right). \quad (5)$$

The figure of merit, in units of µm² hr⁻¹, increases as ion-beam current increases, decreases as resolution degrades, and collapses to $\alpha^{-1}I^{1-\beta}\overline{m}_{\text{s}}$ in the absence of a sacrificial mask, where $\mathcal{F}_{\text{SR}}(\zeta, \mathcal{S})$ must equal unity. In this way, the figure of merit enables direct comparison of conventional milling of a bilayer before removal of the sacrificial mask, to super-resolution milling through a sacrificial mask after removal. For our experimental system, the mask increases the figure of merit by a factor ranging from 1.8 to 2.9, due to better resolution with a mean factor of 2.1 ± 0.9 at a constant volume throughput (Figure 8b, Table S13). Extrapolation to low values of ion-beam current suggests higher factors of improvement, whereas extrapolation to high values of ion-beam current imply a limit to improvement by a sacrificial mask.

Equation (5) yields two nonobvious insights into the effects of physical selectivity and milling depth on process quality (Figure S17). First, our model predicts that, for a constant value of milling depth, figure of merit increases as physical selectivity decreases. This implies that softer masks hasten the milling process, improving volume throughput with little cost to lateral super-resolution but some cost to milling range. Second, our model predicts that, for a constant physical selectivity, figures of merit increase as milling depth decreases. This implies that thicker masks yield nanostructures with narrower edges than thinner masks, improving lateral resolution and milling range with some cost to throughput. Other bilayer materials might optimize physical selectivity for particular objectives. These new insights into the resolution–throughput tradespace motivate experimental tests of the improvements in volume throughput that sacrificial masking enables.

Lens array fabrication
Our new understanding of the resolution–throughput tradespace of focused-ion-beam machining predicts that sacrificial masking can increase the throughput of device fabrication at a higher value of ion-beam current. This improvement of throughput is relative to equivalent devices resulting from directly milling a workpiece at a lower value of ion-beam current. To test this prediction, we mill two arrays of Fresnel lenses that are nominally equivalent, either a single lens directly into silica with an ion-beam current of 26 pA, or 75 lenses through a sacrificial mask of chromia and into silica with an ion-beam current of 2600 pA, in an equal time of 3.75 h. Without a chromia mask to dissipate the accumulation of positive charge[17b], direct milling of silica requires a modification of the fabrication process, such as simultaneous exposure by an electron beam or flood gun, which may or may not be available, to compensate for this effect and achieve comparable results (Table S14). We perform a control experiment to test this effect, milling single lenses directly into silica at an ion-beam current of 2600 pA, with and without simultaneous exposure by an electron beam. Fresnel lenses are simple to design[35] but their structural complexity and sharp transitions in surface relief make them challenging to machine and measure at small scales[36]. For both reasons, Fresnel lenses are suitable targets for novel fabrication and functional characterization. In testing the dose delivery to fabricate the Fresnel lenses, we develop methods to trim the dose around the peaks of concentric ridges to tune the super-resolution effect, and to exhaust the mask in the milling process (Note S10).

In our application, the function of Fresnel lenses is to focus incident light into spots of a certain size at a certain distance away from the surface of a microscope coverslip, enabling projection standards for, and functional characterization by, optical microscopy. These primary metrics of performance are nearly equivalent for both lens arrays. The single lens that we mill directly into silica at an ion-beam current of 26 pA and with electron-beam exposure to achieve the best results serves as a reference. In comparison, the projection distances of the lenses that we mill through chromia are equal to within 0.14 % $^{+1.4\,\%}/_{-1.6\,\%}$ and the apparent spot sizes are equal to within −0.27 % $^{+0.88\,\%}/_{-0.78\,\%}$ (**Figure 9**, Figure S19). In this way, the reference lens and lens array perform similarly as projection standards. In contrast, the projection distances and apparent spot sizes of the two control lenses differ significantly from the reference lens. Direct milling into silica at an ion-beam current of 2600 pA with electron-beam exposure results in a lens with a projection distance that differs by 3.46 % $^{+3.27\,\%}/_{-3.14\,\%}$ and an apparent spot size that differs by 14.38 % $^{+0.04\,\%}/_{-0.04\,\%}$ from the reference lens (Figure S20, Table S14). Without the electron beam, charge accumulation severely impairs the milling process, degrading the quality of the resulting lens structure so much as to compromise its performance. There is also a 30 % to 40 % reduction in the transmittance of the lenses that we mill through chromia, which is potentially attributable to the intermixing of chromia and silica and potentially remediable by pattern transfer to an optical polymer[37]. The similar performance of the reference lens and the lens that we mill through chromia demonstrates an improvement in fabrication throughput by a factor of 75, testing the new tradespace and showing a significant benefit from the paradigm shift.

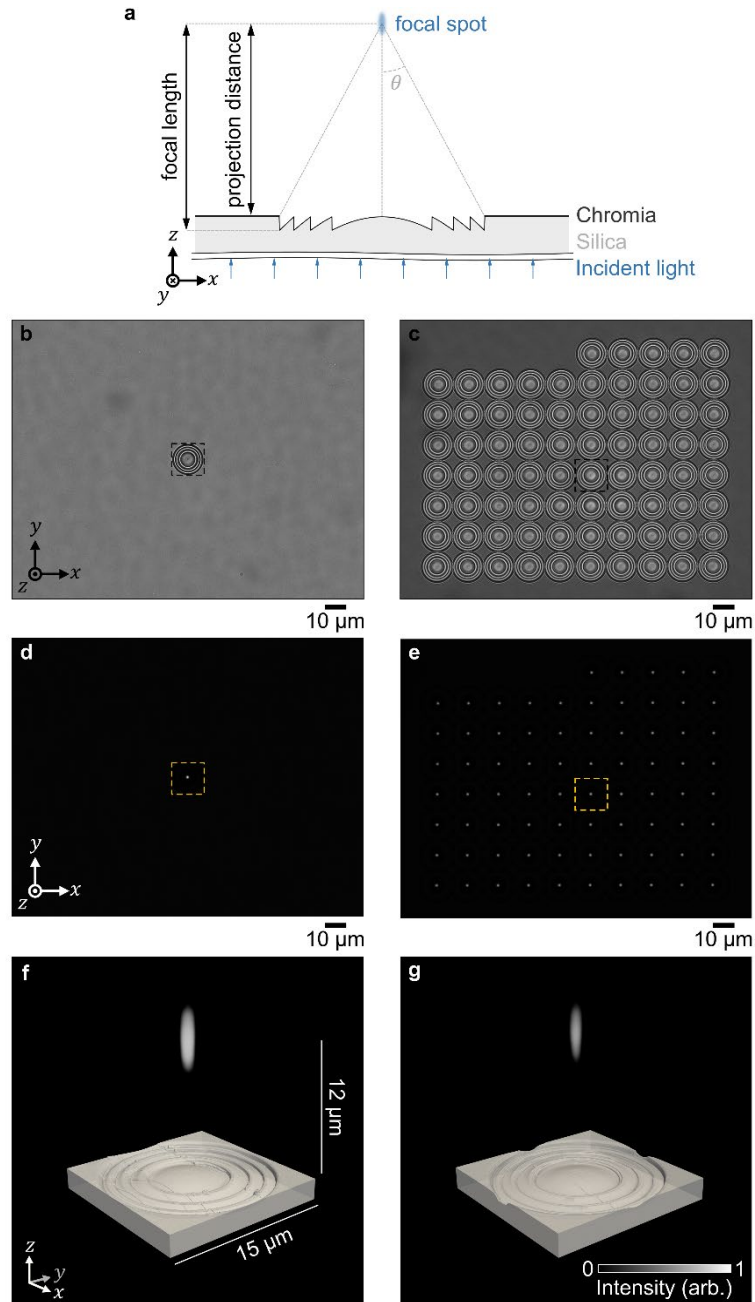

**Figure 9.** Fresnel lens arrays. (a) Schematic showing the design of a single Fresnel lens. (b, c) Brightfield optical micrographs at best focus of the top of the central zone of Fresnel lenses that we mill (b) directly into silica with an ion-beam current of 26 pA and (c) through chromia and into silica with an ion-beam current of 2600 pA. Different background intensities result from the absence or presence of chromia. (d, e) Brightfield optical micrographs at best focus of the spots from the lenses in (b) and (c). (f, g) Volume reconstruction of the focal spots from brightfield optical micrographs that we register to surface reconstructions from atomic-force micrographs of the Fresnel lenses in (b) and (c).

## Conclusions

Focused-ion-beam machining occurs within a tradespace of lateral resolution and volume throughput. At the root of this tradespace is a power-law dependence of lateral resolution on ion-beam current, such that fine features mill slowly. This limits the utility of what is otherwise a powerful process to directly form complex nanostructures without serial lithography, yielding the conventional view of the focused ion beam as that of a machine tool for prototyping rather than for manufacturing. In this study, we comprehensively and systematically investigate how a sacrificial mask can radically improve the resolution–throughput tradespace, using a chromia film to mask a silica substrate from gallium ions. Chromia proves to be a multifunctional mask, facilitating in-line metrology of ion-beam focus without issues from dielectric charging, enabling theoretical and experimental studies of the super-resolution effect, and being chemically suitable for a variety[23a] of processes and applications. This general process may extend to other bilayer materials and ion sources. Building on our characterization and application of these useful properties, we develop foundational theory for the spatial masking process and super-resolution effect, deriving an analytic model that generally predicts our experimental results and provides key insight into the parameter space of the process. Our tractable theory informs a paradigm shift from empirical to engineering design of masking and milling processes. Through measurements that are independent of our model and its simplifying approximations, we experimentally demonstrate super-resolution factors of up to 6 ± 2, and improvements to volume throughput of at least factors of 42 ± 2 for our smallest mean super-resolution factor and with minimal extrapolation. For our larger super-resolution factors, further extrapolation to equal values of resolution at lower values of ion-beam current implies a higher volume throughput by two to three orders of magnitude. An effect of this scale is analogous to the gain mechanism of chemical amplification of resist materials, in which effective thresholding improves contrast in photolithography, or increases in sensitivity improve areal throughput of electron-beam lithography[38]. We test these predictions by the fabrication of Fresnel lenses, increasing throughput by a factor of 75 for functionally equivalent lenses that we mill through chromia and into silica relative to lenses that we mill directly into silica. Further work could develop the Fresnel lens array into a novel standard for projecting focused spots into a sample to perform magnification and distortion calibration in the presence of sample aberrations[39]. Further work is also necessary to explore this advantage and optimize the tradespace for different objectives, such as fabricating other types of diffractive optics[36c], electromagnetic metasurfaces[40], or nanofluidic molds[11, 41]. However, our results show a temporal advantage that is so decisive as to enable a paradigm shift from conventional machining to ultrarapid prototyping of complex structures for research and development, and even to commercial manufacturing of device arrays that would conventionally require hundreds to thousands of hours of machine time. Moreover, our chromia characterization and model development can impact sample preparation for materials characterization[6] and biological imaging[7], enabling prediction and optimization of the super-resolution and volume throughput that are latent to masking and milling processes for volumetric imaging. In general, our new insights into the resolution–throughput tradespace extend the utility of focused-ion-beam machining from fundamental science to commercial products.

## Methods
### Experimental
*Fabrication*

To form a bilayer to study the spatial masking process, we thermally oxidize p-type silicon substrates to produce a silica film, over which we deposit a sacrificial film of chromia. We mill arrays of pits to test the focus of the ion beam and arrays of squares to test the spatial masking process, using an electron–ion beam system with a focused beam of gallium cations at an accelerating voltage of 30 kV and ion-beam currents ranging from approximately 80 pA to 800 pA (Table S2). We mill all structures at the eucentric height of our system and at normal incidence to the ion-beam column, which requires a stage tilt of 0.9075 rad (52°) with respect to the electron-beam column. After milling and characterizing the structures resulting from these patterns, we remove the sacrificial chromia mask by immersion in chromium etchant.

We fabricate arrays of Fresnel lenses on silica coverslips. We estimate the ion doses for radial profiles of Fresnel lenses from design depths and mean milling rates of chromia and silica. We use the radial profiles to interpolate the ion doses for milling positions in a Cartesian coordinate system. To test direct fabrication, we mill a single Fresnel lens with an ion-beam current of approximately 26 pA in a time of 3.75 h, while scanning the electron beam over the patterning area at an electron-beam current of 100 pA to compensate for the accumulation of positive charge from ion irradiation. In a control experiment, we mill single Fresnel lenses with an ion-beam current of approximately 2600 pA in a milling time of 0.04 h with and without charge compensation by an electron-beam current of 6400 pA. We deposit a sacrificial film of chromia onto one silica coverslip and mill an array of 75 Fresnel lenses with an ion-beam current of approximately 2600 pA in a similar time of 3.75 h.

*Characterization*

We measure the thickness of silica films by ellipsometry, and the thickness of chromia films by ellipsometry, transmission electron microscopy, and X-ray diffractometry.

We measure surface topography of all structures that we fabricate by atomic-force microscopy. We measure the surface topography of complex test-structures before and after removal of the sacrificial mask. In all measurements by atomic-force microscopy, we use probes that consist of silicon-nitride cantilevers with a length of 27 μm, a resonant frequency of approximately 1400 kHz, a force constant of 17 N m$^{-1}$, and carrying a silicon tip with nominal front angle of 261 mrad ± 44 mrad (15° ± 2.5°), a nominal back angle of 446 mrad ± 44 mrad (25° ± 2.5°), a nominal radius of 5 nm, and a maximum radius of 12 nm.

We perform X-ray diffractometry with incident X-rays that correspond to copper Kα transitions, which have an energy of 1.29 fJ (8.04 keV) and a wavelength of 0.1506 nm. We fit an X-ray reflectivity model[26] to the resulting data to measure density, surface roughness, and thickness. We fit all models to data by damped least-squares estimation with uniform weighting. We apply the Scherrer equation to estimate crystallite size.

We perform scanning electron microscopy of secondary electrons incident on a through-lens detector with an acceleration voltage of 5 kV, an electron-beam current of approximately 100 pA, a working distance of 4.2 mm, at normal incidence.

We characterize the projection distances and apparent spot sizes of Fresnel lens by brightfield optical microscopy, using an objective lens with a nominal magnification of 50×, a numerical aperture of 0.95, and trans-illumination at wavelengths ranging from 420 nm to 510 nm and a peak wavelength of approximately 460 nm. We use an array of subwavelength apertures with a pitch of 5001.45 nm ± 1.08 nm to calibrate[42] the mean pixel size of all optical micrographs as 126.82 nm ± 0.03 nm. We record a series of micrographs of each lens and spot in focal volumes of approximately 258 µm by 258 µm by 40 µm, scanning through focus in increments of approximately 40 nm. We measure the vertical positions of best focus of lens surfaces by finding local maxima of a ninth-order polynomial model of image contrast in a range of images that include the top surface of the central zones of the lenses. We measure the vertical positions of best focus of spots by fitting symmetric bivariate Gaussian models to the spot images and finding local minima of a ninth-order polynomial model of the apparent standard deviations. We calculate the projection distance as the difference between the two positions of best focus.

*In-line Resolution Metrology*
We correlate measurements of test pits by scanning electron micrography and atomic force microscopy, enabling quantitative assessment of pit radius by scanning electron microscopy as a measure of effective lateral patterning resolution just prior to the fabrication of complex test-structures. Our analysis correlates the maximum signals of intensity of secondary electron scattering of pits from scanning electron micrographs with the negative of the second derivative of their surfaces from atomic force micrographs, the latter of which requires second-order differentiation with respect to lateral position. We propagate uncertainties by Monte-Carlo simulations (Note S4, Table S3).

*Milling Responses*
We measure the depth, surface roughness, and edge width of square features in the checkerboard test-patterns by analyzing atomic force micrographs. We level all atomic force micrographs and propagate all uncertainties from atomic force microscopy by Monte-Carlo simulations (Table S3). We use an error function to empirically model the z position $z_{\text{before}}(s)$ in sections of atomic force micrographs that transition from the bottom of a feature that we mill to the top of the chromia surface before chromia removal, Equation (6),

$$z_{\text{before}}(s) = \frac{d}{2}\left[\text{erf}\left(\frac{s-s_0}{\sqrt{2}\sigma_{\text{edge}}}\right) + 1\right] + c, \qquad (6)$$

and an error function that we truncate at the zero plane to empirically model the same features after chromia removal, Equation (7),

$$z_{\text{after}}(s) = \begin{cases} d \cdot \text{erf}\left(\frac{s-s_0}{\sqrt{2}\sigma_{\text{edge}}}\right) + c & s \leq s_0 \\ c & s > s_0 \end{cases}, \qquad (7)$$

where $d$ is the depth of the feature, $s_0$ is the location of the edge, $\sigma_{\text{edge}}$ is the standard deviation of the Gaussian function, and $c$ is a constant. We approximate the width of edges as the 95 % coverage interval of the width of the error functions, which correspond to $w_{\text{before}} = 4\sigma_{\text{edge}}$ before chromia removal and $w_{\text{after}} = 2\sigma_{\text{edge}}$ after chromia removal. We repeat this measurement along the widths of edges for each square feature in the checkerboard patterns.

Theoretical
*Ion Penetration*
We perform three simulations of ion penetration into the bilayer using Stopping and Range of Ions in Matter (SRIM) software (Table S7).[20] In three separate simulations, we input material densities, elemental compositions, and a chromia thickness of either 63.5 nm, corresponding to milling bulk chromia, 16 nm, corresponding to the z position at which the chromia–silica interface evidently begins to influence the milling response of chromia, according to our experimental analysis of milling responses, and 0 nm, corresponding to milling bulk silica. In each Monte-Carlo simulation, 100000 gallium ions, each with a landing energy of 4.81 fJ or 30 keV, bombard either a chromia on silica target or a silica target at normal incidence. We compute the resulting spatial distributions of gallium ions in the target materials to guide our interpretation of the nanostructure topographies that we measure by atomic force microscopy.

*Spatial Masking*
We derive an analytic expression for super-resolution factors that result after milling a uniform semi-infinite region through a sacrificial mask and into an underlying substrate in one dimension. For simplicity, we assume that the mask has an average milling rate $\bar{m}_{\text{m}}$ and a thickness $z_{\text{m}}$ and that the substrate has an average milling rate $\bar{m}_{\text{s}}$ and a final depth $z_{\text{s}}$. A Gaussian approximation of the distribution of current density of the ion beam leads to a model of the spatial profile of the ion dose along the x direction, $D(x)$, as an error function of the form,

$$D(x) \cong D_0\left[1 - \frac{1}{2}\left(\text{erf}\left(\frac{x-x_0}{\sqrt{2}\sigma}\right) + 1\right)\right], \qquad (8)$$

where $D_0$ is the dose of the ion beam, $x_0$ is the center position of the edge of the pattern, which we assign to be zero, and $\sigma$ is the effective standard deviation of the Gaussian profile of the ion beam. The dose necessary to mill through the mask and to a depth, $z_s$, into the substrate is the sum of the dose necessary to mill through the mask, $D_m$, and the dose necessary to mill the nanostructure, $D_s$. Accordingly, $D_0 = D_m + D_s = z_m \bar{m}_m^{-1} + z_s \bar{m}_s^{-1}$. Milling depths, $z_m$ and $z_s$, have units of length and milling rates, $\bar{m}_m$ and $\bar{m}_s$, have units of volume per current per second or µm³ nA⁻¹ s⁻¹. The quotient of measurements of length and milling rate yields a value with units of dose, nA s µm⁻², which we convert to pC µm⁻².

The extent to which the ion beam mills the underlying substrate corresponds to the value of the x position of the dose in Equation (8) that exceeds the dose necessary to mill through the mask. In other words, the x position at which the mask begins to screen the tail of the ion beam occurs where $D(x) = D(x_m) = D_m$ (Figure S12),

$$D_m = \frac{z_m}{\bar{m}_m} = \left[\frac{z_m}{\bar{m}_m} + \frac{z_s}{\bar{m}_s}\right]\left[1 - \frac{1}{2}\left(\text{erf}\left(\frac{x_m}{\sqrt{2}\sigma}\right) + 1\right)\right] \\ = \frac{z_m \bar{m}_s + z_s \bar{m}_m}{\bar{m}_m \bar{m}_s}\left[1 - \frac{1}{2}\left(\text{erf}\left(\frac{x_m}{\sqrt{2}\sigma}\right) + 1\right)\right]. \quad (9)$$

Two dimensionless parameters simplify Equation (9)—the ratio of the milling depth of the substrate to the thickness of the mask, $\zeta = z_s z_m^{-1}$, and the ratio of the milling rate of the substrate to the milling rate of the mask, $\mathcal{S} = m_s m_m^{-1}$, which is the physical selectivity. Making these substitutions in Equation (9) and solving the resulting expression for $x_m$ yields the position at which the mask begins to screen the tail of the ion beam,

$$x_m(\sigma, \mathcal{S}, \zeta) = \sqrt{2}\sigma\, \text{erf}^{-1}\left(1 - \frac{2\mathcal{S}}{\mathcal{S} + \zeta}\right) = \sqrt{2}\sigma\, \text{erfc}^{-1}\left(\frac{2\mathcal{S}}{\mathcal{S} + \zeta}\right). \quad (10)$$

We compute the super-resolution factor as the ratio of the edge width resulting from a low value of ion-beam current, $w_{\text{low}}$, to the edge width resulting from a high value of ion-beam current, $w_{\text{high}}$ – after removal of the sacrificial film, exploiting the favorable yield of secondary electrons of chromia to consistently focus the ion beam and circumventing the need for continuous control of current. We define the edge widths of the nanostructures to be the 95 % coverage interval of the width of the error function, which extends to $\pm 2\sigma_{\text{low}}$ on either side of the center position of the edge resulting from low current, where $\sigma_{\text{low}}$ is the effective standard deviation of the Gaussian profile of the beam resulting from a low current. The extents of the edge resulting from high current before removal of the sacrificial film are similarly $\pm 2\sigma_{\text{high}}$, but after removal of the sacrificial mask layer, the edge width decreases from $4\sigma_{\text{high}}$ to a final width of $w_{\text{high}} = x_m(\sigma_{\text{high}}, \mathcal{S}, \zeta) + 2\sigma_{\text{high}}$. The ratio of widths from either case yields the super-resolution factor, which is valid for $x_m(\sigma_{\text{high}}, \mathcal{S}, \zeta) > -2\sigma_{\text{high}}$,

$$w_{\text{low}} = 4\sigma_{\text{low}}, \quad (11)$$

$$w_{\text{high}} = \sqrt{2}\sigma_{\text{high}} \text{erfc}^{-1}\left(\frac{2\mathcal{S}}{\mathcal{S} + \zeta}\right) + 2\sigma_{\text{high}}, \quad (12)$$

and

$$\mathcal{F}_{\text{SR}}(\sigma_{\text{low}}, \sigma_{\text{high}}, \zeta, \mathcal{S}) = \frac{w_{\text{low}}}{w_{\text{high}}} = \left(\frac{\sigma_{\text{low}}}{\sigma_{\text{high}}}\right)\left(\frac{2}{\frac{1}{\sqrt{2}}\text{erfc}^{-1}\left(\frac{2\mathcal{S}}{\mathcal{S} + \zeta}\right) + 1}\right). \quad (13)$$

Next, we apply the experimental constraint of sparse and discontinuous values of ion-beam current to Equation (13), to model the effect of removal of the sacrificial mask on edge width. Rather than analyzing the patterning resolution resulting from two different values of current that yield similar edge widths – either in the absence of a mask for low current or after removal of the mask for high current – we constrain the current to a single value, such that $\sigma_{\text{low}} = \sigma_{\text{high}} = \sigma$, and analyze the effect of the mask on the edge profile. This constraint yields an expression that is applicable to our experimental system,

$$\mathcal{F}_{\text{SR}}(\zeta, \mathcal{S}) = \frac{w_b}{w_a} = \frac{4\sigma}{x_m(\sigma, \mathcal{S}, \zeta) + 2\sigma} = \frac{2}{\frac{1}{\sqrt{2}}\text{erfc}^{-1}\left(\frac{2\mathcal{S}}{\mathcal{S} + \zeta}\right) + 1}. \quad (14)$$

*Line–Space Patterns*

We simulate the focused-ion-beam milling of periodic lines and spaces to estimate the limitations of the process to form dense arrays in the presence and absence of a sacrificial mask (Figure S15, Figure S16). We apply our knowledge of the vertical milling response of our bilayer to predict structures resulting from patterns of ion dose with pitch, $p$, ranging from $1\sigma$ to $10\sigma$ and duty cycle, $\delta$, ranging from 0 % to 100 %. For each set of parameters, we compute vertical profiles in silica that result from either direct exposure, $z_{\text{direct}}(x)$,

$$z_{\text{direct}}(x) = m_{\text{silica}}\left(D_{\text{ls}}(x, \sigma, \zeta_{\text{design}}, p, \delta)\right), \tag{15}$$

where $m_{\text{silica}}$ is the milling rate of bulk silica and $D_{\text{ls}}(x)$ is a spatial pattern of ion dose resulting from positioning a Gaussian ion beam with standard deviation $\sigma$, to form lines and spaces of design depth, $\zeta_{\text{design}}$ – or from milling through a sacrificial mask, $z_{\text{masking}}(x)$,

$$z_{\text{masking}}(x) = m_{\text{bilayer}}\left(D_{\text{ls}}(x, \sigma, \zeta_{\text{design}}, p, \delta)\right), \tag{16}$$

where $m_{\text{bilayer}}$ is the milling rate of the chromia and silica bilayer. Both profiles are simplifications that neglect the dependence of incidence angle on milling rate, any effects of redeposition, and any defocus of the ion-beam due to charging. We compute the fraction of predicted depth to design depth as a metric of fabrication feasibility. We define a tolerance of 0.975 to identify values of minimum pitch at which patterning is feasible and compare values of minimum pitch for direct milling and milling in the presence of a sacrificial mask.

*Temporal Efficiency*
We calculate values of the temporal efficiency, which we derive analytically (Note S8), as functions of milling depth after normalization by mask thickness and of physical selectivity, by Equation (4) and using Monte-Carlo methods to propagate uncertainties from physical selectivity, and of the scaling exponent of our effective lateral patterning resolution, which we estimate from fits of a power law model to edge widths as a function of ion-beam current (Table S3). Using parameters from such random sampling, we simulate values of temporal efficiency for $\tilde{z}$ ranging from $10^{-1}$ to $10^4$ and compute values of milling depth where $\eta_\tau = 1$ by interpolation to define maximum milling depths for five mean values of physical selectivity ranging from 0.1 to 10 in increments with exponential spacing. We repeat this process $10^4$ times to construct distributions of maximum milling depth. Similarly, we calculate distributions of maximum milling depth of our system using our analytic expression for temporal efficiency and propagating uncertainties from atomic force microscopy, normal distributions of mask thickness, uncertainties from physical selectivity, and of the exponent of the power-law model of our effective lateral patterning resolution (Table S11, Table S3). We use kernel density estimation[43] with a Gaussian kernel to estimate the joint probability density of physical selectivity and maximum milling depth.

*Throughput*
We calculate values of volume throughput for equivalent values of resolution and super-resolution in two steps. We fit power-law models to experimental values of resolution or super-resolution as a function of volume throughput. For each value of super-resolution, we set the power-law model of the resolution trend equal to the value of super-resolution and invert the resolution model, extrapolating outside of our range of experimental values, to solve for volume throughput. We refer to the results of such extrapolation as values of equivalent throughput (Table S13).

## Supporting Information
Supporting Information is available from the Wiley Online Library or from the corresponding author.

## Acknowledgements
The authors acknowledge helpful comments from Lindsay C. C. Elliot, Henri J. Lezec, Jacob M. Majikes, and Adam L. Pintar, and inspiration on silicate formation from Sven Larsen. A. C. M. acknowledges support of a National Research Council Research Associateship. S.M.S. supervised the study. A.C.M. and S.M.S. designed the study with contributions from all authors. K.-T.L. prepared silica substrates. B.R.I. optimized the deposition of chromia. A.C.M. performed atomic force microscopy. A.C.M and K.S. performed X-ray diffraction experiments and analysis. K.S. performed transmission electron microscopy. A.C.M., J.S.V., and S.M.S. developed the in-line metrology method. A.C.M. and J.S. performed focused-ion-beam milling and scanning electron microscopy. A.C.M. and S.M.S. performed SRIM simulations. A.C.M., J.S.V., and S.M.S. developed the theoretical model of super-resolution. A.C.M. and S.M.S. designed the Fresnel lenses. C.R.C. and A.C.M. performed optical microscopy. A.C.M. and S.M.S. designed the statistical analysis and prepared the manuscript with contributions from all authors.

## Conflict of interest
The authors declare no financial or non-financial conflicts of interests.

# Supporting Information *for*
# Unmasking the resolution–throughput tradespace of focused-ion-beam machining

Andrew C. Madison[1] John S. Villarrubia[1] Kuo-Tang Liao[1,2] Craig R. Copeland[1] Joshua Schumacher[3] Kerry Siebein[3] B. Robert Ilic[1,3] J. Alexander Liddle[1] and Samuel M. Stavis[1,*]

## Index[4,5]

[1]Microsystems and Nanotechnology Division, National Institute of Standards and Technology, Gaithersburg, Maryland, USA. [2]Maryland Nanocenter, University of Maryland, College Park, Maryland, USA. [3]CNST NanoFab, National Institute of Standards and Technology, Gaithersburg, Maryland, USA. *e-mail: samuel.stavis@nist.gov. [4]We report uncertainties as 95% coverage intervals, or we note otherwise. [5]We fit models to data by damped least-squares with uniform weighting, or we note otherwise.

**Note S1**. Fabrication methods

We form a silica layer on substrates of p-type silicon with a crystallographic orientation of ⟨100⟩ by thermal oxidation in a furnace at atmospheric pressure, at a temperature of 1100 °C, with an oxygen flow rate of 1000 mL min$^{-1}$ (1000 sccm) and a ratio of hydrogen to oxygen of 1.85. We sputter-deposit chromia, $Cr_2O_3$, of at least 99.8% purity on the silica layer, with a deposition power of 400 W, under argon at a pressure of $1.3 \times 10^{-3}$ Pa ($9.8 \times 10^{-6}$ Torr), and with a deposition rate of $0.215 \pm 0.005$ nm s$^{-1}$, for 285 s.

We use an electron–ion beam system with a focused beam of gallium cations at an accelerating voltage of 30 kV, ion-beam currents ranging from $82 \pm 1$ pA to $796 \pm 4$ pA, which we measure prior to fabrication by deflecting the ion beam into a Faraday cup, a working distance between the final lens and the substrate surface of 4.2 mm, and normal incidence with respect to the sample (Table S2). Uncertainties of ion-beam currents are conservative estimates of 100% coverage intervals. We propagate uncertainties by Monte-Carlo methods.[1] We fit manufacturer specifications of the half-width at half-maximum of the focused ion beam as a function of ion-beam current to the power-law model in Equation (1).

We approximate the radius of our focused ion beam from the values of ion-beam current that we measure prior to fabrication (Table 11, Table S3). We use the radii from this power-law approximation to achieve an overlap of the ion beam of at least 50% between neighboring positions in the patterns.

For pattern control, we use text files with matrices that define lateral positions, dwell times, and Boolean parameters for beam deflection. We mill two test patterns. The first is a 10 by 10 array of points with dwell times that increase linearly across columns. We write this pattern in multiple passes ranging in total dwell time from 0.05 to 5 s to form approximately Gaussian pits (Table S2). The second test pattern resembles a darkening checkerboard, being a 20 by 20 array of adjacent squares of 1 by 1 μm, with ion doses that alternate between 0 pC μm$^{-2}$ and a value that increases from 0 to approximately 1000 pC μm$^{-2}$ across the columns and rows of the array in uniform increments of approximately 5 pC μm$^{-2}$. We choose this dose increment to target depth increments of 1 nm between adjacent squares in the checkerboard pattern on the basis of tests of the milling rate of silica.[2] We remove the sacrificial chromia mask by immersion in a mixture of nitric acid, ceric acid, and water with respective volumetric fractions of 6%, 16%, and 78% for 5 min.

We mill Fresnel lenses into silica coverslips with a thickness of approximately 170 μm, a root-mean-square surface roughness of less than 0.8 nm, and a surface quality with a scratch/dig specification of 20/10. We sputter-deposit a sacrificial mask of chromia with a thickness of $103 \pm 2$ nm onto a silica coverslip. We estimate the ion doses for radial profiles of Fresnel lenses from design depths and mean milling rates of chromia and silica (Note S10). We calculate the curvature of the lens design and trim the ion dose where the curvature drops below a critical value set by the inverse of the lateral extent of the ion beam (Note S11). We use the resulting radial profiles to interpolate values of ion dose for milling positions for each lens in a Cartesian coordinate system. We mill a single Fresnel lens directly into a silica coverslip with an ion-beam current of $26.3 \pm 0.3$ pA in a time of 3.75 h. We mill 75 Fresnel lenses in a square array through the sacrificial chromia mask and into silica with an ion-beam current of $2600 \pm 30$ pA in a time of 3.75 h. In this process, we aim to exhaust the mask, milling through its entire thickness to pattern each Fresnel lens and obviating the need for subsequent removal of the chromia by chemical etching.

**Note S2**. Characterization methods

We measure the thickness of the silica layer by ellipsometry. We measure the surface topography of the silica substrates before and after depositing the chromia mask. For all atomic force micrographs, we image regions of interest of approximately 25 by 25 μm with a line-scan resolution of 2048 points and at a rate of approximately 0.67 Hz. We use silicon-nitride cantilevers with a length of 27 μm, a resonant frequency of approximately 1400 kHz, a force constant of 17 N m$^{-1}$, and carrying a silicon tip with nominal front angle of $261 \pm 44$ mrad (15° $\pm$ 2.5°), a nominal back angle of $446 \pm 44$ mrad (25° $\pm$ 2.5°), a nominal radius of 5 nm, and a maximum radius of 12 nm. Assuming this geometry, the width of the probe is nominally $75 \pm 7$ nm at a depth of 100 nm, which is comparable to the smallest radii of test pits that we measure. Such a geometry imposes an upper limit on the aspect ratio of features that we expect to measure accurately. We input these manufacturer specifications of the probe tip geometry into a certainty-map algorithm to identify and ignore data in atomic force micrographs that exhibit artifacts from tip convolution.[3] We estimate uncertainties of atomic force microscopy by a combination of manufacturer specifications and previous tests (Table S3).[2]

We perform scanning electron microscopy with an acceleration voltage of 5 kV, a nominal electron-beam current of 100 pA, a working distance of 4.2 mm, and at normal incidence with respect to the microscope stage. In all scanning electron micrographs, the imaging mode is of secondary electrons incident on a through-lens detector. Before ion exposure, we record scanning electron micrographs of the chromia mask at a magnification of 250000× $\pm$ 7500×, which corresponds to a horizontal field width of $597 \pm 18$ nm. This uncertainty is a 100% coverage interval per the microscope specification.

We prepare a cross section of the chromia and silica layers for transmission electron microscopy by *ex situ* lift-out.[4] The dimensions of the cross section are approximately 10 μm in length, 5 μm in width, and 100 nm in depth. We image the cross section by brightfield transmission electron microscopy at an acceleration voltage of 300 kV.

We determine the solid-state of chromia by X-ray diffraction from 0.35 to 1.40 rad in increments of 0.1 mrad with a total reflection critical angle of 7.0 mrad. The incident X-rays correspond to copper Kα transitions, which have an energy of 1.29 fJ (8.04 keV) and a wavelength of 0.1506 nm. We fit an X-ray reflectivity model[5] to the resulting data to measure density, surface roughness, and thickness. We fit all models to data by damped least-squares estimation with uniform weighting. We estimate the sizes of crystallites present in X-ray diffractometry data using the Scherrer equation.

For in-line resolution metrology, we measure pits by scanning electron microscopy at a magnification of 5000× $\pm$ 150×, which corresponds to a horizontal field width of $29.8 \pm 0.9$ μm. We measure the surface topography of complex nanostructures that we mill through chromia and into silica by atomic force microscopy before and after removal of the sacrificial mask.

We record brightfield micrographs of Fresnel lenses to characterize the projection distances and apparent sizes of focal spots from each lens through a focal volume of approximately 258 by 258 µm by 40 µm. We axially section the focal volume into 1001 micrographs, scanning vertically in increments of 40 ± 10 nm. For these micrographs, a light-emitting diode trans-illuminates the samples with a wavelength range from 420 to 510 nm. An objective lens with a nominal magnification of 50× and a numerical aperture of 0.95 collects transmission through air immersion. A tube lens projects the images onto a complementary metal–oxide–semiconductor (CMOS) camera with 2048 by 2048 pixels, each with an on-chip size of 6.5 by 6.5 µm. A mean factor of 2.0, per the specification of the camera manufacturer, converts from photoelectrons to analog-to-digital units. We operate the camera at a sensor temperature of -10 °C by thermoelectric and water cooling, without on-board correction of pixel noise, and in fast-scan mode, and we calibrate the imaging system for these parameters.

We calibrate the imaging system with an aperture array that has a pitch of 5001.45 ± 1.08 nm, determining a mean pixel size of 126.82 ± 0.03 nm.[6] We determine positions of best focus of lens surfaces by finding local maxima of a ninth-order polynomial model of image contrast in a range of images that we identify by inspection to include the top surface of the central zones of the lenses. We determine positions of best focus of images of focal spots by fitting symmetric Gaussian models to image data and minimizing a ninth-order polynomial model that we fit to the resulting standard deviations in a range of images that we identify by inspection. We determine projection distance as the difference between the two positions of best focus.

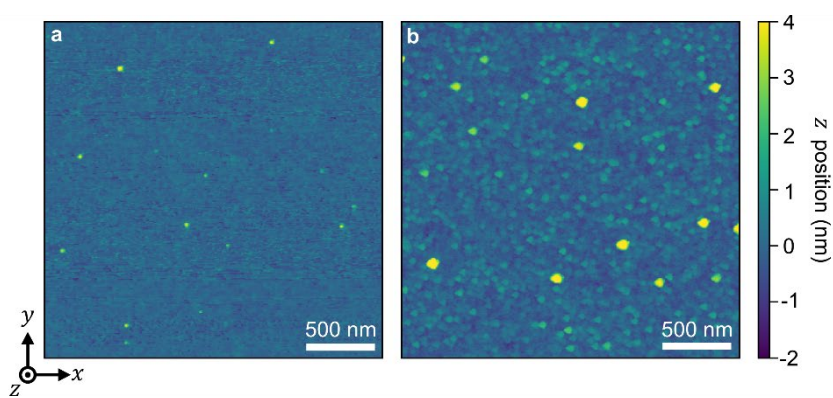

**Figure S1.** Silica and chromia. Atomic force micrographs showing a) silica before deposition of chromia and b) chromia before ion exposure. Comparison of the micrographs indicates that the areal densities of silica asperities and chromia patches with lateral dimensions exceeding 50 nm are comparable, as well as a convolution artifact of the probe tip in (b).

**Table S1**. Material properties

| Material | Composition | Function | Measurement method | Density (g cm$^{-3}$) | Thickness (nm) or (µm) | Root-mean-square roughness (nm) |
|---|---|---|---|---|---|---|
| Chromia | Cr$_2$O$_3$ | Sacrificial mask | Atomic force microscopy | – | – | 0.6 ± 0.2 |
| | | | Transmission electron microscopy | – | 63 ± 2 nm | – |
| | | | X-ray diffraction | 5.3 ± 0.1 | 65 ± 3 nm | 1.5 ± 0.4 |
| Silica | SiO$_2$ | Working material | Atomic force microscopy | – | – | 0.3 ± 0.2 |
| | | | X-ray diffraction | 2.2 ± 0.1 | 488 ± 2 nm | 0.4 ± 0.4 |
| Silicon | Si | Substrate | Atomic force microscopy | – | – | 0.3 ± 0.2 |
| | | | Manufacturer specification | 2.3 | 525 ± 25 µm | – |

Uncertainties of density, thickness, and roughness by X-ray diffraction are conservative estimates of 100% coverage intervals.

**Table S2.** Milling parameters

| Ion-beam current (pA) | Feature | Elements in array | Lateral extent of array (μm by μm) | Pitch of milling positions (nm) | Number of milling positions | Dwell time (μs) | Number of passes | Total milling time (s) |
|---|---|---|---|---|---|---|---|---|
| 82 ± 1 | pits | 10 by 10 | 18 by 18 | 12.3 | 100 | 100 | 3052 | 275.2 |
| 219 ± 2 | pits | 10 by 10 | 18 by 18 | 19.8 | 100 | 100 | 1526 | 137.6 |
| 407 ± 2 | pits | 10 by 10 | 18 by 18 | 26.8 | 100 | 100 | 611 | 55.0 |
| 773 ± 3 | pits | 10 by 10 | 18 by 18 | 36.7 | 100 | 100 | 306 | 27.5 |
| 83 ± 1 | squares | 20 by 20 | 20 by 20 | 12.4 | 1312200 | 100 | 59 | 1241.9 |
| 227 ± 1 | squares | 20 by 20 | 20 by 20 | 20.2 | 472424 | 100 | 14 | 434.2 |
| 421 ± 3 | squares | 20 by 20 | 20 by 20 | 27.3 | 253504 | 100 | 22 | 228.9 |
| 796 ± 4 | squares | 20 by 20 | 20 by 20 | 37.2 | 133136 | 100 | 56 | 118.6 |
| 26 ± 0.3 | lenses | 1 by 1 | 15 by 15 | 7.0 | 3414413 | 50 | 155 | 13500 |
| 2600 ± 30 | lenses | 7.5 by 10 | 150 by 120 | 66.4 | 2791725 | 50 | 166 | 13500 |
| 2600 ± 30 | lenses | 1 by 1 | 15 by 15 | 66.4 | 36473 | 50 | 143 | 144 |

Uncertainties of ion-beam current are conservative estimates of 100% coverage intervals.
The milling time of pits includes a 0.4 μs pause to unblank and blank the ion beam before and after milling at each beam position.

**Table S3.** Statistical variables

| Variable | Type of variable | Symbol | Distribution | Type[7] of evaluation | Value | Units |
|---|---|---|---|---|---|---|
| Localization uncertainty in SEM micrographs | Uncertainty | $u_{loc,\,SEM}$ | Normal | B | Mean: $s_{0,\,SEM}$, SD: $\sigma_{loc,\,SEM}$ | nm |
| Localization uncertainty in AFM micrographs | Uncertainty | $u_{loc,\,AFM}$ | Normal | B | Mean: $s_{0,\,AFM}$, SD: $\sigma_{loc,\,AFM}$ | nm |
| Standard deviation of Gaussian filter | Dimension | $\sigma_G$ | Uniform | A | Range: 10 to 30 | nm |
| Window length of Savitzky-Golay filter | Dimension | $\omega_{SG}$ | Uniform | A | Range: 13 to 27, odd values | pixels |
| Secondary electron scattering intensity | Dimension | $I_{SE}$ | Normal | B | Mean: $I_{SE}(s(\theta))$, SD: $\sigma_{SE}$ | arb. |
| Magnification uncertainty in SEM micrographs | Uncertainty | $u_{mag}$ | Uniform | B | Range: $-0.03 a_{SEM}$ to $0.03 a_{SEM}$ | – |
| Mean pixel size of SEM micrographs | Dimension | $a_{SEM}$ | Delta | A | 4.9 | nm |
| Calibration errors of atomic force microscope | Uncertainty | $u_{cal}$ | Normal | A | Mean: 0, SD: $0.0025 \cdot z$ | nm |
| Position errors from surface roughness | Uncertainty | $u_{rough}$ | Normal | A | Mean: 0, SD: 0.030 | nm |
| Position errors from flatness | Uncertainty | $u_{flat}$ | Normal | A | Mean: 0, SD: 0.065 | nm |
| Position uncertainty due to AFM probe tip | Dimension | $u_{tip}$ | Uniform | B | Range: $-0.5 \cdot r_{tip}$ to $0.5 \cdot r_{tip}$ | nm |
| Maximum nominal radius of AFM probe tip | Dimension | $r_{tip}$ | Delta | A | 5 | nm |
| Radius of pits in SEM micrographs | Dimension | $r_{p,\,SEM}$ | Empirical | B | $u_{loc,\,SEM}, \omega_{SG}, \sigma_G, I_{SE}, u_{mag}$ | nm |
| Radius of pits in AFM micrographs | Dimension | $r_{p,\,AFM}$ | Empirical | B | $u_{loc,\,AFM}, \omega_{SG}, \sigma_G, u_{cal}, u_{rough}, u_{flat}, u_{tip}$ | nm |
| X-ray diffraction intensity | Dimension | $I_{XRD}$ | Normal | B | Mean: $I_{XRD}$, SD: $\sigma_{XRD}$ | arb. |
| Chromia density | Dimension | $\rho_m$ | Normal | A | Mean: 5.3, SD: 0.05 | g cm$^{-3}$ |
| Silica density | Dimension | $\rho_s$ | Normal | A | Mean: 2.2, SD: 0.05 | g cm$^{-3}$ |
| Root-mean-square roughness of chromia | Dimension | $R_{q,\,m}$ | Normal | A | Mean: 1.5, SD: 0.4 | nm |
| Root-mean-square roughness of silica | Dimension | $R_{q,\,s}$ | Normal | A | Mean: 0.4, SD: 0.2 | nm |
| Depth of features in AFM micrographs | Dimension | $z_s$ | Empirical | B | $u_{cal}, u_{rough}, u_{flat}, u_{tip}$ | nm |
| Chromia thickness | Dimension | $z_m$ | Normal | A | Mean: 63, SD: 1 | nm |
| Milling rates of chromia on silica | Dimension | $m_i$ | Normal | B | Table S6 | µm$^3$ nA$^{-1}$ s$^{-1}$ |
| Intercepts of milling responses | Dimension | $b_i$ | Normal | B | Table S6 | nm |
| Mean milling rate of chromia | Dimension | $\bar{m}_m$ | Normal | B | Mean: 0.15, SD: 0.02 | µm$^3$ nA$^{-1}$ s$^{-1}$ |
| Mean milling rate of silica | Dimension | $\bar{m}_s$ | Normal | B | Mean: 0.24, SD: 0.02 | µm$^3$ nA$^{-1}$ s$^{-1}$ |
| Ion-beam current | Dimension | $I$ | Normal | A | Table S2 | pA |
| Coefficient of power law of ion-beam HWHM | Dimension | $\alpha_{beam}$ | Normal | B | Mean: 1.43, SE: 0.12 | nm pA$^{-\beta}$ |
| Exponent of power law of ion-beam HWHM | Dimension | $\beta_{beam}$ | Normal | B | Mean: 0.49, SE: 0.01 | – |
| Coefficient of power law of width of step edges | Dimension | $\alpha$ | Normal | B | Mean: 60, SD: 13 | nm pA$^{-\beta}$ |
| Exponent of power law of width of step edges | Dimension | $\beta$ | Normal | B | Mean: 0.20, SD: 0.05 | – |
| Widths of edges before chromia removal | Dimension | $w_b$ | Empirical | B | $u_{cal}, u_{rough}, u_{flat}, u_{tip}$ | nm |
| Widths of edges after chromia removal | Dimension | $w_a$ | Empirical | B | $u_{cal}, u_{rough}, u_{flat}, u_{tip}$ | nm |
| Effective lateral resolution or super-resolution | Dimension | $\mathcal{R}$ | Empirical | B | $\alpha, \beta, \mathcal{F}_{SR}$ | nm |
| Super-resolution factor | Dimension | $\mathcal{F}_{SR}$ | Empirical | B | $u_{cal}, u_{rough}, u_{flat}, u_{tip}$ | – |
| Physical selectivity of chromia and silica | Dimension | $\mathcal{S}$ | Empirical | B | $u_{cal}, u_{rough}, u_{flat}, u_{tip}$ | – |
| Temporal efficiency | Dimension | $\eta_\tau$ | Empirical | B | $u_{cal}, u_{rough}, u_{flat}, u_{tip}$ | – |
| Temporally efficient milling range | Dimension | $\zeta_{max}$ | Empirical | B | $u_{cal}, u_{rough}, u_{flat}, u_{tip}, \beta$ | – |
| Volume throughput | Dimension | $\mathcal{V}$ | Empirical | B | $u_{cal}, u_{rough}, u_{flat}, u_{tip}, \bar{m}_m, \bar{m}_s, i$ | µm$^3$ hr$^{-1}$ |
| Figure of merit of focused-ion-beam milling | Dimension | $\mathcal{M}$ | Empirical | B | $\mathcal{R}, \mathcal{V}$ | µm$^2$ hr$^{-1}$ |

SEM = scanning electron microscopy
SD = standard deviation
AFM = atomic force microscopy
HWHM = half width at half maximum
SE = standard error
We treat the bulk milling rate of silica as its mean value.
We extend the conventional evaluation Type from uncertainties to dimensions.

**Note S3**. Image formation

We observe a near equality of test-pit radii corresponding to the maximum signals of secondary electrons in scanning electron micrographs and the maximum convexity of surface profiles in atomic force micrographs. Secondary electrons from test pits form images with two main features of dark centers within bright rings, both of which are within the outer rims of the test pits (Figure 3c). In a previous study,[8] similar signals resulted from test pits in silicon ⟨111⟩ and glassy carbon. However, the authors of this previous study removed the bright rings by a threshold and approximated the dark central features with a symmetric bivariate Gaussian function. This approximation would be inaccurate for our scanning electron micrographs and would lead to errors in measurements of pit radius. To better understand the empirical correlation that we use for in-line resolution metrology, we investigate the influence of the surface topography on the theoretical signal of secondary electrons, using fundamental concepts of image formation in scanning electron microscopy. A symmetric bivariate Gaussian surface, which has deviations in z position of approximately 5% in comparison to the surface profiles of our test pits (Figure 3c), is a reasonable approximation of our pit topography for theoretical analysis. Our calculations lead to a new interpretation of the image data.

We expect three effects to dominate image contrast in scanning electron micrographs of pits. First, tilt contrast results from the dependence of the secondary-electron yield on the tilt angle, $\phi$, of the local surface-normal relative to the incident electron-beam. The secant of the tilt angle, $\sec\phi$, is a common approximation of secondary-electron emission due to surface topography.[9] Second, shadow contrast results from surfaces of the pit that reabsorb secondary electrons, obscuring access to the top surface of the sample where an extraction bias pulls electrons toward the detector, reducing signals that originate within the basin of the pit.[4, 9b, 10] Third, material contrast results from local variations in secondary-electron yield due to the presence of the chromia–silica bilayer, gallium dopants from the ion beam, and redeposition during sputtering, further modulating the superimposition of tilt and shadow effects. Although diffusion effects are also relevant, we expect a diffusion length of order $10^0$ nm for secondary electrons in our system. Additionally, we expect the diffusion effect to contribute more to image formation for surface regions of higher slope than for surface regions of lower slope. Therefore, we expect that our analysis of tilt contrast captures relevant spatial information from the diffusion effect.

To investigate the effects of pit topography on image formation, we compute tilt and shadow contrast for images of the surface of a Gaussian pit, $S(x,y)$, with an aspect ratio that is comparable to that of the test pits. With diameters ranging from approximately 160 to 360 nm and corresponding depths ranging from approximately 30 to 100 nm, the aspect ratios of test pits vary from approximately 0.2 to 0.3. For a path along the x direction and through the center of the Gaussian pit, we calculate the primary signal, $\sec\phi = \sqrt{1+(\partial S/\partial x)^2}$, resulting from tilt contrast. For simplicity, we exploit the axial symmetry of the pit and the orientation of the path of interest in the x direction to ignore the y component of the surface in our calculation of the primary signal. We model shadow effects by calculating the accessibility of the surface of the pit to the zero plane, which corresponds to the flat surface above the pit. For any point on the zero plane, the accessibility is the solid angle that a hemisphere subtends, $2\pi$ sr, as any secondary electron escaping the surface is free to move toward a detector above the surface. The concavity of the surface of a Gaussian pit reduces the accessibility from a maximum value of $2\pi$ sr far from the center of the pit to a minimum value at the center of the pit (Figure S2a).

During imaging, an extraction field of approximately 17 kV m$^{-1}$ pulls electrons toward an in-lens detector. However, the extraction field changes the electron energy by less than 1 part in 1000, so that straight lines are good approximations of electron trajectories within submicrometer test pits. This suggests that accessibility approximates shadow contrast from the surface of such pits. We calculate the accessibility, $\mathcal{A}(S, \boldsymbol{x}_i)$, of the $i^{th}$ point on a path through the center of $S$ in three dimensions by integrating the intervisibility function over the solid angle that the hemisphere subtends, $\Omega$, in equation (S1),

$$\mathcal{A}(S, \boldsymbol{x}_i) = \frac{1}{\pi} \int_\Omega v(S, \boldsymbol{x}_i, \boldsymbol{r}(\omega)) d\omega, \tag{S1}$$

where the intervisibility,[11] $v(S, \boldsymbol{r}(\omega), \boldsymbol{x}_i)$, is either one or zero if $S$ occludes a ray, $\boldsymbol{r}(\omega)$, emanating from a point on the surface profile, $\boldsymbol{x}_i = (x_i, y_i, z_i)$, in direction $\omega = (\theta, \varphi)$, where $\theta$ is the polar angle and $\varphi$ is the azimuthal angle. We compute the intervisibility as

$$v(S, \boldsymbol{r}(\omega), \boldsymbol{x}_i) = 0 \begin{cases} 1, z_r > z_i + \dfrac{\|\boldsymbol{r}(\omega)\|}{\|\boldsymbol{x}_k - \boldsymbol{x}_i\|}(z_k - z_i) \text{ for all } \boldsymbol{x}_k \\ 0, \text{otherwise} \end{cases}, \tag{S2}$$

where $\boldsymbol{x}_k = (x_k, y_k, z_k)$ are intermediate points on the surface of the pit between $\boldsymbol{x}_i$ and the projection of $\boldsymbol{r}(\omega)$ onto the surface, $z_r$ is the $z$ position of the terminus of $\boldsymbol{r}(\omega)$. We use Monte-Carlo integration to compute Equation (S1) along $\boldsymbol{x}_i$. We set the length of $\boldsymbol{r}(\omega)$ to span the computational domain and achieve uniform sampling of $\omega$ over $\Omega$ in $10^5$ random directions by $\theta = \cos^{-1}(\mathcal{U}(0,1))$ and $\varphi = 2\pi\mathcal{U}(0,1)$, where $\mathcal{U}(0,1)$ represents a uniform distribution of with lower bound of zero and an upper bound of one. Lastly, we calculate the second spatial derivative of $S$ and a factor of the signal intensity resulting from topographic contrast, $I_{\text{topography}}$, as the product of the surface-tilt and shadowing effects,

$$I_{\text{topography}} \propto \frac{\mathcal{A}(S, \boldsymbol{x}_i)}{2\pi} \sec \phi \,. \tag{S3}$$

Several results of this new analysis indicate a limit of tilt and shadow contrast to predict rings of high intensity in secondary electron images of Gaussian pits. First, the primary signals from tilt contrast appear as peaks in signal intensity at positions where the first spatial derivative, rather than the second spatial derivative, of $S$ are maximal (Figure S2b). Next, the accessibility of $S$ exhibits a concave structure, transitioning from maximum values of $2\pi$ sr outside of the pit to a minimum value of approximately $\pi$ sr in the center of the pit (Figure S2c). In comparison, the negative of the second spatial derivative of $S$ shows two maxima, each of which indicates positions of maximum convexity of $S$, and a global minimum in the center of the pit, indicating a position of maximum concavity (Figure S2d). We interpret these positions of maximum convexity as possible locations for material contrast to arise due to the implantation of gallium, interspersing of chromia and silica, and redeposition of the various materials on the surface of the pit. Last, the product of the tilt and shadow effects predicts a complex response that rises slowly near the outer rim of the pit and falls to a global minimum in the center of the pit (Figure S2e). The discrepancies between the product of tilt and shadow contrast, and the negative second derivative of $S$, also suggest that material contrast affects the formation of images of Gaussian pits. Further study is necessary to fully understand the cause of the empirical correlation of electron signals and pit curvature, which nonetheless enables in-line resolution metrology.

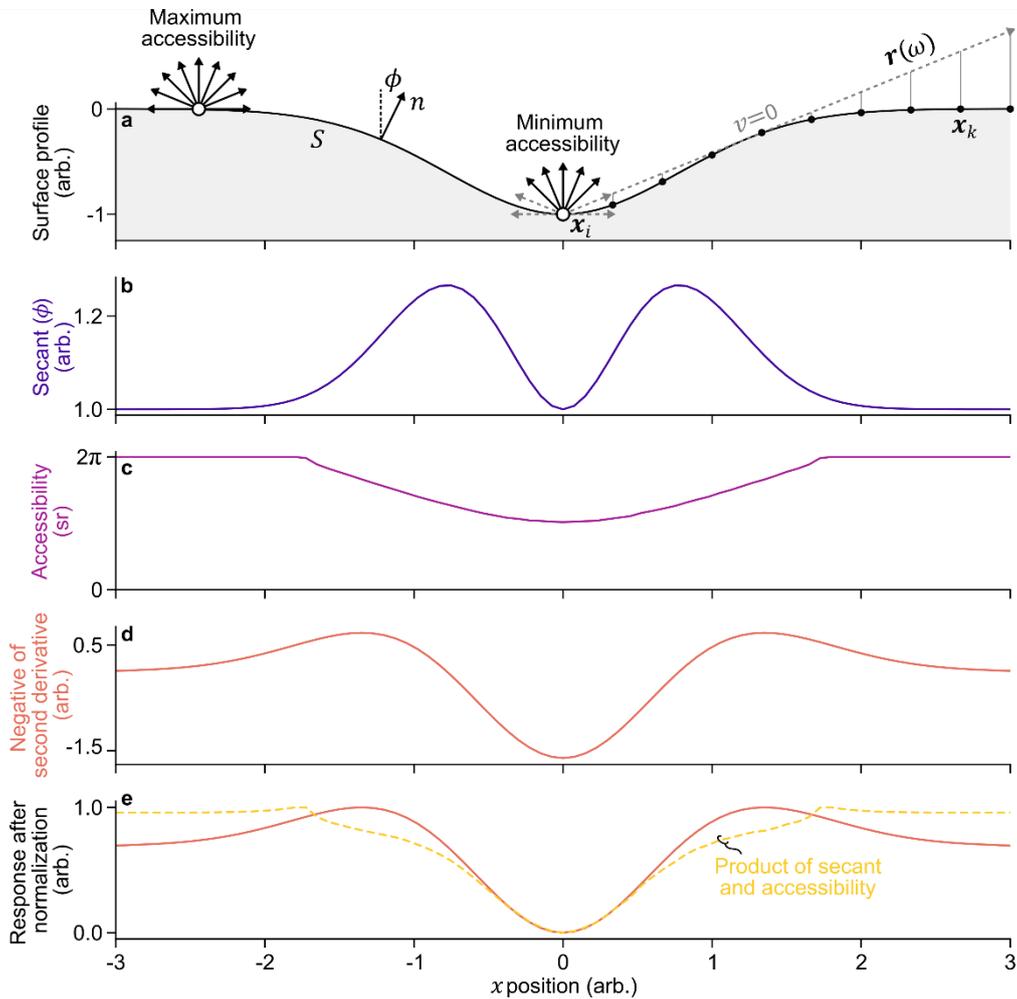

**Figure S2.** Image formation. Plots showing a) an arbitrary surface profile of a Gaussian pit, $S$, b) the secant of the tilt angle, $\phi$, of the surface profile normal, $n$, c) accessibility of the surface profile, d) the negative of the second derivative of the surface profile, and e) an overlay of (solid orange line) the negative of the second derivative and (yellow dash line) the product of $\sec \phi$ and accessibility. Black arrows in (a) indicate rays that escape the surface and gray dash arrows indicate rays that the surface occludes. Abrupt changes in (c) are artifacts from discretization of the computational domain and numerical integration.

**Note S4.** In-line resolution metrology

To measure pit radii in both scanning electron micrographs and atomic force micrographs, we localize pit centers using empirical model approximations and then localize maxima in radial sections of regions of interest of each pit. We propagate uncertainty through our analysis by perturbing position and image data in Monte-Carlo simulations of our measurements.

Our analysis begins by thresholding each micrograph and computing centroids of simply connected regions to obtain the approximate positions of each pit. To localize the center of each pit in each imaging mode, we fit model approximations to image data. In images of pits in atomic force micrographs, we approximate the depths by an asymmetric bivariate function, $G(x,y)$,

$$G(x,y) = A_G \cdot \exp\left\{-\left(\frac{1}{2(1-\rho^2)}\left[\frac{(x-x_0)^2}{\sigma_x^2} - 2\rho\frac{(x-x_0)(y-y_0)}{\sigma_x\sigma_y} + \frac{(y-y_0)^2}{\sigma_y^2}\right]\right)\right\} + c, \tag{S4}$$

where $A_G$ is the amplitude, $(x_0, y_0)$ is the center position of the pit in the x and y directions, $\sigma_x$ and $\sigma_y$ are the standard deviations in the x and y directions, $\rho$ is the correlation coefficient between the x and y directions, and $c$ is a constant background. In images of pits in scanning electron micrographs, we approximate the secondary-electron intensity as the sum of two functions. First, the Gaussian function in Equation (S4) approximates the intensity minimum at the center of a pit. Second, an elliptical annulus, $E(x,y)$, with the Gaussian profile in Equation (S5), approximates the ring of maximum intensity that circumscribes the center of the pit,

$$E(x,y) = A_E \cdot \exp\left\{-\left(\frac{1}{2\omega^2(1-\rho^2)}\left[\frac{(x-x_0)^2}{\sigma_x^2} - 2\rho\frac{(x-x_0)(y-y_0)}{\sigma_x\sigma_y} + \frac{(y-y_0)^2}{\sigma_y^2} - 1\right]^2\right)\right\} + c, \tag{S5}$$

where $A_E$ is the amplitude, $\omega$ is the width of the annulus. We estimate the localization uncertainties of the center of pit images by perturbing image and position data of either micrograph, fitting models to the resulting data, and extracting the resulting center positions. Repeating this process $10^3$ times for each pit, we construct distributions of center positions from which we calculate standard deviations as localization uncertainties of center positions for each micrograph of each pit. We perturb values of position within regions of interest with random noise from uniform distributions, which correspond to lateral uncertainties from the probe tip for atomic force micrographs, or from magnification errors for scanning electron micrographs. We perturb values of the z position and intensity of secondary-electron scattering in regions of interest with random noise from normal distributions, which respectively correspond to errors in calibration, flatness, and various scanning artifacts for atomic force micrographs, or to estimates of the standard deviation of the background intensity of secondary-electron scattering for scanning electron micrographs, which we measure on the borders of regions of interest. This process of perturbing position and image data forms the basis of our Monte-Carlo simulation.

After localizing each pit, we align the array of pits in each micrograph by rotation of the micrographs so that the rows of pits in each micrograph are horizontal, corresponding to an angle of 0 rad. We then extract regions of interest of 1 by 1 μm and concentric with each pit from each micrograph. A symmetric bivariate Gaussian filter with an isotropic standard deviation ranging from 10 to 30 nm smooths image data within the region of interest (Table S3, Figure S3, Figure S4). To evaluate uncertainties of pit locations, we perturb the initial values of the center position of a pit with random noise from normal distributions of the localization uncertainty in the x and y directions. We excise one-dimensional sections of length 500 nm from the center position of the pit at an angle, $\theta$, with respect to the horizontal direction of each image, for angles ranging from 0 to $2\pi$ rad. As before, we apply our Monte-Carlo simulation approach, perturbing the position and image data of these radial sections to propagate uncertainty through our analysis.

Both atomic force micrographs and scanning electron micrographs comprise discrete data with independent sources of noise that degrade images of the pits. Such noise complicates differentiation and may contribute to inaccurate measurements of the positions of extrema of the second derivative. To address this issue, we apply a Savitzky-Golay[12] digital smoothing filter with a cubic polynomial and a window length ranging from 13 to 27 pixels to smooth the one-dimensional sections and to enable second-order differentiation of the sections from atomic force micrographs (Table S3, Figure S3, Figure S4). We vary the window length of the filter to evaluate both random and systematic effects with respect to mean radius (Figure S4). Optimal window length depends primarily upon the signal-to-noise ratio present within an image and may vary with different imaging conditions.

After smoothing and differentiation, the sections from each micrograph include a single global maximum, with a position corresponding to the pit radius of the section. We model the local vicinity of the maximum with a quartic polynomial to localize the maximum of each radial section, $s(\theta)$. We define the angle-dependent radius of the pit in scanning electron micrographs, $r_{p,\text{SEM}}(\theta)$, to be the distance between the center of the pit, $s_{0,\text{SEM}}$, and the position of the maximum of the intensity of secondary electron scattering in scanning electron micrographs,

$$r_{p,\text{SEM}}(\theta) = \arg\max_{s(\theta)}\{\boldsymbol{F}_{\text{SG}}[\boldsymbol{F}_G[\mathcal{N}(\mathcal{I}_{\text{SE}}(s(\theta)), \sigma_{\text{SE}}^2), \sigma_G^2], \omega_{\text{SG}}]\} - \mathcal{N}(s_{0,\text{SEM}}, \sigma_{\text{loc, SEM}}^2), \tag{S6}$$

where $\boldsymbol{F}_{\text{SG}}$ denotes the application of the Savitzky-Golay filter with window length, $\omega_{\text{SG}}$, $\boldsymbol{F}_G$ denotes the application of the Gaussian filter with standard deviation $\sigma_G$, $\mathcal{N}$ denotes a normal distribution, $\mathcal{I}_{\text{SE}}(s(\theta))$ is the intensity of secondary electron scattering on the section, $\sigma_{\text{SE}}$ is the standard deviation of the background intensity of secondary electron scattering, $\sigma_{\text{loc, SEM}}$ is the localization uncertainty of the center of the pit, and the x and y components of the section on the sampling radius, $r_s$, $s(\theta) = (r_s\cos\theta, r_s\sin\theta)$, are each subject to uncertainty from a uniform distribution of $\mathcal{U}(-0.5\delta_{\text{mag}}, 0.5\delta_{\text{mag}})$ from the uncertainty of the magnification of the scanning electron micrographs, $\delta_{\text{mag}} = 0.03 a_{\text{SEM}}$, where $a_{\text{SEM}}$ is the mean value of pixel size of the scanning electron micrograph. Similarly, we define the angle-dependent radius of the pit in scanning electron micrographs, $r_{\text{p, AFM}}(\theta)$, to be the distance between the center of the pit, $s_{0,\text{ AFM}}$, and the position of the maximum value of convexity of the pit in atomic force micrographs,

$$r_{\text{p, AFM}}(\theta) = \underset{s(\theta)}{\arg\max}\left\{-\frac{\partial^2}{\partial s^2}\boldsymbol{F}_{\text{SG}}[\boldsymbol{F}_{\text{G}}[\mathcal{N}(z(s(\theta)), \sigma_{\text{cal}}^2) + \mathcal{N}(0, \sigma_{\text{r}}^2) + \mathcal{N}(0, \sigma_{\text{f}}^2), \sigma_{\text{G}}^2], \omega_{\text{SG}}]\right\} - \mathcal{N}(s_{0,\text{ AFM}}, \sigma_{\text{loc, AFM}}^2), \quad (S7)$$

where $z(s(\theta))$ is the z position on the section, $\sigma_{\text{cal}} = 0.0025z$ accounts for a 0.5% systematic error from calibration of the atomic force microscope, $\sigma_{\text{r}}$ accounts for uncertainty from the configuration of scan rate, scan resolution, and probe tip,[2] $\sigma_{\text{f}}$ accounts for flatness errors, and the x and y components of the section, $s(\theta) = (r_s\cos\theta, r_s\sin\theta)$, are subject to uncertainty from a uniform distribution of $\mathcal{U}(-0.5r_{\text{tip}}, 0.5r_{\text{tip}})$ from the probe tip, where $r_{\text{tip}}$ is the radius of the tip.

For each imaging mode, we repeat these measurements 30 times for angles ranging from 0 to $2\pi$ rad in increments of approximately 100 mrad. To propagate uncertainty, we perturb the position and image data with random noise from uncertainty parameters each time, resulting in 1860 measurements of radius for each pit, and a total of 18600 measurements for each set of 10 replicates. We record measurements of radius from all Monte-Carlo simulations, compiling distributions of radii for all replicates in each micrograph. Table S3 summarizes all statistical variables that are relevant to the Monte-Carlo simulation of the measurements of pit radii.

Our in-line measurements and topographic analyses of test pits are highly relevant for focusing and calibrating electron–ion beam systems. Moreover, our methods of image analysis are broadly applicable to localizing features and inferring dimensions of similar nanostructures in scanning electron micrographs.

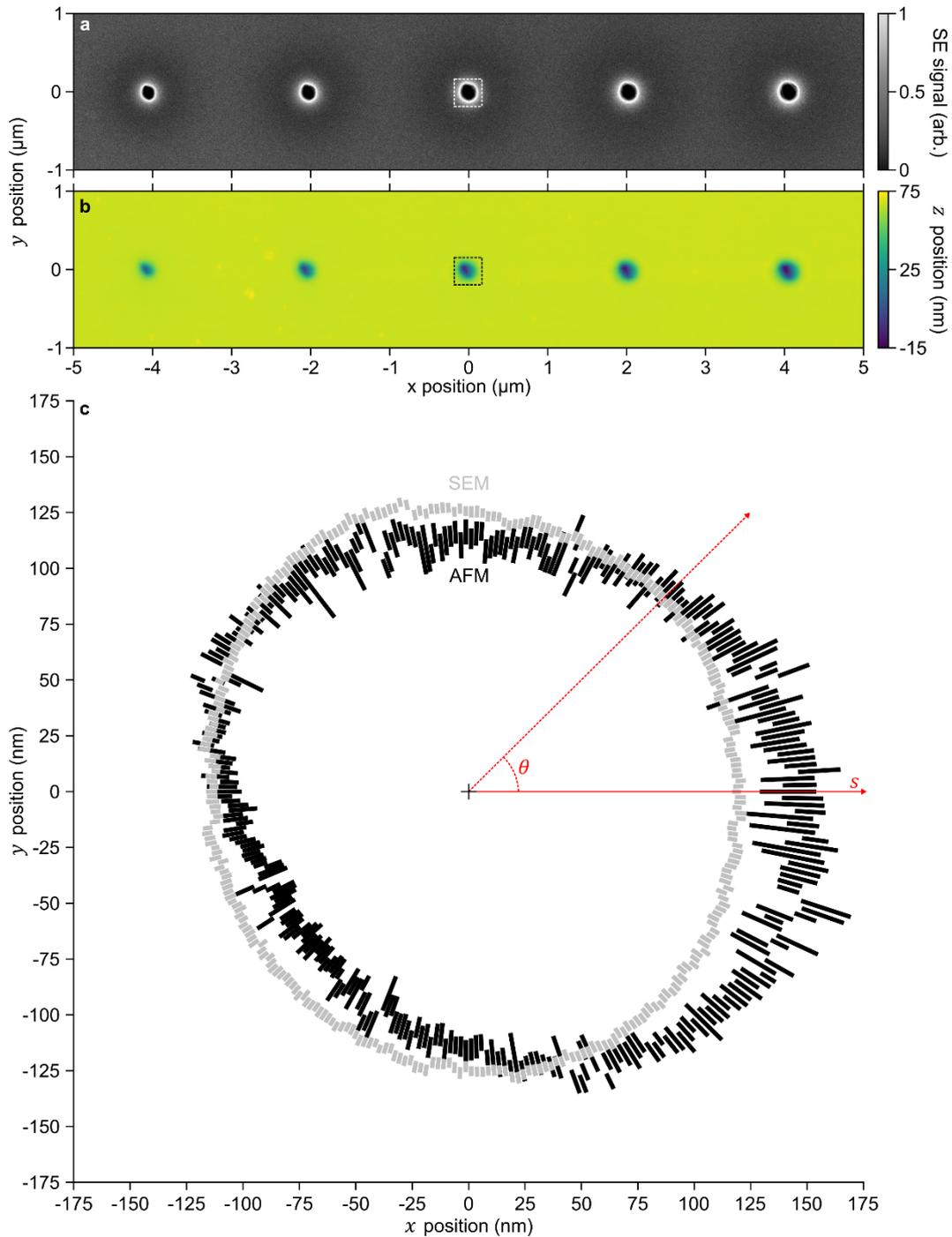

**Figure S3.** Correlative measurements. a) Scanning electron and b) atomic force micrographs showing the same representative array of pits that we mill with a focused ion beam with an ion-beam current of 219 ± 2 pA for dwell times ranging from 0.25 to 1.25 s in increments of 0.25 s. Dash boxes indicate the same region of interest in (c). c) Plots showing measurements of pit radius by (black) AFM and (gray) SEM for section angles ranging from 0 to $2\pi$ rad in increments of 20 mrad. The mean radius of the pit is 122 ± 4 nm by atomic force microscopy and 122 ± 1 nm by scanning electron microscopy. Bars indicate radius distribution widths as 95% coverage intervals.

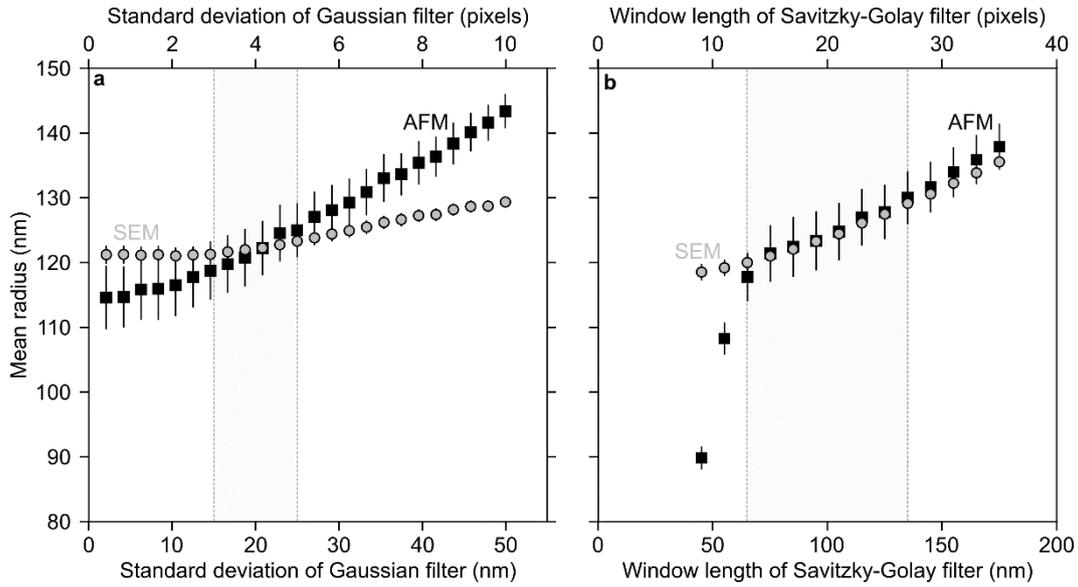

**Figure S4.** Filter parameters. a, b) Plots showing measurements of mean radius of a pit by (gray circles) scanning electron microscopy (SEM) and (black squares) atomic force microscopy (AFM) as functions of a) the standard deviation of the symmetric bivariate Gaussian filter and b) the window length of the Savitzky-Golay digital smoothing filter for the central pit in Figure S3a,b. Single values of the standard deviation of the Gaussian filter in (a) pair with a uniform distribution of the window length of the Savitzky-Golay filter in 1240 Monte-Carlo simulations of pit radius to isolate the effect of the value of standard deviation on measurements of pit radius. Similarly, single values of the window length of the Savitzky-Golay filter in (b) pair with a uniform distribution of the standard deviation of the Gaussian filter in 1240 Monte-Carlo simulations of pit radius to isolate the effect of the value of window length on measurements of pit radius. Gray dash lines indicate experimental ranges of the standard deviation of the Gaussian filter and the window length of the Savitzky-Golay filter (Table S3). These plots show the potential of each filter parameter to bias the measurements. Our simultaneous use of a range of values for each filter parameter yields measurements of mean pit radius in the center of these ranges with uncertainties of approximately 1 nm for scanning electron microscopy and 4 nm for atomic force microscopy.

**Note S5.** Milling Responses

We measure the depth, surface roughness, and edge width of the square features of checkerboard test-structures by analysis of atomic force micrographs. We level the atomic force micrographs and propagate uncertainties by Monte-Carlo simulations (Table S3). For each square feature, we extract image data from regions of interest of size 0.36 by 0.36 μm, or 30 by 30 pixels, and concentric with the center of the square. We perturb values of the z position with random noise from normal distributions to account for errors in calibration, flatness, and scanning artifacts for atomic force micrographs. We compute the mean depth and root-mean-square value of surface roughness of the flat of each square. For each edge in the scanning dimension of a square feature, we extract image data from five sections within regions of interest of size 1 by 0.36 μm, or 82 by 30 pixels. For each section, we propagate uncertainties from measurements by Monte-Carlo simulation and account for errors from systematic deviations in residuals from fits of models of step edges to data with leave-one-out jack-knife resampling.[13] We iteratively exclude one data point from the section and perturb all remaining values of position within this section with random noise from a uniform distribution corresponding to the lateral uncertainty from the radius of the probe tip. We perturb values of z position with random noise from normal distributions to account for errors from calibration, flatness, and scanning artifacts. We use an error function to empirically model the z position, $z_{\text{before}}(s)$, in sections of atomic force micrographs that transition from the bottom of a feature that we mill to the top of the chromia surface before chromia removal, Equation (S8),

$$z_{\text{before}}(s) = \frac{d}{2}\left[\text{erf}\left(\frac{s - s_0}{\sqrt{2}\sigma_{\text{edge}}}\right) + 1\right] + c, \quad (S8)$$

and we use an error function that we truncate at the zero plane to empirically model the same features after chromia removal, Equation (S9),

$$z_{\text{after}}(s) = \begin{cases} d \cdot \text{erf}\left(\frac{s - s_0}{\sqrt{2}\sigma_{\text{edge}}}\right) + c & s \leq s_0 \\ c & s > s_0 \end{cases}, \quad (S9)$$

where $d$ is the depth of the feature, $s_0$ is the location of the edge, $\sigma_{\text{edge}}$ is the standard deviation of a Gaussian, $c$ is a constant, and $s$ is subject to uncertainty from a uniform distribution of $\mathcal{U}(-0.5r_{\text{tip}}, 0.5r_{\text{tip}})$ from the radius of the probe tip. We approximate the width of edges as the 95% coverage interval of the width of the error functions, which correspond to $w_{\text{before}} = 4\sigma_{\text{edge}}$ before chromia removal and $w_{\text{after}} = 2\sigma_{\text{edge}}$ after chromia removal. We repeat this measurement 820 times for each section, retaining all fit parameters. This sampling results in a total of 8200 measurements of the widths of edges for each square feature in the checkerboard patterns.

**Table S4**. Uncertainties and errors

| Ion-beam current (pA) | Measurement uncertainty | | Estimate of error | |
|---|---|---|---|---|
| | Uncertainty of mean radius by atomic force microscopy (nm) | Uncertainty of mean radius by scanning electron microscopy (nm) | Root-mean-square residual to line of equality (nm) | Root-mean-square error (nm) |
| 82 ± 1 | 3.1 ± 0.6 | 1.0 ± 0.1 | 3.7 ± 1.7 | 3.8 ± 1.8 |
| 219 ± 2 | 4.1 ± 1.9 | 1.0 ± 0.1 | 3.6 ± 1.6 | 6.7 ± 1.6 |
| 407 ± 2 | 6.5 ± 3.8 | 1.0 ± 0.1 | 3.4 ± 1.6 | 4.9 ± 1.1 |
| 773 ± 3 | 7.4 ± 4.4 | 1.0 ± 0.1 | 5.5 ± 2.6 | 5.8 ± 1.4 |

Uncertainties of ion-beam current are conservative estimates of 100% coverage intervals.

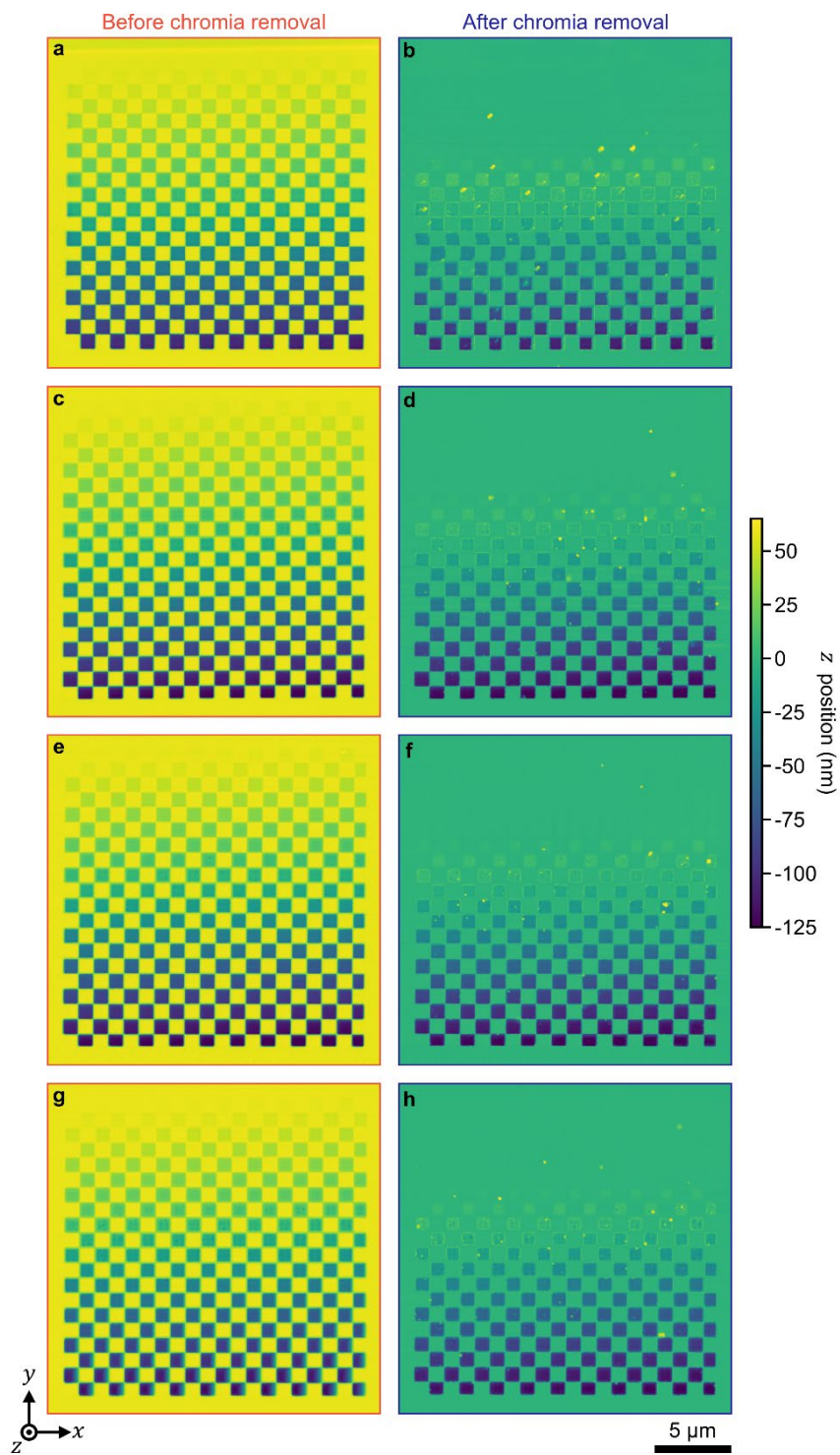

**Figure S5.** Complex test-structures. a-h) Atomic force micrographs showing checkerboard structures before and after chromia removal for several values of ion-beam current: (a,b) 83 ± 1 pA; (c,d) 227 ± 1 pA; (e,f) 421 ± 3 pA; and (g,h) 796 ± 4 pA. Uncertainties of ion-beam current are conservative estimates of 100% coverage intervals.

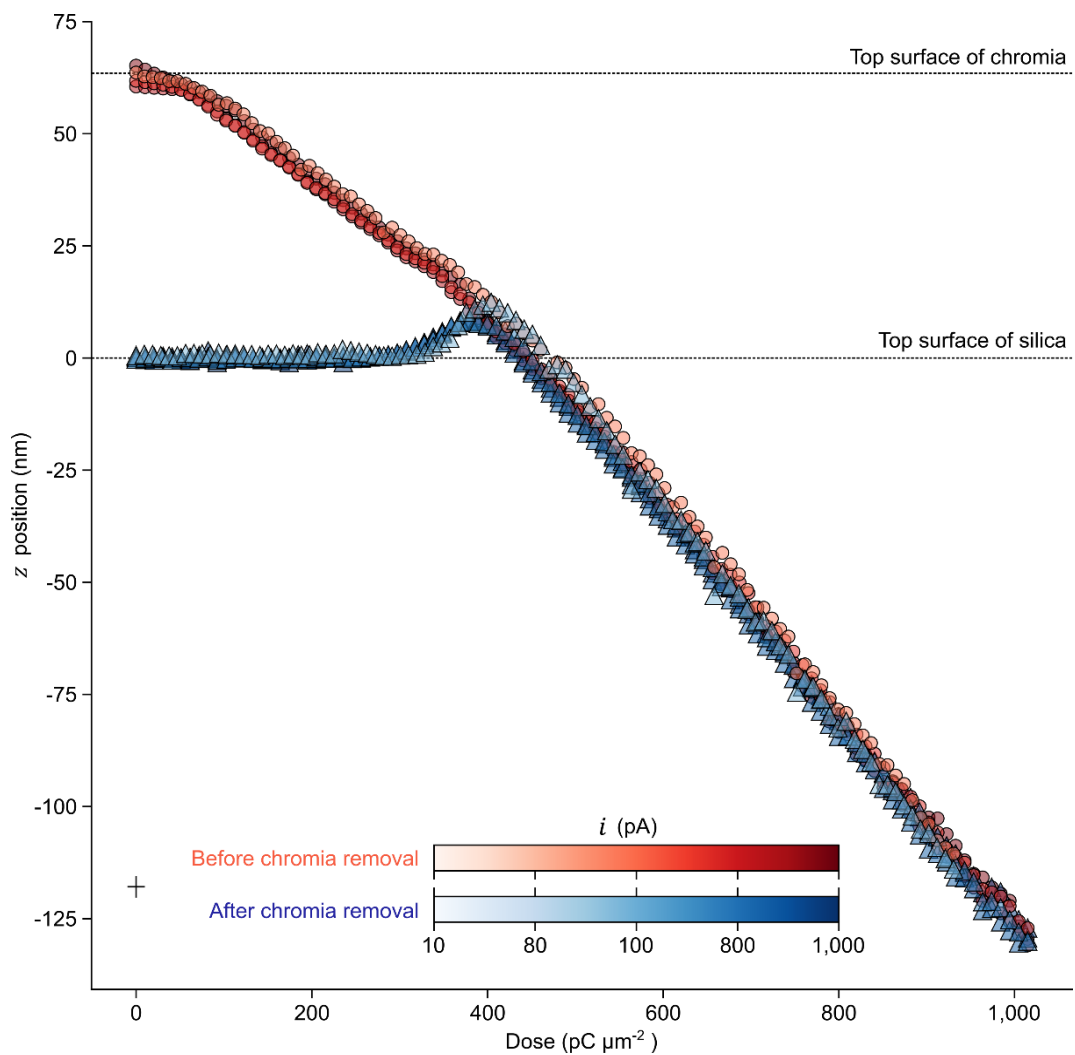

**Figure S6**. Vertical control. Plot showing milling responses of chromia and silica for a range of ion-beam currents (red circles) before and (blue triangles) after chromia removal. The color codes are logarithmic. The lone black cross near the lower left corner of the plot indicates a 100% coverage interval of dose and a 95% coverage interval of depth.

**Table S5**. Depth increments

| | Depth increment before removal of chromia mask | | Depth increment after removal of chromia mask | |
|---|---|---|---|---|
| Ion-beam current (pA) | Mean (nm) | Standard deviation (nm) | Mean (nm) | Standard deviation (nm) |
| 83 ± 1 | 1.11 ± 0.22 | 1.23 ± 0.18 | 1.19 ± 0.24 | 0.94 ± 0.14 |
| 227 ± 1 | 1.13 ± 0.16 | 0.99 ± 0.10 | 1.18 ± 0.16 | 0.93 ± 0.10 |
| 421 ± 3 | 1.14 ± 0.12 | 0.91 ± 0.08 | 1.17 ± 0.14 | 0.99 ± 0.08 |
| 796 ± 4 | 1.14 ± 0.12 | 0.98 ± 0.06 | 1.16 ± 0.12 | 0.97 ± 0.06 |

Uncertainties of ion-beam current are conservative estimates of 100% coverage intervals.

**Table S6**. Milling responses

| Segment | Material | Dose offset (pC µm$^{-2}$) | z position (nm) | Milling rate (µm$^3$ nA$^{-1}$ s$^{-1}$) | Intercept (nm) |
|---|---|---|---|---|---|
| – | – | 0 ± 0 | 63 ± 1 | – | – |
| 1 | Top surface of silica | 69 ± 6 | 59 ± 1 | 0.05 ± 0.03 | 66 ± 2 |
| 2 | Bulk silica | 366 ± 10 | 16 ± 3 | 0.15 ± 0.04 | 101 ± 10 |
| 3 | Chromia–silica interface | 531 ± 10 | -16 ± 4 | 0.19 ± 0.04 | 126 ±11 |
| 4 | Bulk chromia | 1020 ± 0 | -132 ± 6 | 0.24 ± 0.04 | 170 ± 14 |

**Table S7**. Gallium penetration

| | | | z direction | | | | x and y directions | | |
|---|---|---|---|---|---|---|---|---|---|
| Figure | z position (nm) | Chromia thickness (nm) | Mean (nm) | Standard deviation (nm) | Skew | Kurtosis | Mean (nm) | Standard deviation (nm) | Skew | Kurtosis |
| Figure 5c-i | 59 | 59 | 13.9 | 5.7 | 0.4 | 0.1 | 0.0 | 4.6 | 0.0 | 0.5 |
| Figure 5c-ii | 16 | 16 | 15.3 | 7.9 | 1.0 | 1.0 | 0.0 | 5.3 | 0.0 | 1.3 |
| Figure 5c-iii | -16 | 0 | 27.6 | 9.1 | 0.4 | 0.1 | 0.0 | 7.0 | 0.0 | 0.4 |

All simulations are of 10$^5$ gallium ions with a landing energy of 4.81 fJ (30 keV) at normal incidence, a chromia, Cr$_2$O$_3$, density of 5.3 g cm$^{-3}$, and a silica, SiO$_2$, density of 2.2 g cm$^{-3}$.

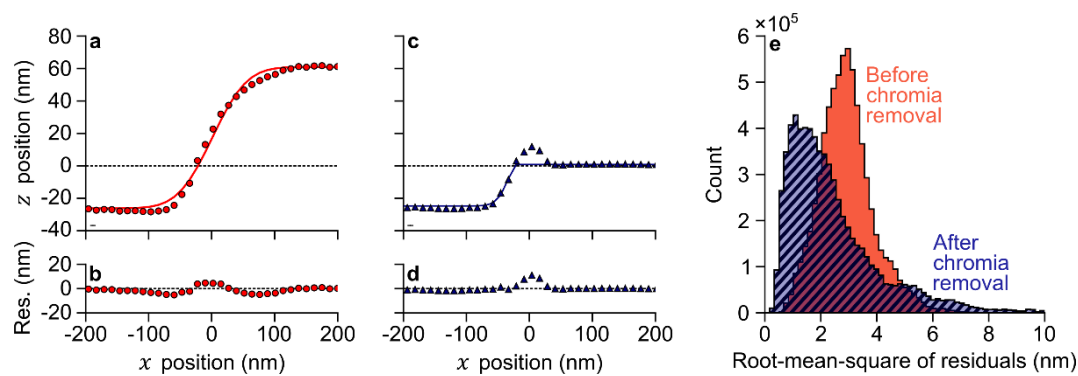

**Figure S7.** Model summary. a-d) Plots showing representative (a,c) model fits and (b,d) residuals (a,b) before, and (c,d) after chromia removal. Overshoot artifact is apparent in (c). The root-mean-square of residuals in (a,b) and (c,d) are respectively 1.9 and 2.0 nm. e) Histograms showing the root-mean-square of residuals of fits of all test structures and ion-beam currents. Lone black bars in (a) and (c) indicate representative uncertainties of $x$ position as 95% coverage intervals. Uncertainties of $z$ position are smaller than the data markers.

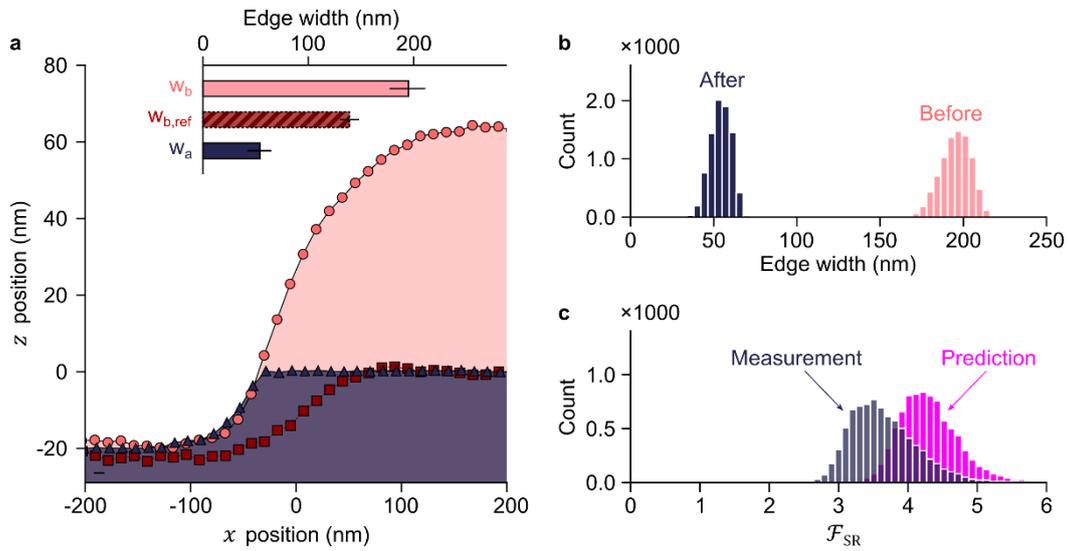

**Figure S8.** Measurement and prediction. a) Plots showing representative edge profiles. The solid regions show profiles resulting from a dose of 536 pC μm$^{-2}$ and from an ion-beam current of 227 ± 1 pA, (red circles) before and (blue triangles) after removal of the chromia mask. The dark red squares are from a feature with a similar depth resulting from a dose of 189 pC μm$^{-2}$ and from an ion-beam current of 227 ± 1 pA. The lone black bar in (a) indicates a representative uncertainty of $x$ position as a 95% coverage interval. Uncertainties of $z$ position are smaller than the data markers. The inset bar chart shows the edge width of each profile with black bars indicating edge width distributions as 95% coverage intervals. b) Histograms showing edge widths (light red) before and (dark blue) after removal of the chromia mask. c) Histograms showing super-resolution factors, $\mathcal{F}_{SR}$, from (dark blue) measurement and (magenta) prediction by the spatial masking model. Additional details are in Table S8.

**Table S8.** Measurement and prediction

| Method | Depth (nm) | Depth after normalization by mask thickness | Edge width (nm) | | | Super-resolution factor ($\mathcal{F}_{SR}$) |
| --- | --- | --- | --- | --- | --- | --- |
| | | | Before removal ($w_b$) | Before removal at reference depth ($w_{b,\,ref}$) | After removal ($w_a$) | |
| Measurement | 17.7 $^{+0.8}/_{-0.9}$ | 0.278 $^{+0.014}/_{-0.014}$ | 195 $^{+16}/_{-18}$ | 139 $^{+9}/_{-9}$ | 54 $^{+11}/_{-12}$ | 3.6 $^{+1.2}/_{-0.7}$ |
| Prediction | – | 0.278 $^{+0.010}/_{-0.010}$ | – | – | – | 4.3 $^{+1.0}/_{-0.7}$ |

Fractional notation denotes 95% coverage intervals that are asymmetric.
The sample size is 830 for all values.

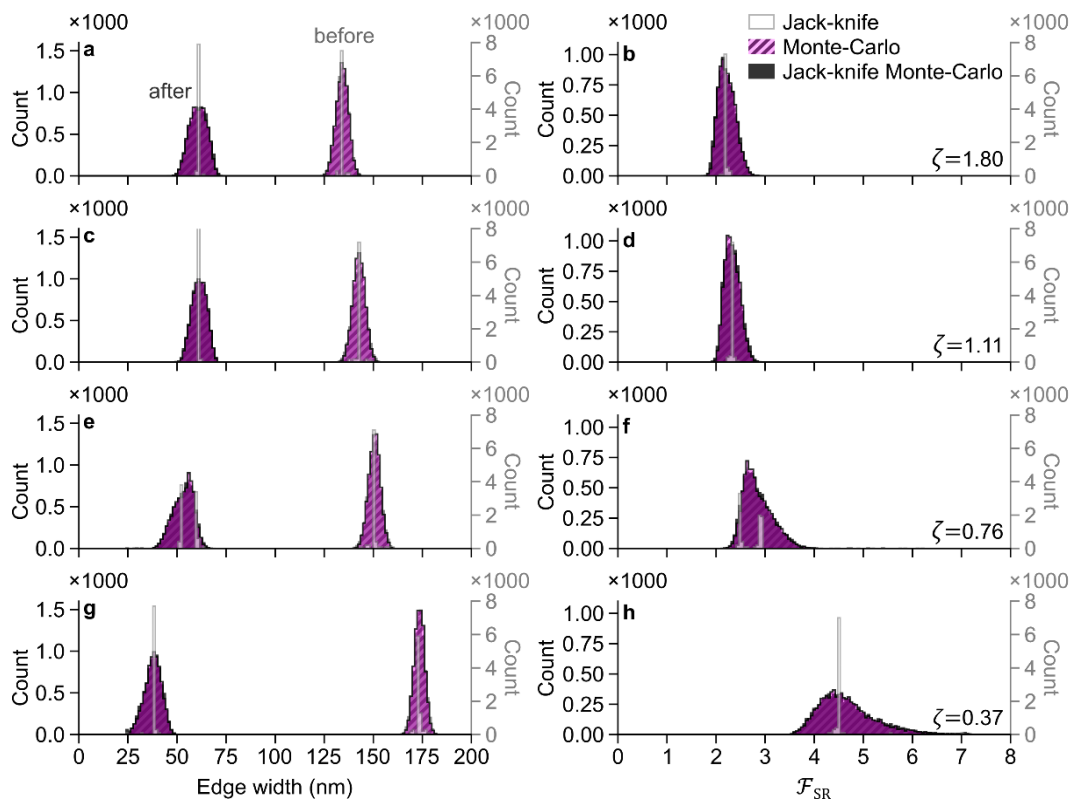

**Figure S9.** Numerical analysis. a,c,e,g) Histograms showing edge widths (light shading) before and (dark shading) after chromia removal and b,d,f,h) histograms showing super-resolution factors for three simulation and resampling methods: (white region with gray outline) jack-knife resampling only; (magenta and black hash region) Monte-Carlo simulation only; and (black region) jack-knife resampling with Monte-Carlo simulation. (a,b) A feature resulting from a dose of 561 ± 2 pC µm$^{-2}$ and having a depth of 24 ± 1 nm. (c,d) A feature resulting from a dose of 663 ± 3 pC µm$^{-2}$ and having a depth of 49 ± 1 nm. (e,f) A feature resulting from a dose of 765 ± 3 pC µm$^{-2}$ and having a depth of 71 ± 2 nm. (g,h) A feature resulting from a dose of 969 ± 4 pC µm$^{-2}$ and having a depth of 114 ± 4 nm. In each panel, counts of distributions from the Monte-Carlo and jack-knife Monte-Carlo methods correspond to vertical axes on the left, and counts of distributions from the jack-knife method correspond to vertical axes on the right. Depths after normalization by mask thickness appear in the lower right corner of panels (b,d,f,h). Additional details are in Table S9.

**Table S9.** Numerical analysis

| Method | Feature number | Samples | Depth | | Super-resolution factor | |
|---|---|---|---|---|---|---|
| | | | Mean (nm) | Mean after normalization (–) | Mean (–) | Standard deviation (–) |
| Jack-knife | 110 | 8300 | 23.7 ± 0.02 | 0.373 ± 0.0004 | 4.531 ± 0.006 | 0.302 ± 0.003 |
| Monte-Carlo | 110 | 8300 | 23.6 ± 0.02 | 0.372 ± 0.0004 | 4.660 ± 0.012 | 0.564 ± 0.008 |
| Jack-knife Monte-Carlo | 110 | 8300 | 23.7 ± 0.02 | 0.373 ± 0.0004 | 4.691 ± 0.014 | 0.618 ± 0.010 |
| Jack-knife | 130 | 8300 | 48.2 ± 0.03 | 0.759 ± 0.0004 | 2.709 ± 0.004 | 0.206 ± 0.004 |
| Monte-Carlo | 130 | 8300 | 48.2 ± 0.03 | 0.760 ± 0.0004 | 2.857 ± 0.006 | 0.290 ± 0.004 |
| Jack-knife Monte-Carlo | 130 | 8300 | 48.2 ± 0.03 | 0.760 ± 0.0004 | 2.862 ± 0.008 | 0.316 ± 0.005 |
| Jack-knife | 150 | 8300 | 70.8 ± 0.03 | 1.115 ± 0.0006 | 2.325 ± 0.001 | 0.026 ± 0.001 |
| Monte-Carlo | 150 | 8300 | 70.8 ± 0.04 | 1.115 ± 0.0006 | 2.336 ± 0.004 | 0.151 ± 0.002 |
| Jack-knife Monte-Carlo | 150 | 8300 | 70.8 ± 0.03 | 1.115 ± 0.0006 | 2.339 ± 0.004 | 0.157 ± 0.002 |
| Jack-knife | 190 | 8300 | 114.3 ± 0.05 | 1.801 ± 0.0008 | 2.212 ± 0.001 | 0.021 ± 0.001 |
| Monte-Carlo | 190 | 8300 | 114.3 ± 0.05 | 1.800 ± 0.0008 | 2.222 ± 0.004 | 0.167 ± 0.002 |
| Jack-knife Monte-Carlo | 190 | 8300 | 114.3 ± 0.05 | 1.800 ± 0.0008 | 2.220 ± 0.004 | 0.172 ± 0.002 |

**Table S10.** Edge widths

| | Ion-beam current (pA) | | | | | | | |
|---|---|---|---|---|---|---|---|---|
| | 83 ± 1 | | 227 ± 1 | | 421 ± 3 | | 796 ± 4 | |
| Metric | Before (nm) | After (nm) | Before (nm) | After (nm) | Before (nm) | After (nm) | Before (nm) | After (nm) |
| 2.5th percentile | 117 | 32 | 116 | 51 | 127 | 64 | 137 | 66 |
| 25th percentile | 133 | 43 | 135 | 67 | 157 | 83 | 165 | 94 |
| Mean | 142 | 54 | 154 | 79 | 174 | 101 | 195 | 113 |
| 75th percentile | 148 | 59 | 171 | 84 | 185 | 105 | 221 | 132 |
| 97.5th percentile | 173 | 87 | 216 | 117 | 241 | 165 | 276 | 179 |

Uncertainties of ion-beam current are conservative estimates of 100% coverage intervals.
Uncertainties of values of percentiles[14] are all less than 1 nm.

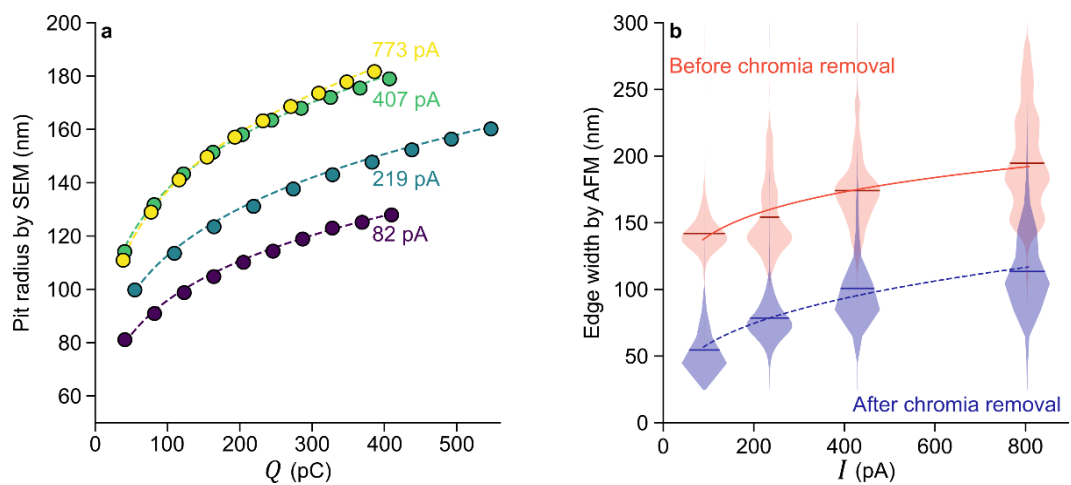

**Figure S10.** Patterning resolution. a) Plot showing pit radius by scanning electron microscopy as a function of charge, $Q$, with fits of power-law models. The non-monotonic trend in pit radius as a function of ion-beam current shows imperfections in the reproduction of ion-beam focus due to manual operation of the system, motivating development of methods for complete automation of ion-beam focusing, which our study of in-line resolution metrology enables. b) Plot showing edge width by atomic force microscopy before chromia removal as a function of ion-beam current, $I$. Violin plots show distributions of edge widths from complex test-structures below the zero plane, corresponding to the metrics in Table S10 for edge widths (red) before and (blue) after removal of chromia. The dark lines in violin plots indicate mean values. The solid and dash lines indicate representative fits of power-law models to data.

**Table S11.** Symbols, values, and definitions

| Symbol | Range of values | Units | Definition |
|---|---|---|---|
| $I$ | 83 to 796 | pA | Ion-beam current |
| $z$ | −150 to 65 | nm | $z$ position of edge profile resulting from exposure of substrate to a focused ion beam |
| $z_m$ | 61 to 65 | nm | Mask thickness |
| $z_s$ | 0 to 150 | nm | Milling depth into the substrate |
| $\zeta$ | 0 to 2.2 | – | Milling depth into the substrate after normalization by mask thickness, $\zeta = |z|z_m^{-1}$ |
| $\bar{m}_m$ | 0.11 to 0.19 | – | Average milling rate of the mask |
| $\bar{m}_s$ | 0.20 to 0.28 | – | Average milling rate of the substrate |
| $\mathcal{S}$ | 1.0 to 2.1 | – | Physical selectivity, the ratio of average milling rates of the substrate, $\mathcal{S} = \bar{m}_s \bar{m}_m^{-1}$ |
| $w_b$ | 116 to 276 | nm | Width of nanostructure edges before removal of the mask |
| $w_a$ | 32 to 179 | nm | Width of nanostructure edges after removal of the mask |
| $\mathcal{F}_{SR}$ | 1 to 6 | – | Super-resolution factor, ratio of widths of step edges before and after removal of the mask, $\mathcal{F}_{SR} = w_b w_a^{-1}$ |
| $\mathcal{R}$ | 32 to 276 | nm | Effective lateral patterning resolution or super-resolution |
| $V_m$ | 0.000 to 0.065 | µm³ | Volume of the mask that the ion-beam mills for each square feature of the complex test-structures |
| $V_s$ | 0.000 to 0.150 | µm³ | Volume of the substrate that the ion-beam mills for each square feature of the complex test-structures |
| $\mathcal{V}$ | 55 to 530 | µm³ hr⁻¹ | Volume throughput of material that the ion-beam mills per unit time |
| $\alpha$ | 34 to 86 | nm pA⁻ᵝ | Coefficient in power-law model of the widths of step edges with respect to ion-beam current |
| $\beta$ | 0.1 to 0.3 | – | Exponent in power-law model of the widths of step edges with respect to ion-beam current |
| $\eta_\tau$ | 10⁻¹ to 10⁵ | – | Temporal efficiency, ratio of milling times in the absence, $t_s$, and presence, $t_s + t_m$, of a mask, $\eta_\tau = t_s(t_s + t_m)^{-1}$ |
| $\mathcal{M}$ | 10² to 10⁵ | µm² hr⁻¹ | Figure of merit for focused-ion-beam milling, $\mathcal{M} = \mathcal{V} \mathcal{R}^{-1}$ |

**Note S6.** Bi-Gaussian approximation

Under the conditions that are necessary to mill a semi-infinite edge, an error function is a good approximation of the total dose pattern from both Gaussian and bi-Gaussian approximations of the current density distribution of a focused ion beam. The bi-Gaussian[15] model, $BG(x; A, \sigma_{\text{core}}, \sigma_{\text{tail}}, \omega, x_i)$, in Equation (S10) accounts for contributions to the total current density from both the core and the tails of the ion beam in a summation of two Gaussian functions,

$$BG(x; A, \sigma_{\text{core}}, \sigma_{\text{tail}}, \omega, x_i) = A\left[\omega \exp\left\{-\frac{(x-x_i)^2}{2\sigma_{\text{core}}^2}\right\} + (1-\omega) \exp\left\{-\frac{(x-x_i)^2}{2\sigma_{\text{tail}}^2}\right\}\right], \quad \text{(S10)}$$

where $A$ is the amplitude of the ion beam dose, $\omega$ is a parameter that varies between 0 and 1 to weight the Gaussian components, $\sigma_{\text{core}}$ is the standard deviation of the core of the ion beam, $\sigma_{\text{tail}}$ is the standard deviation of the tail of the ion beam, and $x_i$ is the milling location of the ion beam. Equation (S10) reduces to a Gaussian model when $\omega = 1$. Models of the ion-beam shape and size inform discrete positioning of the ion beam during the direct-write process of focused-ion-beam milling. In particular, the full-width at half-maximum of the core of the ion beam is a common approximation of the diameter of the ion beam, $d_{\text{beam}} = 2\sqrt{2\ln(2)}\sigma_{\text{core}}$. The diameter of the ion beam and the overlap, $o$, define the pitch or spacing of adjacent milling positions in a rectilinear coordinate system, $\Delta x = (1-o)d_{\text{beam}}$. As such, a semi-infinite dose pattern comprises a series of milling positions that occur with equal spacing, $\Delta x$, on the half-line in one-dimensional space. As previous studies[8, 16] show, an overlap greater than or equal to 0.35 ensures the delivery of a uniform dose in the bulk of the half-line.

Our study of spatial masking motivates an inquiry into the effect of the bi-Gaussian approximation of the profile of a focused ion beam on the edge of a semi-infinite dose pattern. In particular, the deviation from an error function of a semi-infinite dose pattern that the bi-Gaussian model imposes is of interest. We model a semi-infinite dose pattern in one dimension, $D(x)$, as a series of discrete milling locations of a focused ion beam that follows the bi-Gaussian approximation,

$$D(x) = \frac{D_0}{D_N} \sum_i^N BG(x; A, \sigma_{\text{core}}, \sigma_{\text{tail}}, \omega, x_i), \quad \text{(S11)}$$

where $D_0$ is the value of dose in the bulk of the half-line and $D_N$ is a factor that normalizes the maximum value of the summation in Equation (S11) to unity. To quantify the deviation of the semi-infinite dose pattern from an error function, we simulate dose patterns for ratios of $\sigma_{\text{tail}}$ to $\sigma_{\text{core}}$ ranging from 1 to 10, a range of weights ranging from 0 to 0.5, an amplitude of the bi-Gaussian model of 1, and an overlap of 0.5 (50%). We fit an error function,

$$D(x) = \frac{D_0}{D_N} \sum_i^N BG(x; A, \sigma_{\text{core}}, \sigma_{\text{tail}}, \omega, x_i) \cong D_0\left[1 - \frac{1}{2}\left(\text{erf}\left(\frac{x-x_0}{\sqrt{2}\sigma_{\text{eff}}}\right) + 1\right)\right], \quad \text{(S12)}$$

the resulting dose patterns, where $\sigma_{\text{eff}}$ is the effective standard deviation of the edge of the dose pattern and $x_0$ is the position of the center of the edge. For each parameterization, we compute the root-mean-square error between each dose pattern and its corresponding fit.

Values of root-mean-square error are less than 2.5% for all parameterizations of the focused-ion-beam shape (Figure S11). This good agreement indicates a general consistency between the bi-Gaussian approximation of the current density distribution of a focused ion beam and the error function model of the semi-infinite dose pattern that results from milling in discrete positions, which supports the validity of our use of the error function in our model of the lateral super-resolution effect.

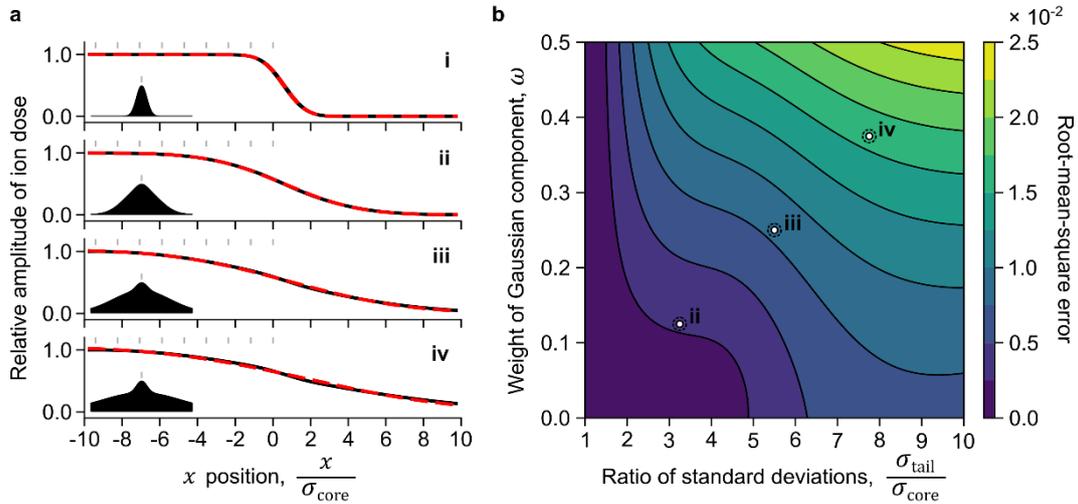

**Figure S11.** Bi-Gaussian approximation. a) Plots showing (solid black line) the relative amplitude of ion dose resulting from milling an array of (gray ticks) discrete positions and (red dash line) fits of the error function in Equation (S12) for four combinations of values of $\sigma_{tail}\sigma_{core}^{-1}$ and $\omega$: (i) $\sigma_{tail}\sigma_{core}^{-1} = 1.00$, $\omega = 1.000$; (ii) $\sigma_{tail}\sigma_{core}^{-1} = 3.25$, $\omega = 0.125$; (iii) $\sigma_{tail}\sigma_{core}^{-1} = 5.5$, $\omega = 0.250$; and (iv) $\sigma_{tail}\sigma_{core}^{-1} = 7.75$, $\omega = 0.375$. The insets of panels (a-i–a-iv) show the spatial profile of the ion beam over the extents of the spatial domain, $\pm 10\sigma_{core}$, for each parameterization. b) Contour plot showing root-mean-square error of fits of the error function to relative dose profiles for a subset of the parameter space of the bi-Gaussian function. The position on the contour plot of the Gaussian profile in (a-i) is beyond the vertical extent of the plot. The root-mean-square error of fits of the error function to the relative dose profile corresponding to the Gaussian profile is zero.

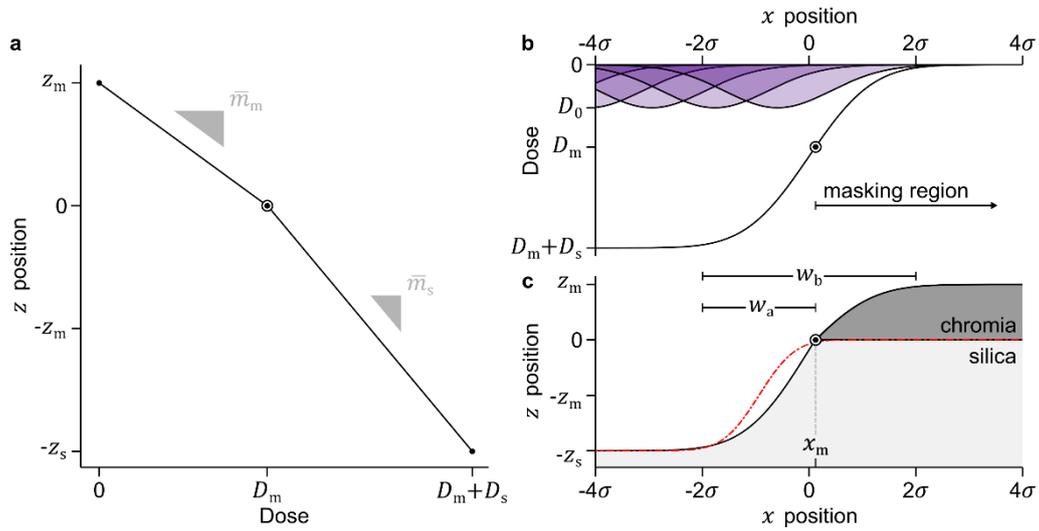

**Figure S12**. Spatial masking. a) Plot showing z position as a function of dose for a simple model of milling in which a sacrificial mask of thickness, $z_m$, and average milling rate, $\bar{m}_m$, screens a substrate with an average milling rate of $\bar{m}_s$. b) Plot showing spatial ion-dose pattern from (violet) discrete positions of an ion beam as a function of x position. c) Plot showing (solid black line) a theoretical surface profile resulting from application of the dose in (b) to the milling model in (a). The red dash-dot line is a theoretical surface profile that results from the application of a dose of similar magnitude to that in (b) but from a lower value of ion current and milling directly into the silica substrate. Consequently, the red profile has a different shape from the black profile. After removal of the chromia mask, both profiles have an edge width of $2\sigma + x_m$. The red profile includes the entire transition region of the sigmoid of the dose profile, whereas after removal, the black profile shows truncation of a sigmoid. The black and the red profiles are simplifications that neglect the dependence of incidence angle on milling rate, any effects of redeposition, and any defocus of the ion-beam due to charging. However, the standard deviation of the ion-beam that yields the black profile, which has the greater value of current, is $\sigma_{high} = \sigma$ while the standard deviation of the ion-beam that yields the red profile is $\sigma_{low} = 0.53\sigma$. An exponent of $\beta = 0.2$ in the power-law model from Equation (1) implies a ratio of the two currents of $(\sigma_{high}\sigma_{low}^{-1})^{1/\beta}$, or approximately 24.

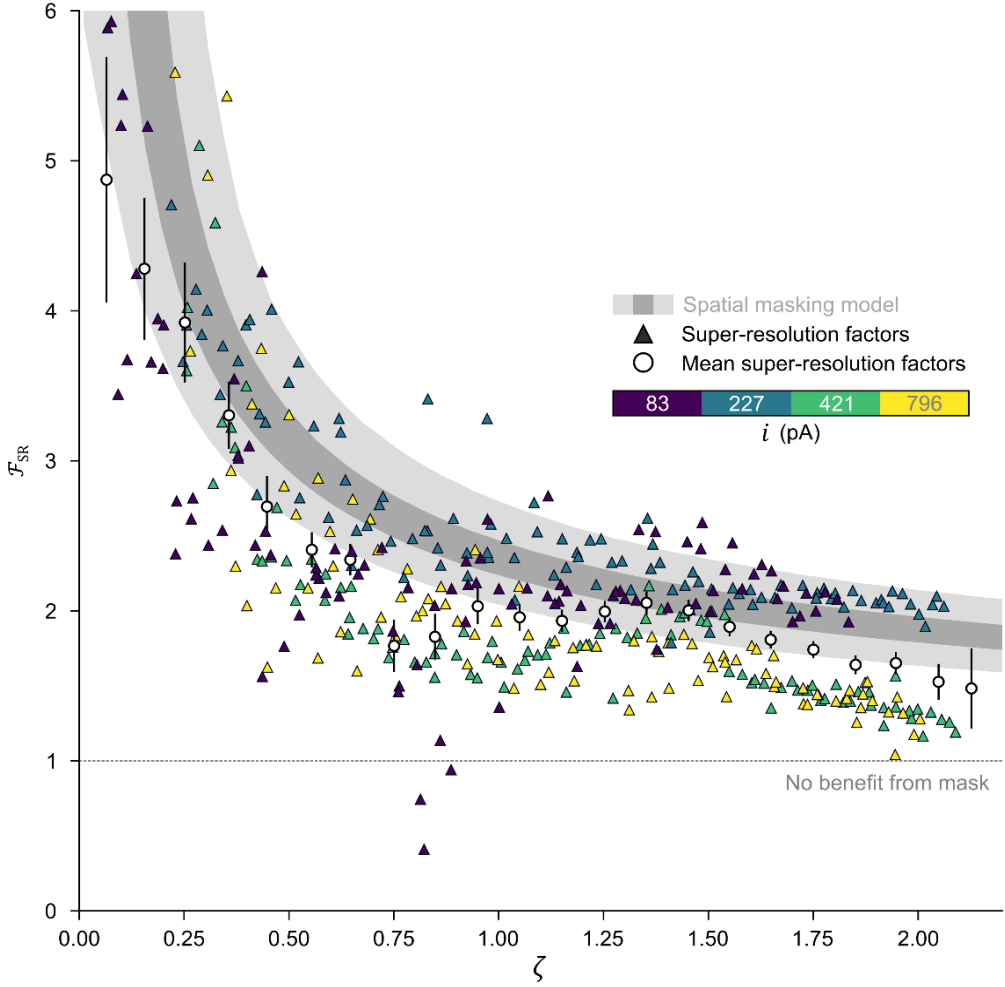

**Figure S13**. Super-resolution. Plot showing values of super-resolution factors, $\mathcal{F}_{SR}$, as a function of the ratio of milling depth into the substrate after normalization by mask thickness, $\zeta$, for features below the zero plane resulting from four values of ion-beam current. The light and dark gray shading respectively indicate the 95% coverage interval and the interval between the lower and upper quartiles of the spatial masking model. The triangles are features that we mill with ion-beam currents that range from approximately 80 to 800 pA. The white circles are mean super-resolution factors from the entire data set. For clarity, we show only a fraction of these values. The black bars indicate uncertainties of mean super-resolution factor as 95% coverage intervals. Uncertainties of $\zeta$ are smaller than the data markers.

**Note S7**. Spatial masking of a line or point
We derive an analytic expression for the spatial limit of a sacrificial mask to screen the tails of ion beam from an underlying substrate during exposure in one dimension. We apply the resulting model to fit the experimental data of Menard and Ramsey,[17] with several simplifying assumptions. We assume that the mask has a bulk milling rate, $m_m$, and a thickness, $z_m$, and similarly, that the substrate has a bulk milling rate, $m_s$, and a final depth, $z_s$, and that the spatial profile of the ion dose along the x direction, $D(x)$, follows a Gaussian function of the form,

$$D(x) = D_0 \exp\left\{-\frac{(x-x_0)^2}{2\sigma^2}\right\}, \tag{S13}$$

where $D_0$ is the dose that is necessary to mill through the mask and into the working material, $x_0$ is the center position of the line scan, which we assign to be 0, and $\sigma$ is the effective standard deviation of the Gaussian profile of the ion beam. Substitution into Equation (S13) of $D_0 = D_m + D_s = z_m m_m^{-1} + z_s m_s^{-1}$, the ratio of milling depth of the substrate to the thickness of the mask, $\zeta = z_s z_m^{-1}$, and the ratio of the milling rate of the substrate to the milling rate of the mask, $\mathcal{S} = m_s m_m^{-1}$, yields the condition for spatial masking of the dose pattern, where

$$D_m = \frac{z_m}{m_m} = \left[\frac{z_m}{m_m} + \frac{z_s}{m_s}\right] \exp\left\{-\frac{x^2}{2\sigma^2}\right\} = \frac{z_m m_s + z_s m_m}{m_m m_s} \exp\left\{-\frac{x^2}{2\sigma^2}\right\}, \tag{S14}$$

which implies

$$\frac{S}{S+\zeta} = \exp\left\{-\frac{x^2}{2\sigma^2}\right\}. \tag{S15}$$

Solving Equation (S15) for $x$ yields the positions at which the mask begins to screen the tails of the ion beam,

$$x = \pm\sigma\sqrt{-2\ln\left(\frac{S}{S+\zeta}\right)}. \tag{S16}$$

The diameter of a pit or the width of a nanochannel resulting after removal of the sacrificial mask is the distance between these positions,

$$w_a = 2\sigma\sqrt{-2\ln\left(\frac{S}{S+\zeta}\right)}. \tag{S17}$$

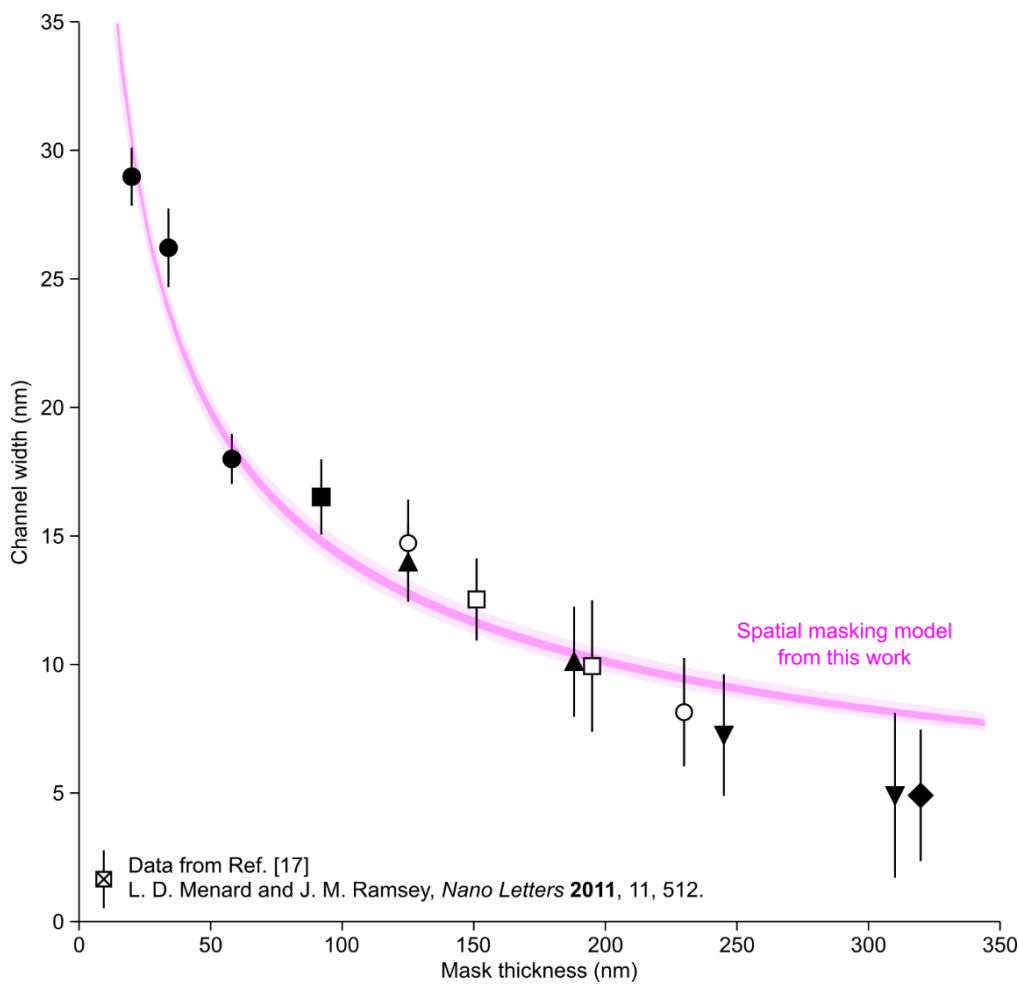

**Figure S14.** Comparison to a previous study. Plot showing widths of channels as a function of the thickness of sacrificial masks of chromium after removal of the mask. The channels result from the spatial masking of line scans of a focused beam of gallium ions, milling through sacrificial chromium masks and into underlying quartz substrates. The data markers show the experimental results from Menard and Ramsey.[17] Using approximate values of experimental parameters from correspondence with these scholars, we bound the values of the effective standard deviation of the ion beam to the range from 20 to 60 nm, the milling depth parameter to the range from 0 to 20 nm, and the value of the physical selectivity parameter to the range from 0 to 3, and we fit our spatial masking model of a line scan in Equation (S17). This fit yields a reduced chi-square statistic, $\chi^2_\nu$, of 5.3, extracts reasonable values of experimental parameters including standard deviation of the ion-beam profile, milling depth, and physical selectivity, and indicates that our spatial masking model is generally applicable beyond the specific system in our current study. The light and dark magenta regions indicate uncertainties of mean channel width respectively as 95% coverage intervals and interquartile ranges of fit results to our spatial masking model. This analysis yields estimates of experimental parameters, including an effective standard deviation of the ion beam of $46\,^{+14\,\text{nm}}_{-15\,\text{nm}}$, a milling depth of $7.8\,^{+3.9\,\text{nm}}_{-4.2\,\text{nm}}$, and a physical selectivity of quartz to chromium of $1.5\,^{+1.0}_{-0.6}$. Uncertainties of these parameters and of the data markers are 95% coverage intervals.

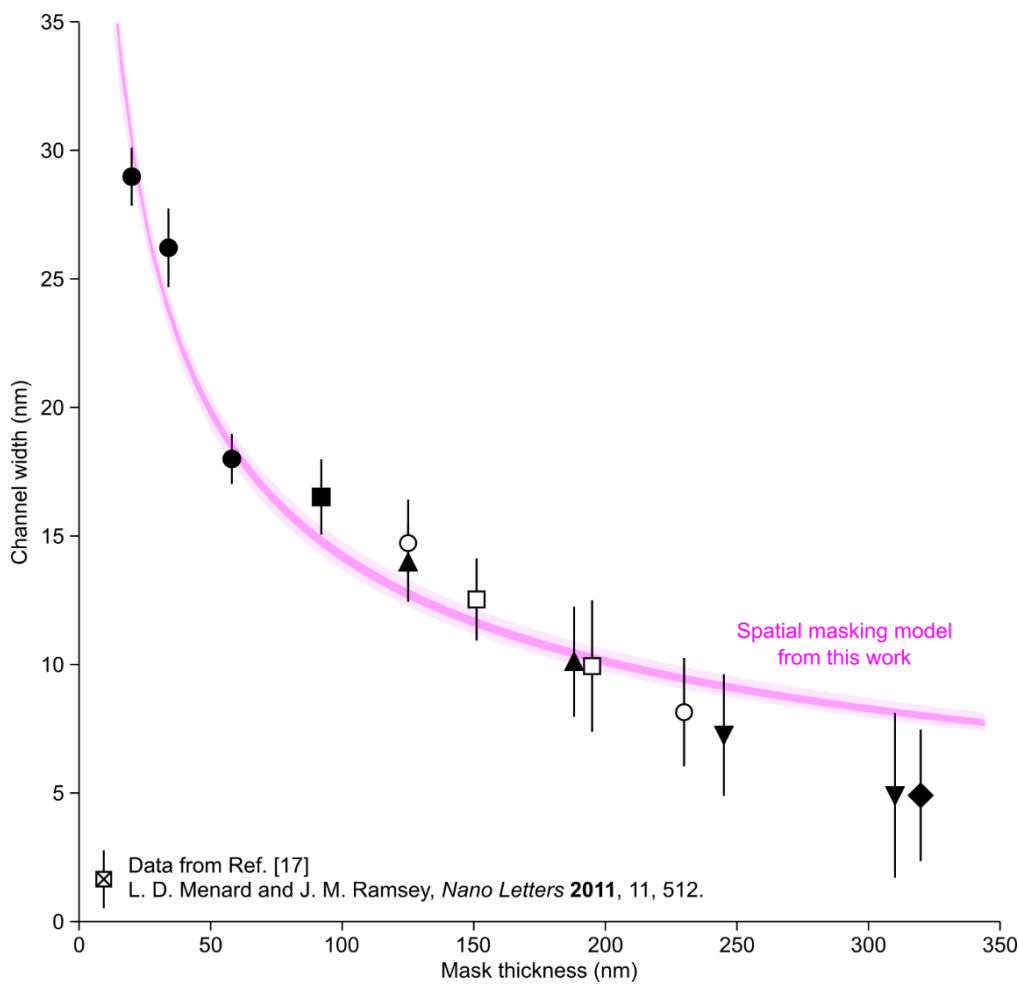

**Figure S14.** Comparison to a previous study. Plot showing widths of channels as a function of the thickness of sacrificial masks of chromium after removal of the mask. The channels result from the spatial masking of line scans of a focused beam of gallium ions, milling through sacrificial chromium masks and into underlying quartz substrates. The data markers show the experimental results from Menard and Ramsey.[17] Using approximate values of experimental parameters from correspondence with these scholars, we bound the values of the effective standard deviation of the ion beam to the range from 20 to 60 nm, the milling depth parameter to the range from 0 to 20 nm, and the value of the physical selectivity parameter to the range from 0 to 3, and we fit our spatial masking model of a line scan in Equation (S17). This fit yields a reduced chi-square statistic, $\chi^2_\nu$, of 5.3, extracts reasonable values of experimental parameters including standard deviation of the ion-beam profile, milling depth, and physical selectivity, and indicates that our spatial masking model is generally applicable beyond the specific system in our current study. The light and dark magenta regions indicate uncertainties of mean channel width respectively as 95% coverage intervals and interquartile ranges of fit results to our spatial masking model. This analysis yields estimates of experimental parameters, including an effective standard deviation of the ion beam of $46\,^{+14\,\text{nm}}_{-15\,\text{nm}}$, a milling depth of $7.8\,^{+3.9\,\text{nm}}_{-4.2\,\text{nm}}$, and a physical selectivity of quartz to chromium of $1.5\,^{+1.0}_{-0.6}$. Uncertainties of these parameters and of the data markers are 95% coverage intervals.

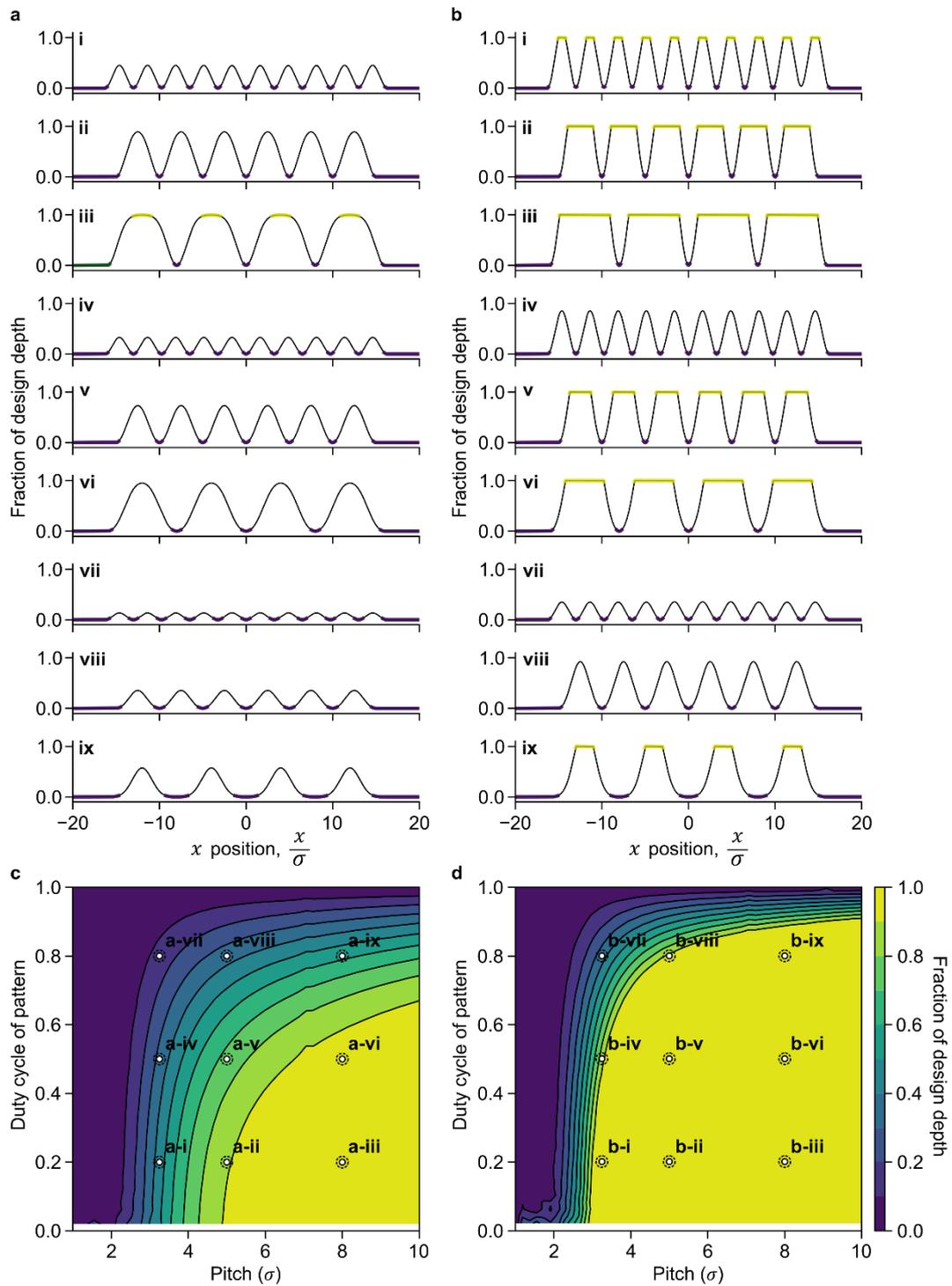

**Figure S15**. Line–space patterns. Plots showing theoretical profiles of lines and spaces that result from a) direct milling of silica and b) milling of silica through a sacrificial mask of chromia for nine combinations of pitch, ranging from 3.25$\sigma$ to 8$\sigma$, and duty cycle, ranging from 20% to 80%. Purple regions indicate the floor of the profile. Yellow regions indicate segments of profiles that exceed a tolerance threshold of 0.975. The profiles in (a) are ideal cases which neglect the dependence of incidence angle on milling rate, any effects of redeposition, and any defocus of the ion-beam due to charging. c) Contour plot showing the fraction of design depths for the set of profiles in panels (a-i–a-ix) that result from the direct milling of silica. d) Contour plot showing the fraction of design depth for the set of profiles in panels (b-i–b-ix) that result from milling of silica through a sacrificial mask of chromia. Our selection of nine combinations of pitch and duty cycle highlights the transition regions in both contour plots. In each contour plot, a sampling artifact, which is inconsequential to our analysis, is aparent for values of pitch of approximately 7$\sigma$.

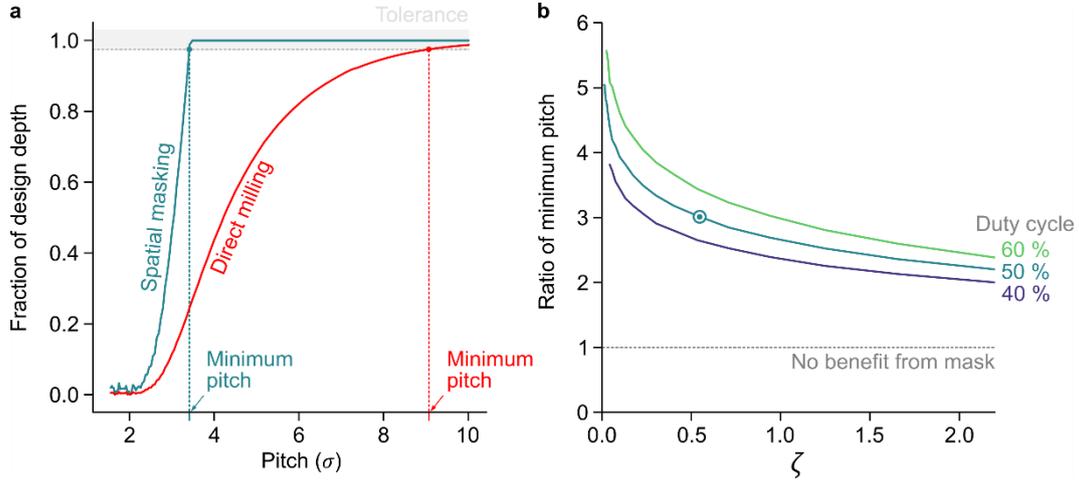

**Figure S16**. Minimum pitch ratio. a) Plot showing fraction of design depth as a function of pitch at a duty cycle of 50% and a design depth of $\zeta \approx 0.5$ for (solid green line) milling of silica with a sacrificial mask of chromia and (solid red line) direct milling of silica. The trends in (a) represent ideal cases that neglect the dependence of incidence angle on milling rate, any effects of redeposition, and any defocus of the ion-beam due to charging. The gray dash line indicates a tolerance threshold of 0.975. The green and red vertical dash lines respectively indicate a minimum pitch for spatial masking of approximately $3\sigma$ and a minimum pitch for direct milling of approximately $3\sigma$, yielding a ratio of minimum pitch of 3. b) Plot showing ratios of minimum pitch as a function of milling depth for duty cycles ranging from 40% to 60%. The green roundel corresponds to the simulation result in (a).

**Note S8**. Temporal efficiency

The time that is necessary to mill a certain nanostructure is typically inversely proportional to ion-beam current. Consequently, simply reducing the ion-beam current to improve lateral patterning resolution can dramatically increase the milling time. Furthermore, multicurrent processes involving a high current to mill coarse structures and a low current to mill fine structures increase the complexity of the process, requiring realignment after switching between the two ion beams, and can still require long milling times, depending on constraints of lateral resolution. These issues motivate our study of the temporal efficiency of sacrificial masking of a focused ion beam.

We derive an analytic expression for temporal efficiency, defining the theoretical condition for which the use of a sacrificial mask and a high value of ion-beam current is faster than milling a structure with similar edge width directly into the substrate by use of a lower value of ion-beam current. We assume that the nominal radius of the ion-beam follows a power law approximation, $r_\text{beam} \cong \alpha I^\beta$, where $\alpha$ is a constant, $I$ is the ion-beam current, and the scaling exponent, $\beta$, typically ranges from 0.3 to 1 for ion-beam currents of less than 10 nA.[18] In the absence of a sacrificial mask, the lateral resolution is, $\mathcal{R} \cong r_\text{beam}$. The presence of a sacrificial mask improves the lateral resolution by a multiplicative super-resolution factor, $\mathcal{F}_\text{SR}$, where $\zeta = |z_s| z_m^{-1}$ is the ratio of the depth of the nanostructure, $z_s$ to the thickness of the mask, $z_m$, and $\mathcal{S} = \bar{m}_s \bar{m}_m^{-1}$ is the physical selectivity of the substrate and the mask, which we define as the ratio of their average milling rates, $\bar{m}_m$ and $\bar{m}_s$, respectively. We consider an equality of lateral resolution from a low value of ion-beam current, $I_\text{low}$, and lateral super-resolution from a high value of ion-beam current, $I_\text{high}$, milling through a sacrificial mask. Then,

$$\mathcal{R} = \alpha I_\text{low}^\beta = \mathcal{F}_\text{SR}^{-1} \alpha I_\text{high}^\beta, \tag{S18}$$

which implies generally that

$$\mathcal{F}_\text{SR}(\sigma_\text{low}, \sigma_\text{high}, \tilde{z}, \mathcal{S}) = \left(\frac{I_\text{high}}{I_\text{low}}\right)^\beta. \tag{S19}$$

The condition of equivalent edge widths for the high and low values of ion-beam current from Equations (11) and (12) imply that

$$w_\text{low} = w_\text{high} = 4\sigma_\text{low} = \sqrt{2}\sigma_\text{high}\text{erf}^{-1}\left(1 - \frac{2\mathcal{S}}{\mathcal{S}+\zeta}\right) + 2\sigma_\text{high} = \sigma_\text{high}\left[\sqrt{2}\,\text{erfc}^{-1}\left(\frac{2\mathcal{S}}{\mathcal{S}+\zeta}\right) + 2\right], \tag{S20}$$

where we apply the identity, $\text{erf}^{-1}(1-x) = \text{erfc}^{-1}(x)$. The power-law approximation of the radius of the ion beam relates our model of super-resolution factor from Equation (13) to values of ion-beam current that are necessary to achieve equivalent lateral resolution,

$$\left(\frac{I_{\text{high}}}{I_{\text{low}}}\right) = \left(\frac{\sigma_{\text{high}}}{\sigma_{\text{low}}}\right)^{\frac{1}{\beta}} = \left(\frac{2}{\frac{1}{\sqrt{2}}\text{erfc}^{-1}\left(\frac{2\mathcal{S}}{\mathcal{S}+\zeta}\right)+1}\right)^{\frac{1}{\beta}} = \mathcal{F}_{\text{SR}}(\zeta,\mathcal{S})^{\frac{1}{\beta}}. \qquad (S21)$$

The time that is necessary to mill a nanostructure of arbitrary rectangular volume, $V_s$, through the sacrificial mask, $t_M$, with a high value of ion-beam current is the sum of the time that is necessary to mill through the chromia mask, $t_m$, and the time that is necessary to mill the underlying substrate, $t_s$,

$$t_M = t_m + t_s \approx \frac{V_m}{\overline{m}_m I_{\text{high}}} + \frac{V_s}{\overline{m}_s I_{\text{high}}} = \frac{z_m l^2}{\overline{m}_m I_{\text{high}}} + \frac{z_s l^2}{\overline{m}_s I_{\text{high}}}, \qquad (S22)$$

where $l^2$ is the area of the rectangular nanostructure and $V_m$ is the volume of the mask above the milling area. For clarity, $\overline{m}_m$ and have $\overline{m}_s$ units of volume per current per second or µm³ nA⁻¹ s⁻¹. Therefore, in Equation (S22), dividing a volume by the product of a milling rate and ion-beam current yields a value with units of time. In contrast, the time that is necessary to mill a similar nanostructure directly into the substrate with a low value of ion-beam current is

$$t_s \approx \frac{V_s}{\overline{m}_s I_{\text{low}}} = \frac{z_s l^2}{\overline{m}_s I_{\text{low}}}. \qquad (S23)$$

We define the temporal efficiency, $\eta_\tau$, to be the ratio of these milling times,

$$\eta_\tau = \frac{t_s}{t_M} = \frac{\frac{z_s l^2}{\overline{m}_s I_{\text{low}}}}{\frac{z_m l^2}{\overline{m}_m I_{\text{high}}} + \frac{z_s l^2}{\overline{m}_s I_{\text{high}}}} = \left(\frac{I_{\text{high}}}{I_{\text{low}}}\right) \frac{z_s \overline{m}_m}{z_m \overline{m}_s + z_s \overline{m}_m}. \qquad (S24)$$

Substitution of $z_s = \zeta z_m$, $\overline{m}_s = \mathcal{S} \overline{m}_m$, and Equation (S21) into Equation (S24) yields an analytic expression for the temporal efficiency, which we argue must be greater than unity for masking to be beneficial,

$$\eta_\tau = \left(\frac{I_{\text{high}}}{I_{\text{low}}}\right) \frac{\zeta}{\mathcal{S}+\zeta} = \mathcal{F}_{\text{SR}}(\zeta,\mathcal{S})^{\frac{1}{\beta}} \frac{\zeta}{\mathcal{S}+\zeta}. \qquad (S25)$$

**Table S12.** Power-law models

| Tradespace | Model parameters | |
|---|---|---|
| | $\alpha$ (nm (µm³ h⁻¹)⁻ᵝ) | $\beta$ (–) |
| Resolution | 76.4 ± 0.1 | 0.154 ± 0.001 |
| Super-resolution | 16.2 ± 0.1 | 0.314 ± 0.010 |

**Note S9.** Milling currents of equivalent resolution
For two power law models, the condition of equivalent resolution is

$$\mathcal{R}_R = \alpha_R \mathcal{V}_R^{\beta_R} = \mathcal{R}_{SR} = \alpha_{SR} \mathcal{V}_{SR}^{\beta_{SR}}. \qquad (S26)$$

Solving for $\mathcal{V}_{SR}$ and accounting for the two different milling rates and thickness of the bilayer gives the volume throughput in terms of a high value of ion-beam current with equaivalent resolution,

$$\mathcal{V}_{SR} = \left(\frac{\mathcal{R}_{SR}}{\alpha_{SR}}\right)^{\frac{1}{\beta_{SR}}} = I_{\text{high}} \left(\frac{z_m + z_s}{\frac{z_m}{\overline{m}_m} + \frac{z_s}{\overline{m}_s}}\right). \qquad (S27)$$

Applying the right side of the power-law equation (S26) gives

$$\mathcal{V}_{SR} = \left(\frac{\alpha_R}{\alpha_{SR}} \mathcal{V}_R^{\beta_R}\right)^{\frac{1}{\beta_{SR}}} = I_{\text{high}} \left(\frac{z_m + z_s}{\frac{z_m}{\overline{m}_m} + \frac{z_s}{\overline{m}_s}}\right). \qquad (S28)$$

Applying the left side of equation (S26) and solving for $I_{\text{high}}$,

$$I_{\text{high}} = \left(\frac{\alpha_R}{\alpha_{SR}}[m_s I_{\text{low}}]^{\beta_R}\right)^{\frac{1}{\beta_{SR}}} \left(\frac{\frac{z_m}{\overline{m}_m} + \frac{z_s}{\overline{m}_s}}{z_m + z_s}\right). \quad (S29)$$

Conversely,

$$I_{\text{low}} = \frac{1}{\overline{m}_s}\left[\frac{\alpha_{SR}}{\alpha_R}\left(I_{\text{high}} \frac{z_m + z_s}{\frac{z_m}{\overline{m}_m} + \frac{z_s}{\overline{m}_s}}\right)^{\beta_{SR}}\right]^{\frac{1}{\beta_R}}. \quad (S30)$$

**Table S13.** Factors of improvement

| | Before chromia removal | | | After chromia removal | | | Factor of improvement | | |
|---|---|---|---|---|---|---|---|---|---|
| Ion-beam current (pA) | Throughput (μm³ hr⁻¹) | Resolution (nm) | Figure of merit (μm² hr⁻¹) | Equivalent throughput (μm³ hr⁻¹) | Super-resolution (nm) | Figure of merit (μm² hr⁻¹) | Throughput | Resolution | Figure of merit |
| 83 ± 1 | 54.5 ± 0.2 | 144.4 ± 0.2 | 385 ± 2 | 0.10 ± 0.02 | 53.9 ± 0.4 | 1112 ± 2 | 528 ± 28 | 2.678 ± 0.016 | 2.890 ± 0.006 |
| 227 ± 1 | 151.0 ± 0.2 | 160.7 ± 0.2 | 986 ± 2 | 1.25 ± 0.04 | 79.2 ± 0.2 | 2058 ± 4 | 119 ± 4 | 2.029 ± 0.006 | 2.088 ± 0.004 |
| 421 ± 3 | 280.7 ± 0.2 | 180.1 ± 0.2 | 1621 ± 2 | 5.52 ± 0.22 | 99.3 ± 0.6 | 3115 ± 6 | 51 ± 3 | 1.814 ± 0.010 | 1.921 ± 0.004 |
| 796 ± 4 | 529.8 ± 0.4 | 202.5 ± 0.2 | 2773 ± 4 | 12.72 ± 0.22 | 112.9 ± 0.2 | 5103 ± 10 | 42 ± 2 | 1.794 ± 0.004 | 1.840 ± 0.006 |

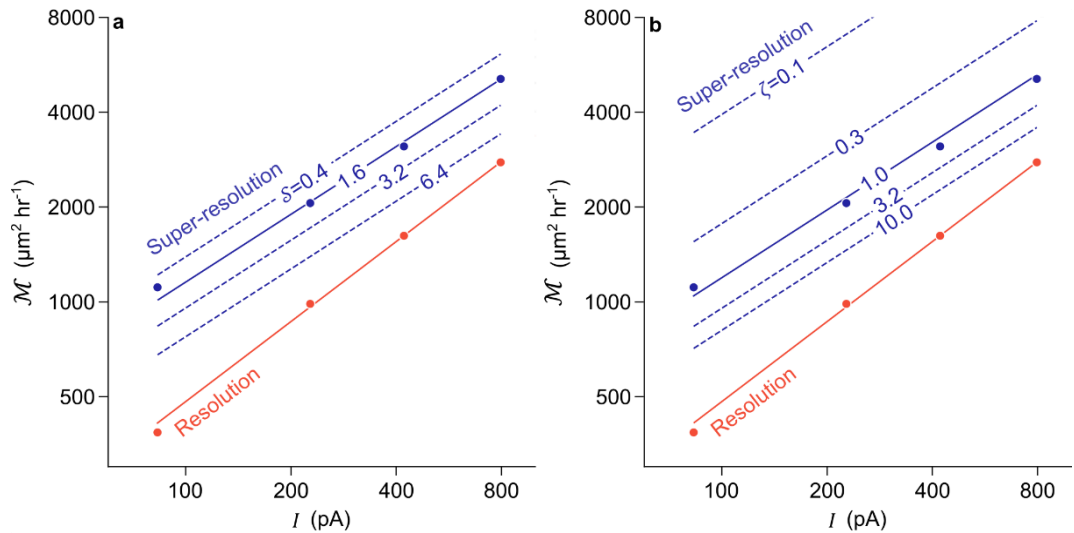

**Figure S17.** Figure of merit. a) Plot showing figure of merit, $\mathcal{M}$, as a function of ion-beam current, $I$, for a constant value of milling depth after normalization by mask thickness of $\zeta = 1$ and values of physical selectivity, $\mathcal{S}$, ranging from 0.4 to 6.4. Solid lines indicate the trends from experimental data and dash lines indicate values that we calculate using the relations in Equations (1) and (5). b) Plot showing figure of merit as a function of ion-beam current for a constant value of $\mathcal{S}$ = 1.6 and values of $\zeta$ ranging from 0.1 to 10.0. Solid lines indicate the trends from experimental data and dash lines indicate values that we calculate using the relation in Equations (1) and (5).

**Note S10**. Design of Fresnel lenses
A plane wave of incident light along the optical axis has a constant phase over the lateral extent of a spherical lens.[19] As a result of diffraction through the lens, the converging wave has the phase retardation distribution,

$$\psi(r) = \frac{2\pi}{\lambda}\left(f - \sqrt{f^2 - r^2}\right). \tag{S31}$$

where $r$ is the radial position orthogonal to the optical axis, $f$ is the focal length of the lens, and $\lambda$ is the wavelength. Fresnel lenses with blazing exploit a sawtooth profile to modify $\psi(r)$ with a modulo $2\pi$ phase structure,

$$\psi_F(r) = \psi(r) + 2m\pi, \tag{S32}$$

for $r_m < r \leq r_{m+1}$ where $r_m = \sqrt{2m\lambda f + (m\lambda)^2}$ is the inner radius of the zone radii.

The phase shift in the converging wave results from differences in the relative optical path length along the lens profile. In the design process, the phase retardation distribution determines the thickness of the Fresnel lens,

$$T(r) = \frac{\lambda}{\Delta n}\left[\frac{\psi_F(r)}{2} + 1\right]. \tag{S33}$$

Fresnel lenses of this design have a numerical aperture of,

$$\mathrm{NA} = \frac{1}{\sqrt{1+\left(\frac{f}{R}\right)^2}} = \sin\theta. \tag{S34}$$

The Rayleigh criterion is,

$$d = 0.61\frac{\lambda}{\mathrm{NA}}. \tag{S35}$$

**Note S11**. Dose trimming
Tests of dose delivery are generally necessary in nanofabrication, and optimization methods can be either computational[15b] or empirical.[20] Our simple model of the bilayer response provides useful insights but ignores many details of the milling process, which would be desirable to account for in a computational optimization of dose delivery. At this state of the art, we develop an empirical method of dose optimization to minimize the differences between design and actual structure. This method involves trimming the ion dose near regions of high negative curvature to improve the masking of features that are sensitive to unintentional ion bombardment. Regions with high aspect ratios and extrema of negative curvature such as pillars, ridges, or blaze peaks are susceptible to such unintentional damage, which is likely to occur if actual and theoretical models of ion-beam profile deviate significantly or if dependences of milling rates on incidence angle are unknown. Reducing the ion dose in regions where the design of the surface specifies a curvature that exceeds a certain threshold can compensate for such discrepancies. The curvature of an axially symmetric surface, $z(r)$, is

$$\kappa(r) = \frac{z''}{(1 + z'^2)^{\frac{3}{2}}}, \tag{S36}$$

where $z' = \frac{\partial z(r)}{\partial r}$ and $z'' = \frac{\partial^2 z(r)}{\partial r^2}$ respectively denote the first and second derivatives of $z(r)$ with respect to $r$. A natural choice for a curvature threshold is the inverse of the lateral extent of the ion-beam profile, which we select to be a value of $-(4\sigma)^{-1}$, corresponding to the maximum negative curvature from the 95% coverage interval of a Gaussian approximation. Figure S18 shows the best result of three tests of trimming the ion dose around the blaze peaks in the milling of a Fresnel lens through chromia and into silica with an ion-beam current of 2600 pA.

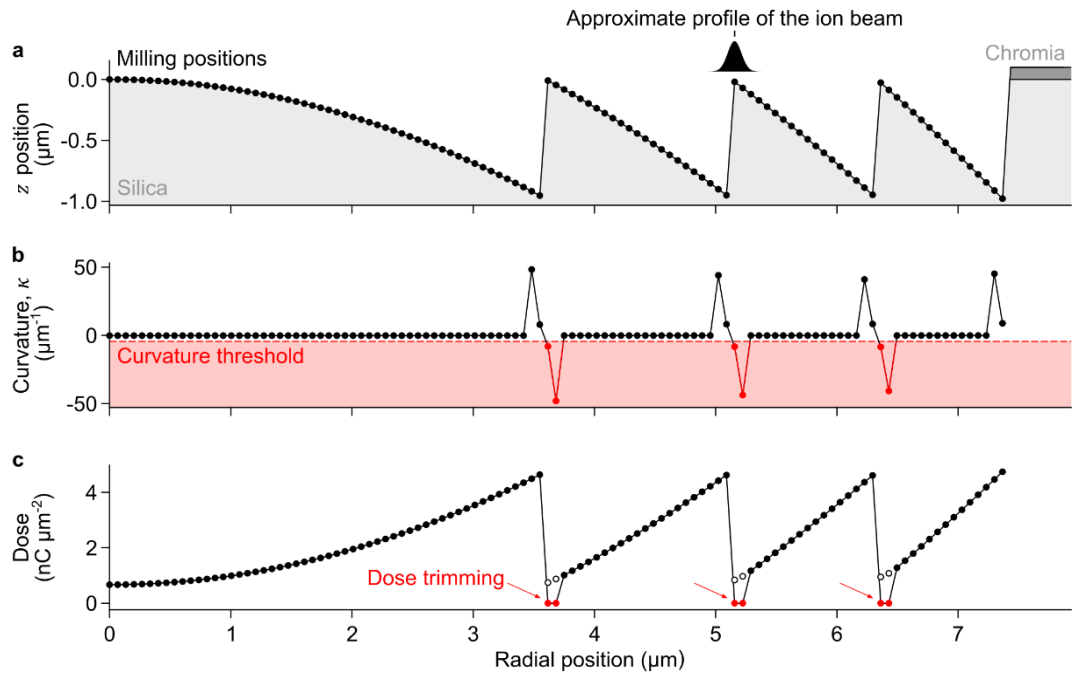

**Figure S18.** Dose trimming. a) Plot showing the radial profile of the design of a Fresnel lens, showing (black circles and solid black line) milling positions and (inset) an approximate profile of a Gaussian ion beam. b) Plot showing the curvature of the profile in (a) and (red dash line) a curvature threshold of $-(4\sigma)^{-1}$. c) Plot showing the radial dose profile resulting from trimming the ion dose at (white circles) milling positions that drop below the curvature threshold.

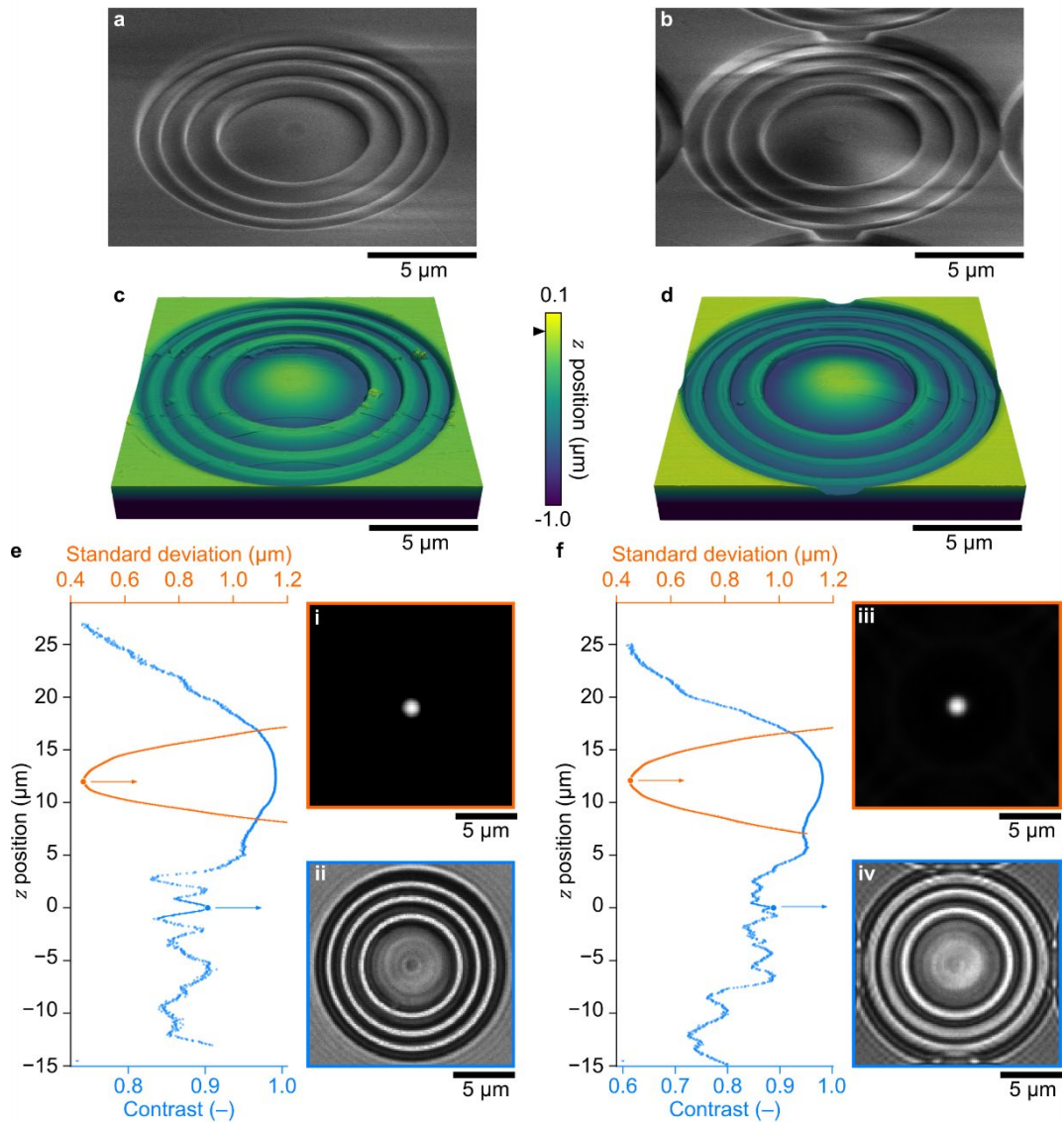

**Figure S19.** Fresnel lens characterization. a) and b) Scanning electron micrographs at a tilt angle of 0.9075 rad (52°) showing Fresnel lenses. c) and d) Atomic force micrographs at the same tilt angle showing the same Fresnel lenses in (a) and (b). e) and f) Plots showing (blue) contrast and (orange) standard deviation of Gaussian model fits to focal spot images from Fresnel lenses as functions of z position. (i-iv) Brightfield optical micrographs showing (i,iii) focal spots and (ii,iv) top surfaces of Fresnel lenses. Panels (a,c,e) correspond to Fresnel lenses that we mill directly into silica with an ion-beam current of 26 pA. Panels (b,d,f) correspond to Fresnel lenses that we mill through chromia and into silica with an ion-beam current of 2600 pA. Artifacts from charging are apparent in (a) and (b). Convolution artifacts of the probe tip are apparent in (c) and (d). The black triangle in (c,d) indicates the zero plane in both atomic-force micrographs and brightfield optical micrographs. Root-mean-square values of surface roughness in the central regions of either lens are approximatly 4 nm. Differences in contrast between (e-ii) and (f-iv) are partially attributable to diffraction effects from multiple lenses in close proximity in (f-iv). Lone blue bars in (e) and (f) indicate representative uncertainties of contrast as 95% coverage intervals. Uncertainties of standard deviation and z position from measurement repeatability are smaller than the data markers.

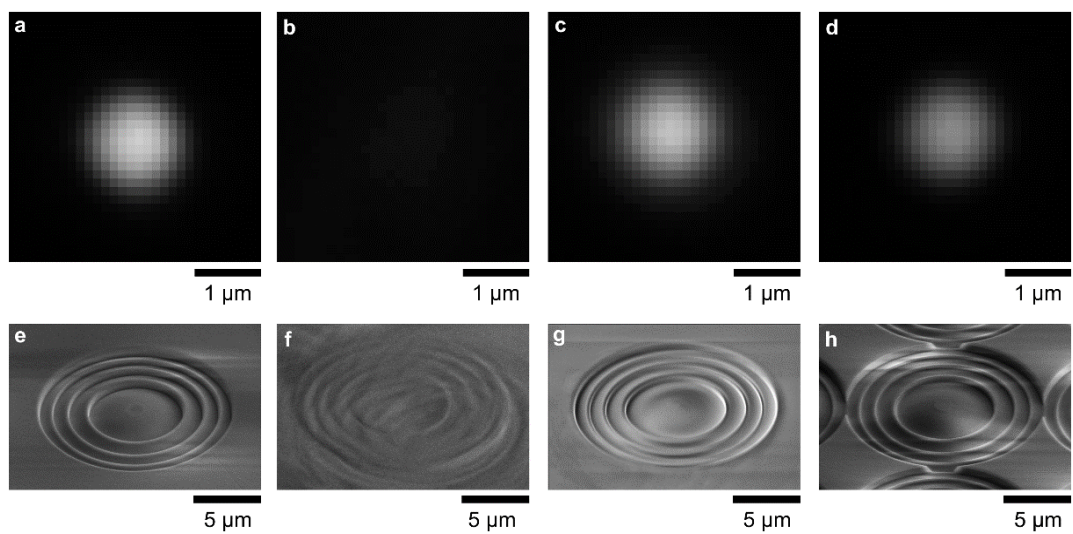

**Figure S20.** Fresnel lens comparison. a-d) Brightfield optical micrographs showing focal spots from Fresnel lenses that we machine under a variety of conditions (Table S14). e-h) Scanning electron micrographs showing the Fresnel lenses that project the focal spots in (a-d). Machining defects from charge accumulation extend beyond the region of interest in (f).

**Table S14.** Fresnel lens parameters

| Material system | Ion-beam current (pA) | FWHM of ion beam (nm) | Electron-beam current (pA) | Number of lenses (–) | Milling time (h) | Lenses per hour (–) | Projection distance (µm) | Apparent standard deviation of focal spot (nm) | Data location (–) |
|---|---|---|---|---|---|---|---|---|---|
| Silica | 26 | 14 | 100 | 1 | 3.75 | 0.27 | 11.97 ± 0.01 | 440.3 ± 0.1 | Fig. S20 a,e |
| Silica | 2600 | 133 | 0 | 1 | 0.04 | 25 | – | – | Fig. S20 b,f |
| Silica | 2600 | 133 | 6400 | 1 | 0.04 | 25 | 12.38 ± 0.40 | 503.6 ± 0.1 | Fig. S20 c,g |
| Chromia on silica | 2600 | 133 | 0 | 75 | 3.75 | 20 | 11.98 ± 0.19 | 439.1 ± 3.9 | Fig. S20 d,h |

FWHM = full width at half maximum
95% coverage intervals for the single lens in silica are measurement uncertainty.
95% coverage intervals for the 75 lenses through chromia in silica are measurement uncertainty and fabrication variability.
Uncertainties in this table are from measurement repeatability only, neglecting systematic effects such as actuator non-linearity and image scale uncertainty. We make the approximation that these systematic effects cancel in a calculation of ratios of distances and standard deviations to determine relative differences with corresponding uncertainties.